\documentclass[a4paper]{article}

\usepackage[utf8]{inputenc}

\usepackage[francais,american]{babel}
\usepackage{amsmath,amstext,amsfonts,amssymb,amsthm,xfrac}
\usepackage{paralist}
\usepackage{url,xspace}
\usepackage[ruled,vlined,fillcomment,noend,linesnumbered]{algorithm2e}
\usepackage{ifthen}
\usepackage{subfig}

\usepackage[shadow]{todonotes}
\usepackage{multirow}
\usepackage{morefloats}
\usepackage{fullpage}

\usepackage{figlatex}
\graphicspath{{fig/}}

\newboolean{isdraft}
\setboolean{isdraft}{true}
\newcommand{\fixme}[1]{\ifthenelse{\boolean{isdraft}}{\textcolor{red}{\textbf{fixme }\textit{#1}}}{}}

\newboolean{withDetails}
\setboolean{withDetails}{false}
\newcommand{\detailedProof}[1]{\ifthenelse{\boolean{withDetails}}{\textcolor{red}{\textit{#1}}}{}}

\newboolean{withEIf}
\setboolean{withEIf}{true}
\newcommand{\exactForm}[1]{\ifthenelse{\boolean{withEIf}}{\textcolor{blue}{\textit{#1}}}{}}
\newcommand{\exactWithDetails}[1]{\ifthenelse{\boolean{withEIf} \and \boolean{withDetails}}{\textcolor{purple}{\textit{#1}}}{}}

\newboolean{withAllFigures}
\setboolean{withAllFigures}{false}
\newcommand{\withMissingFigures}[1]{\ifthenelse{\boolean{withAllFigures}}{{#1}}{}}

\usepackage{tikz-timing}
\usetikztiminglibrary[new={char=Q,reset char=R}]{counters}

\newcommand{\period}{\ensuremath{T}\xspace}

\newcommand{\ttopt}{\ensuremath{T_{\text{opt}}}\xspace}
\newcommand{\Cr}{\ensuremath{C}\xspace}
\newcommand{\Cp}{\ensuremath{C_{p}}\xspace}
\newcommand{\D}{D\xspace}
\newcommand{\R}{R\xspace}
\newcommand{\p}{p\xspace}

\newcommand{\recall}{\ensuremath{r}\xspace}
\newcommand{\precision}{\ensuremath{p}\xspace}
\newcommand{\trust}{\ensuremath{q}\xspace}
\newcommand{\muP}{\ensuremath{\mu_{P}}\xspace}
\newcommand{\muNP}{\ensuremath{\mu_{NP}}\xspace}
\newcommand{\munew}{\ensuremath{\mu_e}\xspace}
\newcommand{\waste}{\ensuremath{\textsc{Waste}}\xspace}

\newcommand{\opt}{\ensuremath{\text{opt}}}
\newcommand{\extr}{\ensuremath{\text{extr}}}

\newcommand{\Tnp}{\ensuremath{T_{\text{R}}}\xspace}
\newcommand{\Tp}{\ensuremath{T_{\text{P}}}\xspace}

\newcommand{\Tlost}[1][]{\ensuremath{T_{\text{lost}#1}}\xspace}
\newcommand{\I}{\ensuremath{I}\xspace}

\newcommand{\EIf}{\ensuremath{\mathbb{E}_{I}^{(f)}}\xspace}

\newcommand{\Wregular}{\ensuremath{W_{\mathit{reg}}}\xspace}
\newcommand{\Time}[1][]{\ensuremath{\textsc{Time}_{\text{#1}}}\xspace}
\newcommand{\Waste}[1][]{\ensuremath{\textsc{Waste}_{\text{#1}}}\xspace}

\newcommand{\Instant}{\textsc{Instant}\xspace}
\newcommand{\Nockpt}{\textsc{NoCkptI}\xspace}
\newcommand{\Withckpt}{\textsc{WithCkptI}\xspace}

\newcommand{\newdaly}{\textsc{RFO}\xspace}
\newcommand{\daly}{\textsc{Daly}\xspace}

\newcommand{\NoPrediction}{\textsc{NoPrediction}\xspace}
\newcommand{\bestper}{\textsc{BestPeriod}\xspace}

\newcommand{\periodicwithNocheckpointduringI}{\Nockpt}
\newcommand{\periodicwithcheckpointduringI}{\Withckpt}
\newcommand{\periodicignoringI}{\Instant}

\newcommand{\bestnoprediction}{\bestper \NoPrediction\xspace}

\newcommand{\bestperiodicwithNocheckpointduringI}{\bestper \Nockpt\xspace}
\newcommand{\bestperiodicwithcheckpointduringI}{\bestper\Withckpt\xspace}
\newcommand{\bestperiodicignoringI}{\bestper\Instant \xspace}

\newcommand{\faultbis}[1]{
\draw[<-, color=red] (#1) -- ($(#1)+(0.2,1.2)$) -- ($(#1)+(0.1,1.4)$) --  ($(#1)+(0.2,2)$) node[above, left] {\scriptsize{fault}};
} 

\newcommand{\legende}[3]{
\draw[thick, <->] ($(#1)+(0,-0.20)$) -- ($(#1)+(#2,-0.20)$) node[below=-0.5pt, midway] {\scriptsize{#3}};
}

\newcommand{\legendelongue}[4]{
\draw[thick,<->] ($(#1)+(0,-0.20)$) -- ($(#1)+(#2,-0.20)$) node[below=-0.5pt, midway] {\scriptsize{#3}};
\draw[draw=none] ($(#1)+(0,-0.80)$) -- ($(#1)+(#2,-0.80)$) node[below=-0pt, midway] {\scriptsize{#4}};
}

\newcommand{\arrowtime}[2]{
\draw[thick, color=black,->] (0,#1) -- (#2,#1) node[below=-0.5pt, ] {\scriptsize{Time}};
}

\newcommand{\predfaultint}[2]{
\draw[thick, color=green,<->] ($(#1)+(0,1.10)$) -- ($(#1)+(#2,1.10)$) node[above=-0.5pt, midway] {\scriptsize{\I}};
\draw[dashed, color=green] ($(#1)$) --  ($(#1)+(0,1.10)$);
\draw[dashed, color=green] ($(#1)+(#2,0)$) --  ($(#1)+(#2,1.10)$);
} 

\newcommand{\promode}[3]{ 
\draw[thick, color=blue] ($(#1,#3)$) --  ($(#1,#3-1.9)$);
\draw[thick, color=blue] ($(#2,#3)$) --  ($(#2,#3-1.9)$);
\draw[draw=none] ($(#1 ,#3-1.7)$) -- ($(#2,#3-1.7)$) node[midway,color=blue] {\scriptsize{Proactive mode}};
}

\newcommand{\regmode}[3]{
\draw[thick, color=blue] ($(#1,#3)$) --  ($(#1,#3-1.9)$);
\draw[thick, color=blue] ($(#2,#3)$) --  ($(#2,#3-1.9)$);
\draw[draw=none] ($(#1 ,#3-1.7)$) -- ($(#2,#3-1.7)$) node[midway,color=blue] {\scriptsize{Regular mode}};
}
\newcommand{\ttrd}{5} 

\title{Checkpointing strategies with prediction windows}

\author{Guillaume Aupy$^{1,2}$,Yves Robert$^{1,2,3}$, Fr\'ed\'eric Vivien$^{2,1}$ and Dounia Zaidouni$^{2,1}$\\
 $1.$ LIP - \'Ecole Normale Sup\'erieure de Lyon\\
 $2.$ Team ROMA - INRIA Rhône-Alpes, France \\
 $3.$ University of Tennessee Knoxville, USA\\
 \url{{Guillaume.Aupy | Yves.Robert | Frederic.Vivien | Dounia.Zaidouni}@ens-lyon.fr}
 }

\begin{document}
\maketitle

\begin{abstract}
This paper deals with the impact of fault prediction techniques on checkpointing strategies.
We suppose that the fault-prediction system provides
prediction windows instead of exact predictions, which dramatically complicates the
analysis of the checkpointing strategies.
We propose a new approach based upon two periodic modes, a regular mode outside prediction windows, 
and a proactive mode inside prediction windows, whenever the size of these windows is large 
enough. We are able to compute the best period for any size of the prediction windows,
thereby deriving the scheduling strategy that minimizes platform waste. In addition, the results 
of this analytical evaluation are nicely corroborated by a comprehensive
set of simulations, which demonstrate the validity of the model and the accuracy of the approach.
\end{abstract}

\section{Introduction}
\label{sec.intro}

In this paper, we assess the impact of fault prediction techniques on checkpointing strategies.
We assume to have jobs executing on a platform subject to faults,
and we let $\mu$ be the mean time between faults (MTBF) of the platform.
In the absence of fault prediction, the standard approach is to take periodic checkpoints, each of
length \Cr, every period of duration \period. In steady-state utilization of the platform,
the value \ttopt of \period that minimizes the expected waste of resource usage due to 
checkpointing is easily approximated as  $\ttopt = \sqrt{2 \mu\Cr }+\Cr$, or 
$\ttopt = \sqrt{2 (\mu +\R)\Cr }+\Cr$ (where \R is the duration of the recovery).
The former expression is the well-known Young's formula~\cite{young74},
while the latter is due to Daly~\cite{daly04}.

Assume now that some fault prediction system is available. Such a system is  characterized by two 
critical parameters, its recall \recall, which is  the fraction of faults that are indeed predicted, and its precision \precision,
which is the fraction of predictions that are correct (i.e., correspond to actual faults).
In the simple case where predictions are exact-date predictions, several recent papers~\cite{GainaruSC12,rr-journal-prediction} 
have independently shown that the optimal checkpointing period becomes $\ttopt = \sqrt{\dfrac{2 \mu\Cr }{1-\recall}}$.
This latter expression is valid only when $\mu$ is large enough and can be seen as an extension of Young's formula where $\mu$ is replaced by $\frac{\mu}{1-\recall}$: faults are replaced by non-predicted faults, 
 and the overhead due to false predictions is negligible. 
A more accurate expression for the optimal checkpointing period is available in~\cite{rr-journal-prediction}.

This paper deals with the realistic case (see~\cite{5958823,LiangZXS07} and Section~\ref{sec.related})
 where the predictor system
does not provide exact dates for predicted events, but instead provides \emph{prediction windows}.
A \emph{prediction window} is a time interval of length \I during 
which the predicted event is likely to happen. Intuitively, one is more at risk during such an interval
than in the absence of any prediction, hence the need to checkpoint more frequently. But with which period?
And what is the size of the prediction window above which it proves worthwhile to use a different (smaller) checkpointing 
period?

The main objective of this paper is to provide a quantitative answer to these questions.
Our key contributions are the following: (i)
The design of several checkpointing policies that account for the different sizes of prediction windows; 
(ii) The analytical characterization of the best policy for each set of parameters; and (iii) The validation of the theoretical results via extensive simulations, for both Exponential and Weibull failure distributions.
It turns out that the analysis of the waste is dramatically more complicated
than when using exact-date predictions~\cite{GainaruSC12,rr-journal-prediction}.

The rest of the paper is organized as follows. First we detail the
framework in Section~\ref{sec.framework}.  In
Section~\ref{sec.intervals} we describe the new checkpointing policies
with prediction windows, and show how to compute the optimal
checkpointing periods that minimize the platform waste.
Section~\ref{sec.simulations} is devoted to simulations.
Section~\ref{sec.related} provides a brief overview of related work.  Finally, we
present concluding remarks in Section~\ref{sec.conclusion}.

\section{Framework}
\label{sec.framework}

\subsection{Checkpointing strategy}
\label{sec.frame.chkpt} 

We consider a \emph{platform} subject to faults. Our work is agnostic
of the granularity of the platform, which may consist either of a
single processor, or of several processors that work concurrently and
use coordinated checkpointing.   \emph{Checkpoints} are taken at regular intervals, or
periods, of length \period. We denote by \Cr the duration of a
checkpoint; by construction, we must enforce that $\Cr \leq \period$.
Useful work is done only during $\period-\Cr$ units of time for every period of length \period,
if no fault occurs. Hence the \emph{waste} due to checkpointing in  a fault-free execution  is $\waste = \frac{\Cr}{\period}$. 
In the following, the  \emph{waste} always denote the fraction of time that the platform is not doing useful work.
 
When a fault strikes the platform, the application is lacking some
resource for a certain period of time of length $\D$, the
\emph{downtime}. The downtime accounts for software rejuvenation
(i.e., rebooting~\cite{875631,1663301}) or for the replacement of the
failed hardware component by a spare one. Then, the application
recovers from the last checkpoint. \R denotes the duration of this
\emph{recovery} time.

\subsection{Fault predictor}

A fault predictor is a mechanism that is able to predict that some faults will take place, 
within some time-interval window. In this paper,
we assume that the predictor is able to generate its predictions early enough so that a \emph{proactive} 
checkpoint can indeed be taken before or during
the event. A first proactive checkpoint will typically be taken just before the beginning of the prediction window,
and possibly several other ones will be taken inside the prediction window, if its size \I is large enough.

 Proactive checkpoints may have a
different length $\Cp$ than regular checkpoints of length $\Cr$. In
fact there are many scenarios. On the one hand, we may well have $\Cp > \Cr$ in scenarios where regular checkpoints are taken at time-steps where the application memory footprint is minimal~\cite{Hong01}; on the contrary, proactive checkpoints are taken according to
predictions that can take place at arbitrary instants. On the other hand,
we may  have $\Cp < \Cr$ in other scenarios~\cite{5542627}, e.g., when the prediction is
localized to a particular resource subset, hence allowing for a
smaller volume of checkpointed data. 
To keep full generality, we deal with two checkpoint sizes in this paper: \Cr for periodic
checkpoints, and \Cp for proactive checkpoints (those taken upon predictions). 

The accuracy of the fault predictor is characterized by two quantities, the \emph{recall}
and the \emph{precision}. The recall \recall is the fraction of faults that are predicted while
the precision \precision is the fraction of fault predictions that are correct.
Traditionally, one defines three types of \emph{events}: (i) \textit{True positive} events are faults that the predictor has been able to predict (let $\textit{True}_P$ be their number); (ii)
\textit{False positive} events are fault predictions that did not materialize as actual faults (let $\textit{False}_P$ be their number);
and (iii)  \textit{False negative} events are faults that were not predicted (let $\textit{False}_N$ be their number).
With these definitions, we have
$\recall = \frac{\textit{True}_P}{\textit{True}_P+\textit{False}_N}$ 
and $\p = \frac{\textit{True}_P}{\textit{True}_P+ \textit{False}_P}$.

In the literature, the \emph{lead time} is the interval between the
date at which the prediction is made available, and the predicted date
of failure (or, more precisely, the beginning of the prediction
window). However, because we do not consider pro-active actions with
different durations (they all have length \Cp), we point out that the distribution of these lead
times is irrelevant to our problem. Indeed, either we have the time to
take a proactive action before the failure strikes or not. Therefore,
if a failure strikes less than $\Cp$ seconds after the prediction is
made available, the prediction was useless.  In other words, predicted
failures that come too early to enable any proactive action should be
classified as unpredicted faults, leading to a smaller value of the
predictor recall and to a shorten prediction window. Therefore, in the
following, we consider, without loss of generality, that all
predictions are made available $\Cp$ seconds before the beginning of
the prediction window. 

\subsection{Fault rates}

The key parameter is $\mu$, the mean
time between faults (MTBF) of the platform.  If the platform is made
of $N$ components whose individual MTBF is $\mu_{\text{ind}}$, then $\mu =
\frac{\mu_{\text{ind}}}{N}$. This result is true regardless of the fault distribution law\cite{rr-journal-prediction}.
In addition to $\mu$, the platform MTBF, 
let $\muP$ be the mean time between predicted events (both true positive and false positive),  and 
let $\muNP$ be the mean time between unpredicted faults (false negative).
Finally, we define the mean time between events as $\munew$ (including all three event types).
The relationships between $\mu$, $\muP$, $\muNP$, and $\munew$ are the following:
\begin{itemize}
\item Rate of unpredicted faults: $\frac{1}{\muNP} = \frac{{1-\recall
    }}{\mu} $, since $1-\recall$ is the fraction of faults that are
  unpredicted;
\item Rate of predicted faults: $ \frac{\recall}{\mu} =
  \frac{\precision}{\muP}$, since $\recall$ is the fraction of faults
  that are predicted, and $\precision$ is the fraction of fault
  predictions that are correct;

\item Rate of events:
  $\frac{1}{\munew}=\frac{1}{\muP}+\frac{1}{\muNP}$, since events are
  either predictions (true or false), or unpredicted faults.
\end{itemize}

\section{Checkpointing strategies}
	\label{sec.intervals}

In this section, we introduce the new checkpointing strategies, and we determine the waste that they induce.
We then proceed to computing the optimal period for each strategy.


\subsection{Description of the different strategies}

We consider the following general scheme:
\begin{compactenum}
\item  While no fault prediction is available, checkpoints are taken
  periodically with period $\period$;
\item When a fault is predicted, we decide whether to take the
  prediction into account or not. This decision is randomly taken:
  with probability \trust, we trust the predictor and take the
  prediction into account, and, with probability $1-\trust$, we ignore
  the
  prediction;
\item If we decide to trust the predictor, we use various strategies,
  depending upon the length \I of the prediction window.\label{enum.strategies}
\end{compactenum}
Before describing the different strategies in the
situation~\eqref{enum.strategies}, we point out that the rationale for
not always trusting the predictor is to avoid taking useless
checkpoints too frequently. Intuitively, the precision $\precision$ of
the predictor must be above a given threshold for its usage to be
worthwhile.  In other words, if we decide to checkpoint just before a
predicted event, either we will save time by avoiding a costly
re-execution if the event does correspond to an actual fault, or we
will lose time by unduly performing an extra checkpoint.  We need a
larger proportion of the former cases, i.e., a good precision, for the
predictor to be really useful.  

Now, to describe the strategies used when we trust a prediction (situation (3)), we define two \emph{modes} for the scheduling algorithm:\\
\textbf{Regular}: This is the mode used when no fault prediction is available, 
or when a prediction is available but we decide to ignore it (with probability $1-\trust$). 
In regular mode, we use periodic checkpointing with period \Tnp. 
Intuitively, \Tnp corresponds to the checkpointing period \period of Section~\ref{sec.frame.chkpt}.\\
\textbf{Proactive}: This is the mode used when a fault prediction is available and
we decide to trust it, a decision taken with probability \trust.  Consider 
such a trusted prediction made with the prediction window $[t_0,t_0+\I]$.
Several strategies can be envisioned:\\
(1) \Instant, for \emph {Instantaneous--} The first strategy is to ignore the time-window and to execute the same algorithm
      as if the predictor had given an exact date prediction at time $t_{0}$. 
      The algorithm interrupts the current period (of scheduled length \Tnp),
      checkpoints during the interval $[t_{0}-\Cp,t_{0}]$, and then returns to regular mode: at time 
      $t_{0}$, it resumes the work needed to complete the interrupted period of the regular mode.\\
(2) \Nockpt, for \emph{No checkpoint during prediction window--} 
       The second strategy is intended for a short prediction window: instead of ignoring it,
       we acknowledge it, but make the decision not to checkpoint during it.
       As in the first strategy, the algorithm interrupts the current period (of scheduled length \Tnp),
      and checkpoints during the interval $[t_{0}-\Cp,t_{0}]$. But here, we return to regular mode
       only at time $t_0+\I$, where we resume the work needed to complete the interrupted period of the regular mode.
       During the whole length of the time-window, we execute work without checkpointing, at the risk
       of losing work if a fault indeed strikes. But for a small value of \I, it may not be
       worthwhile to checkpoint during the prediction window (if at all possible, since there is no choice if $\I < \Cp$).\\
(3) \Withckpt, for \emph{With checkpoints during prediction window--} 
       The third strategy is intended for a longer prediction window and assumes that $\Cp \leq \I$:
       the algorithm interrupts the current period (of scheduled length \Tnp),
      and checkpoints during the interval $[t_{0}-\Cp,t_{0}]$, but now also decides
       to take several checkpoints during the prediction window. 
       The period \Tp of these checkpoints in proactive mode
       will presumably be shorter than \Tnp, to take into account the higher fault probability.
       In the following, we analytically compute the optimal
       number of such periods. But we take at least one period here, hence one checkpoint, which implies $\Cp \leq I$.
       We return to regular mode either right after the fault strikes within the time window 
       $[t_0,t_0+\I]$, or at time $t_0+\I$ if no actual fault happens within this window. 
       Then, we resume the work needed to complete the interrupted period of the regular mode.
    The third strategy is the most complex to describe, and the
complete behavior of the corresponding scheduling algorithm is shown in Algorithm~\ref{algo.proactive}.

Note that, for all strategies, we insert some additional work for the particular case where there is not 
enough time to take a checkpoint before entering proactive mode (because a checkpoint for the regular mode is currently on-going). 
We account for this work as idle time in the expression of the 
waste, to ease the analysis. Our expression of the waste is thus an upper bound.

\begin{algorithm}
  \If{fault happens}{After downtime, execute recovery\;
    Enter \emph{regular} mode\;}
  \If{in \emph{proactive} mode for a time greater than or equal to \I\label{algo.proactive.Ilimit}}{Switch
  to \emph{regular} mode}
  \If{Prediction made with interval $[t, t+I]$ \textbf{and} prediction
    taken into account}{%
    Let $t_\Cr$ be the date of the last checkpoint under
    \emph{regular} mode to start no later than $t-\Cp$\;
    \If({(enough time for an extra checkpoint)}) {$t_\Cr+\Cr < t-\Cp$}{Take a checkpoint starting at   
     time $t-\Cp$\label{algo.proactive.addC}}
    \Else ({(no time for the extra checkpoint)}){
      Work in the time interval $[t_\Cr+\Cr, t]$\label{algo.proactive.wait}
    }
    $\Wregular \leftarrow \max \left (0, t-\Cp - (t_\Cr + \Cr ) \right )$ \label{algo.proactive.wreg}\;
    Switch to \emph{proactive} mode at time $t$\;
  }
  \While{in \emph{regular} mode and no predictions are made and no faults
    happen}{
    Work for a time \Tnp-\Wregular-\Cr and then checkpoint\label{algo.proactive.completion}\;
    $\Wregular \leftarrow 0$\;
  }
  \While{in \emph{proactive} mode and no faults happen}{
    Work for a time \Tp-\Cp and then checkpoint\;
  }
\caption{\Withckpt.\label{algo.proactive}}
\end{algorithm}

\begin{figure*}
\centering
\scalebox{0.9}{\input{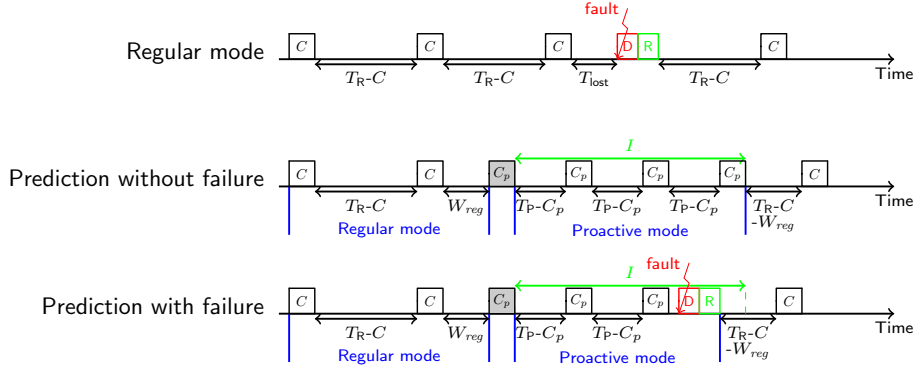}}
	\caption{Outline of Algorithm~\ref{algo.proactive} (strategy \Withckpt). }
\end{figure*}

\subsection{Strategy \Withckpt}
\label{sec.time-withckpt}
\label{sec-opt-int}

In this section we evaluate the execution time under heuristic
\Withckpt. To do so, we partition the whole execution into time
intervals defined by the presence or absence of events.  An interval
starts and ends with either the completion of a checkpoint or of a
recovery (after a failure). To ease the analysis, we make a
simplifying hypothesis: we assume that \emph{at most} one event, failure or
prediction,  occurs within any interval of length
$\Tnp+\I+\Cp$.  In particular, this implies that a prediction or an
unpredicted fault always take place during the
regular mode. 

  We list below the four types of
intervals, and evaluate their respective average length, together with the average
work completed during each of them (see Table~\ref{tab.summary.modes}
for a summary):
\begin{compactenum}
\item \textbf{Two consecutive regular checkpoints with no intermediate
    events.} The time elapsed between the completion of the two
  checkpoints is exactly \Tnp, and the work done is exactly
  $\Tnp-\Cr$.
\item \textbf{Unpredicted fault.} Recall that, because of the simplifying
  hypothesis, the fault happens in regular mode. Because instants where the
  fault strikes and where the last checkpoint was taken are independent, 
  on average the fault strikes at time $\Tnp/2$. A
  downtime of length $\D$ and a recovery of length $\R$ occur before the
  interval completes. There is no work done.
\item \textbf{False prediction.} Recall that it happens in regular mode. There are two cases:
  \begin{compactenum}
  \item \textbf{Taken into account.} This happens with probability
    $\trust$. The interval lasts $\Tnp+\Cp+\I$, since we take a proactive
    checkpoint and spend the time $\I$ in proactive mode. The work
    done is $(\Tnp-\Cr)+(\I-\frac{\I}{\Tp}\Cp)$.
  \item \textbf{Not taken into account.} This happens with probability
    $1-\trust$. The interval lasts $\Tnp$ and the work done is $\Tnp-\Cr$.
  \end{compactenum}

  Considering both cases with their probabilities, the average time
  spent is equal to: $\trust(\Tnp+\Cp+\I) + (1-\trust)\Tnp=
  \Tnp+\trust(\Cp+\I)$. The average work done is: $\trust
  (\Tnp-\Cr+\I-\frac{\I}{\Tp}\Cp) + (1-\trust)(\Tnp-Cr) =
  \Tnp-\Cr+\trust(\I-\frac{\I}{\Tp}\Cp)$.

\item \textbf{True prediction.} Recall that it happens in regular mode. There are two cases:
  \begin{compactenum}
  \item \textbf{Taken into account.} Let \EIf be the average time at which a
    fault occurs within the prediction window (the time at
    which the fault strikes is certainly correlated to the starting
    time of the prediction window; \EIf may not be equal to $\I/2$).
    Up to time \EIf, we work and checkpoint in proactive mode, with period \Tp. In addition, 
    we take a proactive checkpoint right before the start
    of the prediction window. Then we spend the time \EIf in proactive
    mode, and we have a downtime and a recovery. Hence, such an
    interval lasts $\Tnp+\Cp+\EIf+\D+\R$ on average. The total work
    done during the interval is $\Tnp-\Cr+ x (\Tp-\Cp)$ where $x$ is the
    expectation of the number of proactive checkpoints successfully taken
    during the prediction window. Here, $x
    \approx \frac{\EIf}{\Tp}-1$. 
  \item \textbf{Not taken into account.}  On average the fault occurs
    at time $\Tnp/2$. The time interval has duration
    $\Tnp/2+\D+\R$, and there is no work done.
  \end{compactenum}
  Overall the time spent is $\trust(\Tnp+\Cp+ \EIf+\D+\R) + (1-\trust)
  (\Tnp/2+\D+\R)$, and the work done is $\trust(\Tnp-\Cr+
  (\frac{\EIf}{\Tp}-1) (\Tp-\Cp)) + (1-\trust) 0$.
\end{compactenum}
So far, we have evaluated the length, and the work done, for each of
the interval types. We now estimate the expectation of the number of
intervals of each type. Consider the intervals defined by an event
whose mean time between occurrences is $\lambda$. On average, during a time
$T$, there will be $T/\lambda$ such intervals. Due to the
simplifying hypothesis, intervals of different types never overlap. Table~\ref{tab.summary.modes}
presents
the estimation of the number of intervals of each type.

\begin{table}
  \centering
  \resizebox{\textwidth}{!}{%
    \begin{tabular}{c|c|c|c}
      Mode & Number of intervals & Time spent & Work done \\
      \hline
      (1) & $w_1$ & \Tnp & $\Tnp-\Cr$ \\[1mm]
      (2) & $w_2=\frac{\Time[Final]}{\muNP}$ & $\Tnp/2 + \D + \R$ & 0 \\[1mm]
      {(3)} &  {$w_3=\frac{(1-\precision)\Time[Final]}{\muP}$} &$\Tnp + \trust (\I + \Cp) $ & $\Tnp-\Cr + \trust (\I - \frac{\I}{\Tp}\Cp)$ \\ 
      \multirow{2}{*}{(4)} &\multirow{2}{*}{$w_4=\frac{\precision\Time[Final]}{\muP}$} &$\trust (\Tnp + \EIf + \Cp) $& \multirow{2}{*}{$\trust \left(\Tnp-\Cr+\left(\frac{\EIf}{\Tp}-1\right) (\Tp-\Cp)\right) $}  \\ &&$+ (1-\trust)\Tnp/2 +\D + \R$&
    \end{tabular}}
  \caption{Summary of the different types of interval for \Withckpt.}
  \label{tab.summary.modes}
\end{table}

We want to estimate the total execution time. To estimate the
time spent within intervals of a given type, we multiply the
expectation of the number of intervals of that type by the expectation of
the time spent in each of them. Of course, multiplying
expectations is correct only if the corresponding random variables are independent.
Nevertheless, we hope that this will lead us to
a good approximation of the expected execution time. We will assess the
quality of the approximation through simulations in
Section~\ref{sec.simulations}. With our assumptions we have:
\begin{multline}\label{eq.timefinal}
  \Time[Final] = w_1 \times \Tnp + 
  w_2 \left(\frac{\Tnp}{2} + \D + \R\right) + 
  w_3\left( \Tnp +\trust (\I + \Cp) \right)\\ +
  w_4\left(\trust (\Tnp + \EIf +
  \Cp) + (1-\trust)\frac{\Tnp}{2} +\D + \R\right)
\end{multline}
We use the same line of reasoning to compute the overall amount of
work done, that must be equal, by definition, to $\Time[base]$, the execution time of the application
without any overhead:
\begin{multline}\label{eq.timebase}
  \Time[base] = w_1 (\Tnp-\Cr) + w_2\times 0 + w_3 \left (\Tnp-\Cr + \trust\left(\I - \frac{\I}{\Tp}\Cp\right)\right)  \\
  + w_4 \left (\trust\left(\Tnp-\Cr+\left(\frac{\EIf}{\Tp}-1\right)(\Tp- \Cp)\right) \right)
\end{multline}
This equation gives the value of $w_1$ as a function of the other
parameters. Looking at Equations~\eqref{eq.timefinal} and
\eqref{eq.timebase}, and at the values of $w_2$, $w_3$, and $w_4$, we
remark that \Time[Final] can be rewritten as a function of \trust, as
follows: $\Time[Final] = \alpha\Time[base] + \beta \Time[Final] +
\trust \gamma \Time[Final]$, that is $\Time[Final] =
\frac{\alpha}{1-\beta-\trust \gamma}\Time[base]$, where neither
$\alpha$, nor $\beta$, nor $\gamma$ depend on \trust.
With a simple differentiation of $\Time[Final]$ with respect to
\trust, we obtain that $\Time[Final]$ is either increasing or
decreasing with \trust, depending on the sign of $\gamma$.
Consequently, in an optimal solution, either $\trust=0$ or
$\trust=1$. This (somewhat unexpected) conclusion is that the
predictor should sometimes be always trusted, and sometimes never, but
no in-between value for \trust will do a better job. Thus we can now
focus on the two functions \Time[Final], the one when $\trust=0$
($\Time[Final]^{\{0\}}$), and the one when $\trust=1$
($\Time[Final]^{\{1\}}$).

From Table~\ref{tab.summary.modes} and Equations~\eqref{eq.timefinal}
and \eqref{eq.timebase},
one can easily see that
\[\Time[Final]^{\{0\}} = \frac{\Tnp}{\Tnp-\Cr}\Time[base] +
\frac{\Time[Final]^{\{0\}}}{\mu}\left ( \frac{\Tnp}{2} + \D + \R\right), \text{
  i.e., that}\]
\begin{equation}
\left (1-\frac{\Cr}{\Tnp}\right)\left ( 1- \frac{\Tnp/2 + \D + \R}{\mu}\right )\Time[Final]^{\{0\}} = \Time[base]
\label{eq.easycase}
\end{equation}
This is exactly the equation from~\cite{rr-journal-prediction} in the case of exact-date predictions that are never taken into account
(a good sanity check!).  When $\trust=1$, we have:
\begin{multline*}
\Time[Final]^{\{1\}} = \Time[base]\frac{\Tnp}{\Tnp-\Cr}\\
-  \frac{\Time[Final]^{\{1\}}}{\muP} \frac{\Tnp}{\Tnp\!-\!\Cr}\left(
  \!\left(\Tnp\!-\!\Cr\right) 
  \!+\! (1\!-\!\precision)\left(\I \!-\!  \frac{\I}{\Tp}\Cp\right)
  \!+\!\precision\left(\frac{\EIf}{\Tp}\!-\!1\right) (\Tp\!-\!\Cp)
\right)\\
 + \frac{\Time[Final]^{\{1\}}}{\muNP} \left ( \frac{\Tnp}{2} + \D + \R \right ) 
  + \frac{(1-\precision)\Time[Final]^{\{1\}}}{\muP} \left ( \Tnp + \I + \Cp \right ) \\
  + \frac{\precision\Time[Final]^{\{1\}}}{\muP} \left ( \Tnp+ \Cp + \EIf + \D + \R  \right ) 
\end{multline*}

  \exactWithDetails{
  \begin{multline*}
    \Time[Final]^{\{1\}} = \Time[base]\frac{\Tnp}{\Tnp-\Cr}\\
    -  \frac{\Time[Final]^{\{1\}}}{\muP} \frac{\Tnp}{\Tnp\!-\!\Cr}\left(
      \!\left(\Tnp\!-\!\Cr\right) 
      \!+\! (1\!-\!\precision)\left(\I \!-\!  \frac{\I}{\Tp}\Cp\right)
      \!+\!\precision\left(\frac{\EIf}{\Tp}\!-\!1\right) (\Tp\!-\!\Cp)
    \right)\\
    + \frac{\Time[Final]^{\{1\}}}{\muNP} \left ( \frac{\Tnp}{2} + \D + \R \right ) 
    + \frac{(1-\precision)\Time[Final]^{\{1\}}}{\muP} \left ( \Tnp + \I + \Cp \right ) \\
    + \frac{\precision\Time[Final]^{\{1\}}}{\muP} \left ( \Tnp+ \Cp + \EIf + \D + \R  \right ) 
  \end{multline*}
  \begin{multline*}
    \Time[Final]^{\{1\}} = \Time[base]\frac{\Tnp}{\Tnp-\Cr}\\
    -  \frac{\recall\Time[Final]^{\{1\}}}{\precision\mu} \frac{\Tnp}{\Tnp\!-\!\Cr}\left(
      \!\left(\Tnp\!-\!\Cr\right) 
      \!+\! (1\!-\!\precision)\left(\I \!-\!  \frac{\I}{\Tp}\Cp\right)
      \!+\!\precision\left(\frac{\EIf}{\Tp}\!-\!1\right) (\Tp\!-\!\Cp)
    \right)\\
    + \frac{(1-\recall)\Time[Final]^{\{1\}}}{\mu} \left ( \frac{\Tnp}{2} + \D + \R \right ) 
    + \frac{(1-\precision)\recall\Time[Final]^{\{1\}}}{\precision\mu} \left ( \Tnp + \I + \Cp \right ) \\
    + \frac{\precision\recall\Time[Final]^{\{1\}}}{\precision\mu} \left ( \Tnp+ \Cp + \EIf + \D + \R  \right ) 
  \end{multline*}
  \begin{multline*}
    \Time[Final]^{\{1\}} = \Time[base]\frac{\Tnp}{\Tnp-\Cr}\\
    -  \frac{\recall\Time[Final]^{\{1\}}}{\precision\mu} \frac{\Tnp}{\Tnp\!-\!\Cr}\left(
      \!\left(\Tnp\!-\!\Cr\right) 
      \!+\! (1\!-\!\precision)\left(\I \!-\!  \frac{\I}{\Tp}\Cp\right)
      \!+\!\precision\left(\frac{\EIf}{\Tp}\!-\!1\right) (\Tp\!-\!\Cp)
    \right)\\
    + \frac{\Time[Final]^{\{1\}}}{\precision\mu}\left( (1-\recall)\precision\left ( \frac{\Tnp}{2} + \D + \R \right ) 
    + (1-\precision)\recall \left ( \Tnp + \I + \Cp \right ) 
    + \precision\recall \left ( \Tnp+ \Cp + \EIf + \D + \R  \right ) \right ) 
  \end{multline*}
  \begin{multline*}
    \Time[Final]^{\{1\}} = \Time[base]\frac{\Tnp}{\Tnp-\Cr}\\
    -  \frac{\recall\Time[Final]^{\{1\}}}{\precision\mu} \frac{\Tnp}{\Tnp\!-\!\Cr}\left(
      \!\left(\Tnp\!-\!\Cr\right) 
      \!+\! (1\!-\!\precision)\left(\I \!-\!  \frac{\I}{\Tp}\Cp\right)
      \!+\!\precision\left(\frac{\EIf}{\Tp}\!-\!1\right) (\Tp\!-\!\Cp)
    \right)\\
    + \frac{\Time[Final]^{\{1\}}}{\precision\mu}\left(
      \precision  (\D + \R) 
      + \recall(\Tnp+\Cp)
      + (1-\recall)\precision\frac{\Tnp}{2} 
    + (1-\precision)\recall \I 
    + \precision\recall \EIf \right ) 
  \end{multline*}
  \begin{multline*}
    \Time[Final]^{\{1\}} = \Time[base]\frac{\Tnp}{\Tnp-\Cr}\\
    -  \frac{\recall\Time[Final]^{\{1\}}}{\precision\mu} \frac{\Tnp}{\Tnp\!-\!\Cr}\left(
       (1\!-\!\precision)\left(\I \!-\!  \frac{\I}{\Tp}\Cp\right)
      \!+\!\precision\left(\frac{\EIf}{\Tp}\!-\!1\right) (\Tp\!-\!\Cp)
    \right)\\
    + \frac{\Time[Final]^{\{1\}}}{\precision\mu}\left(
      \precision  (\D + \R) 
      + \recall\Cp
      + (1-\recall)\precision\frac{\Tnp}{2} 
    + (1-\precision)\recall \I 
    + \precision\recall \EIf \right ) 
  \end{multline*}
  \begin{multline*}
    \Time[Final]^{\{1\}} = \Time[base]\frac{\Tnp}{\Tnp-\Cr}\\
    -  \frac{\recall\Time[Final]^{\{1\}}}{\precision\mu} \frac{\Tnp}{\Tnp-\Cr}\left(
       (1-\precision)\left(1 -  \frac{\Cp}{\Tp}\right)\I
      +\precision\left(\frac{\EIf}{\Tp}-1\right) \Tp\left(1-\frac{\Cp}{\Tp}\right)
    \right)\\
    + \frac{\Time[Final]^{\{1\}}}{\precision\mu}\left(
      \precision  (\D + \R) 
      + \recall\Cp
      + (1-\recall)\precision\frac{\Tnp}{2} 
    + (1-\precision)\recall \I 
    + \precision\recall \EIf \right ) 
  \end{multline*}
  \begin{multline*}
    \Time[Final]^{\{1\}} = \Time[base]\frac{\Tnp}{\Tnp-\Cr}\\
    -  \frac{\recall\Time[Final]^{\{1\}}}{\precision\mu} \frac{\Tnp}{\Tnp-\Cr}\left(
       \left(1 -  \frac{\Cp}{\Tp}\right)\left((1-\precision)\I
      +\precision\left(\EIf-\Tp\right)
    \right)\right)\\
    + \frac{\Time[Final]^{\{1\}}}{\precision\mu}\left(
      \precision  (\D + \R) 
      + \recall\Cp
      + (1-\recall)\precision\frac{\Tnp}{2} 
    + (1-\precision)\recall \I 
    + \precision\recall \EIf \right ) 
  \end{multline*}
  \begin{multline*}
    \Time[Final]^{\{1\}}\left( 1
      +  \frac{\recall}{\precision\mu} \frac{\Tnp}{\Tnp-\Cr}\left(
        \left(1 -  \frac{\Cp}{\Tp}\right)\left((1-\precision)\I
          +\precision\left(\EIf-\Tp\right)
        \right)\right)\right.\\\left.
      - \frac{1}{\precision\mu}\left(
        \precision  (\D + \R) 
        + \recall\Cp
        + (1-\recall)\precision\frac{\Tnp}{2} 
        + (1-\precision)\recall \I 
        + \precision\recall \EIf \right )
    \right)\\
    = \Time[base]\frac{\Tnp}{\Tnp-\Cr}
  \end{multline*}
  \begin{multline*}
    \Time[Final]^{\{1\}}\left(
      \frac{\recall}{\precision\mu} \left(
        \left(1 -  \frac{\Cp}{\Tp}\right)\left((1-\precision)\I
          +\precision\left(\EIf-\Tp\right)
        \right)\right)\right.\\
    \left.+\frac{\Tnp-\Cr}{\Tnp}\left(1
      - \frac{1}{\precision\mu}\left(
        \precision  (\D + \R) 
        + \recall\Cp
        + (1-\recall)\precision\frac{\Tnp}{2} 
        + (1-\precision)\recall \I 
        + \precision\recall \EIf \right )\right)\right)\\
    = \Time[base]
  \end{multline*}
  \begin{multline*}
     \frac{\Time[base]}{\Time[Final]^{\{1\}}} = \left(
      \frac{\recall}{\precision\mu} \left(
        \left(1 -  \frac{\Cp}{\Tp}\right)\left((1-\precision)\I
          +\precision\left(\EIf-\Tp\right)
        \right)\right)\right.\\
    \left.+\frac{\Tnp-\Cr}{\Tnp}\left(1
      - \frac{1}{\precision\mu}\left(
        \precision  (\D + \R) 
        + \recall\Cp
        + (1-\recall)\precision\frac{\Tnp}{2} 
        + (1-\precision)\recall \I 
        + \precision\recall \EIf \right )\right)\right)  
  \end{multline*}}
\noindent After a little rewriting we obtain:
  \begin{multline*}
     \frac{\Time[base]}{\Time[Final]^{\{1\}}} = 
      \frac{\recall}{\precision\mu} 
        \left(1 -  \frac{\Cp}{\Tp}\right)\left((1-\precision)\I
          +\precision\left(\EIf-\Tp\right)
        \right)\\
    +\!\left(1-\frac{\Cr}{\Tnp}\right)\!\!\left(1
      - \frac{1}{\precision\mu}\!\left(\!
        \precision  (\D + \R) 
        + \recall\Cp
        + (1\!-\!\recall)\precision\frac{\Tnp}{2} 
        + \recall \left(\!(1\!-\!\precision)\I 
        \!+\! \precision \EIf \right )\!\!\right )\!\!\right)
  \end{multline*}
  Finally, the waste is equal by definition to
  $\frac{\Time[Final]-\Time[base]}{\Time[Final]}$. Therefore, we have:
  \begin{multline}\label{eq.waste.q1}
     \Waste = 1- 
      \frac{\recall}{\precision\mu} 
        \left(1 -  \frac{\Cp}{\Tp}\right)\left((1-\precision)\I
          +\precision\left(\EIf-\Tp\right)
        \right)\\
    -\!\left(1\!-\!\frac{\Cr}{\Tnp}\right)\!\!\left(1
      \!-\! \frac{1}{\precision\mu}\!\left(\!
        \precision  (\D + \R) 
        \!+\! \recall\Cp
        \!+\! (1\!-\!\recall)\precision\frac{\Tnp}{2} 
        \!+\! \recall \left(\!(1\!-\!\precision)\I 
        \!+\! \precision \EIf \right )\!\!\right )\!\!\right)
  \end{multline}

\subsubsection*{Waste minimization}

When $\trust=0$, the optimal period can readily be computed from
Equation~\eqref{eq.easycase} and we derive that the optimal period is
$\sqrt{2(\mu -(\D+\R))\Cr}$. This defines a periodic policy we call
\newdaly, for Refined First-Order approximation.  We now minimize the
waste of the strategy where $\trust=1$.  In order to compute the
optimal value for $\Tp$, we identify the fraction of the waste in
Equation~\eqref{eq.waste.q1} that depends on $\Tp$. We can rewrite
Equation~\eqref{eq.waste.q1} as:
\[
\Waste^{\{1\}} = \alpha + \frac{\recall}{\precision\mu}
\left(\left(\left(1-\precision\right)\I+\precision\EIf\right) \frac{\Cp}{\Tp} + \precision\Tp \right )
\]
where $\alpha$ does not depend on $\Tp$. The waste is thus minimized
when $\Tp$ is equal to
$\Tp^{\extr}=\sqrt{\frac{\left( (1-\precision)\I+\precision\EIf\right)\Cp}{\precision}}$.
Note that we always have to enforce that $ \Tp^{\extr}$ is larger than \Cp and does not exceed \I, and we may have to
round its values accordingly in some extreme cases.

In order to compute the optimal value for $\Tnp$, we identify the
fraction of the waste in Equation~\eqref{eq.waste.q1} that depends on
$\Tnp$. We can rewrite Equation~\eqref{eq.waste.q1} as:
\begin{multline}\label{eq.waste.fnofTnp}
\!\!\Waste^{\{1\}} = \beta \!+\!
\frac{\Cr}{\Tnp}\left(1\!-\!\frac{1}{\precision\mu}\left(\!\precision(\D\!+\!\R)
  \!+\!
  \recall\left(\!\Cp\!+\!\left(\!1\!-\!\precision\right)\I\!+\!\precision\EIf\!\right)\!\right)\!\!\right)\!+\!\frac{1\!-\!\recall}{\mu} \frac{\Tnp}{2}\!\!  
\end{multline}
where $\beta$ does not depend on $\Tnp$ because $\Tp^{\opt}$ does not
depend on $\Tnp$. Therefore, $\Waste^{\{1\}}$ is minimized when $\Tnp$
is equal to%
\begin{equation}
\Tnp^{\extr} = 
\sqrt{\frac{2\Cr\left(\precision\mu-\left(\precision(\D+\R)   +
        \recall\left(\Cp+\left(\left(1-\precision\right)\I+\precision\EIf\right)\right)\right)\right)}{\precision(1-\recall)}}\label{eq:Tr.extr}
\end{equation}
Recall that we must always enforce that $\Tnp^{\extr}$ is always greater than \Cr.

One can note that when $\recall=0$, this means that none of
the prediction predicts an actual fault, and we obtain the same period
than without a predictor.  
Finally, if we assume that, on average, fault strikes at the middle of
the prediction window, i.e., $\EIf=\frac{I}{2}$, we obtain simplified values:
$$\Tp^{\extr}=\sqrt{\frac{(2-\precision)\I\Cp}{\precision}} \text{ and }
\Tnp^{\extr} = 
\sqrt{\frac{2\Cr\left(\precision\mu-\left(\precision(\D+\R) +
        \recall\left(\Cp+\left(1-\frac{\precision}{2}\right)\I\right)\right)\right)}{\precision(1-\recall)}}$$

\subsection{Strategy \Nockpt}
\label{sec.time-nockpt}

In this section we evaluate the execution time under heuristic
\Nockpt. The analysis is rather similar to that of \Withckpt, the only
differences being, obviously, in the presence of true and false
predictions:
\begin{compactenum}\setcounter{enumi}{2}
\item \textbf{False prediction.} There are two cases:
  \begin{compactenum}
  \item \textbf{Taken into account.} This happens with probability
    $\trust$. The interval lasts $\Tnp+\Cp+\I$, since we take a
    proactive checkpoint and spend the time $\I$ in proactive mode
    (here, working without checkpointing). The work done is
    $(\Tnp-\Cr)+\I$.
  \item \textbf{Not taken into account.} This happens with probability
    $1-\trust$. The interval lasts $\Tnp$ and the work done is $\Tnp-\Cr$.
  \end{compactenum}

  Considering both cases with their probabilities, the average time
  spent is equal to: $\trust(\Tnp+\Cp+\I) + (1-\trust)\Tnp=
  \Tnp+\trust(\Cp+\I)$. The average work done is: $\trust
  (\Tnp-\Cr+\I) + (1-\trust)(\Tnp-Cr) = \Tnp-\Cr+\trust\I$.

\item \textbf{True prediction.} There are two cases:
  \begin{compactenum}
  \item \textbf{Taken into account.} Let \EIf be the average time at
    which a fault occurs within the prediction window. We take a
    proactive checkpoint right before the start of the prediction
    window. Then we spend the time \EIf in proactive mode working
    without checkpointing, and we have a downtime and a
    recovery. Hence, such an interval lasts $\Tnp+\Cp+\EIf+\D+\R$ on
    average. The total work done during the interval is $\Tnp-\Cr$.
  \item \textbf{Not taken into account.}  On average the fault occurs
    at time $\Tnp/2$. The time interval has duration
    $\Tnp/2+\D+\R$, and there is no work done.
  \end{compactenum}
  Overall the time spent is $\trust(\Tnp+\Cp+ \EIf+\D+\R) + (1-\trust)
  (\Tnp/2+\D+\R)$, and the work done is $\trust(\Tnp-\Cr) +
  (1-\trust) 0$.
\end{compactenum}
So far, we have evaluated the length, and the work done, for each of
the interval types. We now estimate the expectation of the number of
intervals of each type as we did for
\Withckpt. Table~\ref{tab.summary.modes.nockpt} presents the
estimation of the number of intervals of each type.

\begin{table}
  \centering
  {%
    \begin{tabular}{c|c|c|c}
      Mode & Number of intervals & Time spent & Work done \\
      \hline
      (1) & $w_1$ & \Tnp & $\Tnp-\Cr$ \\[1mm]
      (2) & $w_2=\frac{\Time[Final]}{\muNP}$ & $\Tnp/2 + \D + \R$ & 0 \\[1mm]
      {(3)} &  {$w_3=\frac{(1-\precision)\Time[Final]}{\muP}$} &$\Tnp + \trust (\I + \Cp) $ & $\Tnp-\Cr + \trust \I$ \\ 
      \multirow{2}{*}{(4)} &\multirow{2}{*}{$w_4=\frac{\precision\Time[Final]}{\muP}$} &$\trust (\Tnp + \EIf + \Cp) $& \multirow{2}{*}{$\trust \left(\Tnp-\Cr\right)$}  \\ &&$+ (1-\trust)\Tnp/2 +\D + \R$&
    \end{tabular}}
  \caption{Summary of the different types of interval for \Nockpt.}
  \label{tab.summary.modes.nockpt}
\end{table}

We estimate the total execution time as for \Withckpt. The formula is
the exact same function of $w_1$, $w_2$, $w_3$, and $w_4$ (but the
values of there four parameters will change as the average work done during some
of the types of intervals changes):
\begin{multline}\label{eq.timefinal.nockpt}
  \Time[Final] = w_1 \times \Tnp + w_2 \left(\frac{\Tnp}{2} + \D +
    \R\right) + w_3\left( \Tnp +\trust (\I + \Cp) \right)\\ +
  w_4\left(\trust (\Tnp + \EIf + \Cp) + (1-\trust)\frac{\Tnp}{2} +\D +
    \R\right)
\end{multline}
We use the same line of reasoning as previously to compute the overall amount of
work done:
\begin{multline}\label{eq.timebase.nockpt}
  \Time[base] = w_1 (\Tnp-\Cr) + w_2\times 0 + w_3 \left (\Tnp-\Cr + \trust\I\right)
  + w_4 \left (\trust\left(\Tnp-\Cr \right) \right)
\end{multline}
This equation gives the value of $w_1$ as a function of the other
parameters. As for \Withckpt, one can easily show that in an optimal
solution, either $\trust=0$ or $\trust=1$. Thus we can now focus on
the two functions \Time[Final], the one when $\trust=0$
($\Time[Final]^{\{0\}}$), and the one when $\trust=1$
($\Time[Final]^{\{1\}}$).

From Table~\ref{tab.summary.modes.nockpt} and
Equations~\eqref{eq.timefinal.nockpt} and \eqref{eq.timebase.nockpt}, one can easily see that
\[\Time[Final]^{\{0\}} = \frac{\Tnp}{\Tnp-\Cr}\Time[base] +
\frac{\Time[Final]^{\{0\}}}{\mu}\left ( \frac{\Tnp}{2} + \D + \R\right), \text{
  i.e., that}\]
\begin{equation}
\left (1-\frac{\Cr}{\Tnp}\right)\left ( 1- \frac{\Tnp/2 + \D + \R}{\mu}\right )\Time[Final]^{\{0\}} = \Time[base]
\label{eq.easycase.nockpt}
\end{equation}
This is exactly the equation from~\cite{rr-journal-prediction} in the
case of exact-date predictions that are never taken into account, what
we had already retrieved with \Withckpt (same sanity check!).  When
$\trust=1$, we have:
\begin{multline*}
\Time[Final]^{\{1\}} = \Time[base]\frac{\Tnp}{\Tnp-\Cr}
-  \frac{\Time[Final]^{\{1\}}}{\muP} \frac{\Tnp}{\Tnp-\Cr}\left(
  \left(\Tnp-\Cr\right) 
  + (1-\precision)\I 
\right)\\
 + \frac{\Time[Final]^{\{1\}}}{\muNP} \left ( \frac{\Tnp}{2} + \D + \R \right ) 
  + \frac{(1-\precision)\Time[Final]^{\{1\}}}{\muP} \left ( \Tnp + \I + \Cp \right ) \\
  + \frac{\precision\Time[Final]^{\{1\}}}{\muP} \left ( \Tnp+ \Cp + \EIf + \D + \R  \right ) 
\end{multline*}

  \exactWithDetails
  {
  \begin{multline*}
    \Time[Final]^{\{1\}} = \Time[base]\frac{\Tnp}{\Tnp-\Cr}
    -  \frac{\recall\Time[Final]^{\{1\}}}{\precision\mu} \frac{\Tnp}{\Tnp-\Cr}\left(
      \left(\Tnp-\Cr\right) 
      + (1-\precision)\I
    \right)\\
    + \frac{(1-\recall)\Time[Final]^{\{1\}}}{\mu} \left ( \frac{\Tnp}{2} + \D + \R \right ) 
    + \frac{(1-\precision)\recall\Time[Final]^{\{1\}}}{\precision\mu} \left ( \Tnp + \I + \Cp \right ) \\
    + \frac{\precision\recall\Time[Final]^{\{1\}}}{\precision\mu} \left ( \Tnp+ \Cp + \EIf + \D + \R  \right ) 
  \end{multline*}
  \begin{multline*}
    \Time[Final]^{\{1\}} = \Time[base]\frac{\Tnp}{\Tnp-\Cr}
    -  \frac{\recall\Time[Final]^{\{1\}}}{\precision\mu} \frac{\Tnp}{\Tnp-\Cr}\left(
      \left(\Tnp-\Cr\right) 
      + (1-\precision)\I
    \right)\\
    + \frac{\Time[Final]^{\{1\}}}{\precision\mu}\left( (1-\recall)\precision\left ( \frac{\Tnp}{2} + \D + \R \right ) 
    + (1-\precision)\recall \left ( \Tnp + \I + \Cp \right ) 
    + \precision\recall \left ( \Tnp+ \Cp + \EIf + \D + \R  \right ) \right ) 
  \end{multline*}
  \begin{multline*}
    \Time[Final]^{\{1\}} = \Time[base]\frac{\Tnp}{\Tnp-\Cr}
    -  \frac{\recall\Time[Final]^{\{1\}}}{\precision\mu} \frac{\Tnp}{\Tnp-\Cr}\left(
      \left(\Tnp-\Cr\right) 
      + (1-\precision)\I
    \right)\\
    + \frac{\Time[Final]^{\{1\}}}{\precision\mu}\left(
      \precision  (\D + \R) 
      + \recall(\Tnp+\Cp)
      + (1-\recall)\precision\frac{\Tnp}{2} 
    + (1-\precision)\recall \I 
    + \precision\recall \EIf \right ) 
  \end{multline*}
  \begin{multline*}
    \Time[Final]^{\{1\}} = \Time[base]\frac{\Tnp}{\Tnp-\Cr}
    -  \frac{\recall\Time[Final]^{\{1\}}}{\precision\mu} \frac{\Tnp}{\Tnp-\Cr}
       (1-\precision)\I\\
    + \frac{\Time[Final]^{\{1\}}}{\precision\mu}\left(
      \precision  (\D + \R) 
      + \recall\Cp
      + (1-\recall)\precision\frac{\Tnp}{2} 
    + (1-\precision)\recall \I 
    + \precision\recall \EIf \right ) 
  \end{multline*}
  \begin{multline*}
    \Time[Final]^{\{1\}}\left( 1
      +  \frac{\recall}{\precision\mu} \frac{\Tnp}{\Tnp-\Cr}
        (1-\precision)\I
        \right.\\\left.
      - \frac{1}{\precision\mu}\left(
        \precision  (\D + \R) 
        + \recall\Cp
        + (1-\recall)\precision\frac{\Tnp}{2} 
        + (1-\precision)\recall \I 
        + \precision\recall \EIf \right )
    \right)\\
    = \Time[base]\frac{\Tnp}{\Tnp-\Cr}
  \end{multline*}
  \begin{multline*}
    \Time[Final]^{\{1\}}\left(
      \frac{\recall}{\precision\mu} (1-\precision)\I
    \right.\\
    \left.+\frac{\Tnp-\Cr}{\Tnp}\left(1
      - \frac{1}{\precision\mu}\left(
        \precision  (\D + \R) 
        + \recall\Cp
        + (1-\recall)\precision\frac{\Tnp}{2} 
        + (1-\precision)\recall \I 
        + \precision\recall \EIf \right )\right)\right)\\
    = \Time[base]
  \end{multline*}
  \begin{multline*}
     \frac{\Time[base]}{\Time[Final]^{\{1\}}} = \left(
      \frac{\recall}{\precision\mu} (1-\precision)\I
    \right.\\
    \left.+\frac{\Tnp-\Cr}{\Tnp}\left(1
      - \frac{1}{\precision\mu}\left(
        \precision  (\D + \R) 
        + \recall\Cp
        + (1-\recall)\precision\frac{\Tnp}{2} 
        + (1-\precision)\recall \I 
        + \precision\recall \EIf \right )\right)\right)  
  \end{multline*}}
\noindent After a little rewriting we obtain:
  \begin{multline*}
     \frac{\Time[base]}{\Time[Final]^{\{1\}}} =
      \frac{\recall}{\precision\mu} (1-\precision)\I
      \\
    +\left(1-\frac{\Cr}{\Tnp}\right)\left(1
      \!-\! \frac{1}{\precision\mu}\left(
        \precision  (\D + \R) 
        \!+\! \recall\Cp
        \!+\! (1\!-\!\recall)\precision\frac{\Tnp}{2} 
        \!+\! \recall \left((1\!-\!\precision)\I 
        \!+\! \precision \EIf \right )\!\right )\!\right)
  \end{multline*}
  Finally, the waste is equal by definition to
  $\frac{\Time[Final]-\Time[base]}{\Time[Final]}$. Therefore, we have:
  \begin{multline}\label{eq.waste.q1.nockpt}
     \Waste = 1- 
      \frac{\recall}{\precision\mu} 
        (1-\precision)\I
        \\
    -\left(1\!-\!\frac{\Cr}{\Tnp}\right)\!\!\left(1
      \!-\! \frac{1}{\precision\mu}\left(
        \precision  (\D + \R) 
        \!+\! \recall\Cp
        \!+\! (1\!-\!\recall)\precision\frac{\Tnp}{2} 
        \!+\! \recall \left((1-\precision)\I 
        \!+\! \precision \EIf \right )\!\!\right )\!\!\right)\!\!
  \end{multline}

\subsubsection*{Waste minimization}~\\

When $\trust=0$, the optimal value for $\Tnp$ is obviously the same
than the one we computed for \Withckpt in the case $\trust=0$.  We now
minimize the waste of the strategy where $\trust=1$.  In order to
compute the optimal value for $\Tnp$, we identify the fraction of the
waste in Equation~\eqref{eq.waste.q1.nockpt} that depends on
$\Tnp$. We can rewrite Equation~\eqref{eq.waste.q1.nockpt} as:
\begin{multline*}
\Waste^{\{1\}} = \beta +
\frac{\Cr}{\Tnp}\left(1\!-\!\frac{1}{\precision\mu}\left(\!\precision(\D\!+\!\R)
  +
  \recall\left(\!\Cp+\left(\!1\!-\!\precision\right)\I+\precision\EIf\!\right)\!\right)\!\!\right) + \frac{1\!-\!\recall}{\mu} \frac{\Tnp}{2}  
\end{multline*}
where $\beta$ does not depend on $\Tnp$. This equation is identical to
Equation~\eqref{eq.waste.fnofTnp} and therefore the value of $\Tnp$
that minimizes the waste is $\Tnp^{\extr}$, the value given by
Equation~\eqref{eq:Tr.extr}.

\subsection{Strategy \Instant}
\label{sec.time-instant}

In this section we evaluate the execution time under heuristic
\Instant. The analysis is very similar to that of \Nockpt. Indeed, we
only focus to the differences between the performance of \Instant and
\Withckpt. The differences happening, obviously, only in the presence
of true and false predictions:
\begin{compactenum}\setcounter{enumi}{2}
\item \textbf{False prediction.} There are two cases:
  \begin{compactenum}
  \item \textbf{Taken into account.} This happens with probability
    $\trust$. The interval lasts $\Tnp+\Cp$, since we fallback to
    regular mode as soon as the proactive checkpoints completes. The
    work done is $\Tnp-\Cr$.
  \item \textbf{Not taken into account.} This happens with probability
    $1-\trust$. The interval lasts $\Tnp$ and the work done is $\Tnp-\Cr$.
  \end{compactenum}

  Considering both cases with their probabilities, the average time
  spent is equal to: $\trust(\Tnp+\Cp) + (1-\trust)\Tnp=
  \Tnp+\trust\Cp$. The average work done is: $\trust (\Tnp-\Cr) +
  (1-\trust)(\Tnp-Cr) = \Tnp-\Cr$.

\item \textbf{True prediction.} There are two cases:
  \begin{compactenum}
  \item \textbf{Taken into account.} Let \EIf be the average time at
    which a fault occurs within the prediction window. We take a
    proactive checkpoint right before the start of the prediction
    window. Then we fallback to the regular mode. After a time \EIf
    the fault strikes. Depending on the size $\I$ of the prediction
    window, and of when the prediction started after the completion of
    the last regular checkpoint three scenarios can happen. Either the
    fault strikes while the heuristic is still trying to complete the
    work of size $\Tnp-\Cr$, or it strikes while the heuristic is
    trying to take the regular checkpoint after that work, or it
    strikes after that regular checkpoint was completed. We
    overestimate the time lost by assuming that we are in one of the
    two former cases, because these are the cases that maximizes the
    amount of work destroyed by a strike. (In some way, this is
    equivalent to assuming that $\I$ is very small with respect to
    $\Tnp$.) The predicted fault and the completion time of the last
    regular checkpoint are independent events. Therefore, on average
    the fault strikes at time $\Tnp/2$. After the fault strikes, the
    downtime and the recovery we complete the period struck by the
    fault.  Then, the interval lasts $\Tnp+\Cp+\EIf+\D+\R$ on
    average. The total work done during the interval is $\Tnp-\Cr$.
  \item \textbf{Not taken into account.}  On average the fault occurs
    at time $\Tnp/2$. The time interval has duration
    $\Tnp/2+\D+\R$, and there is no work done.
  \end{compactenum}
  Overall the time spent is $\trust(\Tnp+\Cp+\EIf+\D+\R) + (1-\trust)
  (\Tnp/2+\D+\R)$, and the work done is $\trust(\Tnp-\Cr)$.
\end{compactenum}
So far, we have evaluated the length, and the work done, for each of
the interval types. We estimate the expectation of the number of
intervals of each type as we did for \Withckpt and for
\Nockpt. Table~\ref{tab.summary.modes.instant} presents the estimation
of the number of intervals of each type.

\begin{table}
  \centering
  {%
    \begin{tabular}{c|c|c|c}
      Mode & Number of intervals & Time spent & Work done \\
      \hline
      (1) & $w_1$ & \Tnp & $\Tnp-\Cr$ \\[1mm]
      (2) & $w_2=\frac{\Time[Final]}{\muNP}$ & $\Tnp/2 + \D + \R$ & 0 \\[1mm]
      {(3)} &  {$w_3=\frac{(1-\precision)\Time[Final]}{\muP}$} &$\Tnp + \trust \Cp $ & $\Tnp-\Cr$ \\ 
      \multirow{2}{*}{(4)} &\multirow{2}{*}{$w_4=\frac{\precision\Time[Final]}{\muP}$} &$\trust (\Tnp + \EIf + \Cp) $& \multirow{2}{*}{$\trust \left(\Tnp-\Cr\right)$}  \\ &&$+ (1-\trust)\Tnp/2 +\D + \R$&
    \end{tabular}}
  \caption{Summary of the different types of interval for \Instant.}
  \label{tab.summary.modes.instant}
\end{table}

We estimate the total execution time as for \Withckpt and \Nockpt:
\begin{multline}\label{eq.timefinal.instant}
  \Time[Final] = w_1 \times \Tnp + w_2 \left(\frac{\Tnp}{2} + \D +
    \R\right) + w_3\left( \Tnp +\trust \Cp \right)\\ +
  w_4\left(\trust (\Tnp + \EIf + \Cp) + (1-\trust)\frac{\Tnp}{2} +\D +
    \R\right)
\end{multline}
We use the same line of reasoning as previously to compute the overall amount of
work done:
\begin{multline}\label{eq.timebase.instant}
  \Time[base] = w_1 (\Tnp-\Cr) + w_2\times 0 + w_3 \left (\Tnp-\Cr\right)
  + w_4 \left (\trust\left(\Tnp-\Cr \right) \right)
\end{multline}
This equation gives the value of $w_1$ as a function of the other
parameters. As with \Withckpt and \Nockpt, one can easily show that in
an optimal solution, either $\trust=0$ or $\trust=1$. Thus we can now
focus on the two functions \Time[Final], the one when $\trust=0$
($\Time[Final]^{\{0\}}$), and the one when $\trust=1$
($\Time[Final]^{\{1\}}$).

From Table~\ref{tab.summary.modes.instant} and
Equations~\eqref{eq.timefinal.instant} and \eqref{eq.timebase.instant}, one can easily see that
\[\Time[Final]^{\{0\}} = \frac{\Tnp}{\Tnp-\Cr}\Time[base] +
\frac{\Time[Final]^{\{0\}}}{\mu}\left ( \frac{\Tnp}{2} + \D + \R\right), \text{
  i.e., that}\]
\begin{equation}
\left (1-\frac{\Cr}{\Tnp}\right)\left ( 1- \frac{\Tnp/2 + \D + \R}{\mu}\right )\Time[Final]^{\{0\}} = \Time[base]
\label{eq.easycase.instant}
\end{equation}
This is exactly the equation from~\cite{rr-journal-prediction} in the
case of exact-date predictions that are never taken into account, what
we had already remarked with \Withckpt and \Nockpt (yet another good sanity check!).  When
$\trust=1$, we have:
\begin{multline*}
\Time[Final]^{\{1\}} = \Time[base]\frac{\Tnp}{\Tnp-\Cr}
-  \frac{\Time[Final]^{\{1\}}}{\muP} \Tnp
 + \frac{\Time[Final]^{\{1\}}}{\muNP} \left ( \frac{\Tnp}{2} + \D + \R \right ) \\
  + \frac{(1-\precision)\Time[Final]^{\{1\}}}{\muP} \left ( \Tnp + \Cp \right ) 
  + \frac{\precision\Time[Final]^{\{1\}}}{\muP} \left ( \Tnp+ \Cp + \EIf + \D + \R  \right ) 
\end{multline*}

\exactWithDetails
  {
  \begin{multline*}
    \Time[Final]^{\{1\}} = \Time[base]\frac{\Tnp}{\Tnp-\Cr}
    -  \frac{\recall\Time[Final]^{\{1\}}}{\precision\mu} \Tnp
    \\
    + \frac{(1-\recall)\Time[Final]^{\{1\}}}{\mu} \left ( \frac{\Tnp}{2} + \D + \R \right ) 
    + \frac{(1-\precision)\recall\Time[Final]^{\{1\}}}{\precision\mu} \left ( \Tnp + \Cp \right ) \\
    + \frac{\precision\recall\Time[Final]^{\{1\}}}{\precision\mu} \left ( \Tnp+ \Cp + \EIf + \D + \R  \right ) 
  \end{multline*}
  \begin{multline*}
    \Time[Final]^{\{1\}} = \Time[base]\frac{\Tnp}{\Tnp-\Cr}
    -  \frac{\recall\Time[Final]^{\{1\}}}{\precision\mu} \Tnp
\\
    + \frac{\Time[Final]^{\{1\}}}{\precision\mu}\left( (1-\recall)\precision\left ( \frac{\Tnp}{2} + \D + \R \right ) 
    + (1-\precision)\recall \left ( \Tnp + \Cp \right ) 
    + \precision\recall \left ( \Tnp+ \Cp + \EIf + \D + \R  \right ) \right ) 
  \end{multline*}
  \begin{multline*}
    \Time[Final]^{\{1\}} = \Time[base]\frac{\Tnp}{\Tnp-\Cr}
    -  \frac{\recall\Time[Final]^{\{1\}}}{\precision\mu} \Tnp\\
    + \frac{\Time[Final]^{\{1\}}}{\precision\mu}\left(
      \precision  (\D + \R) 
      + \recall(\Tnp+\Cp)
      + (1-\recall)\precision\frac{\Tnp}{2} 
    + \precision\recall \EIf \right ) 
  \end{multline*}
  \begin{multline*}
    \Time[Final]^{\{1\}} = \Time[base]\frac{\Tnp}{\Tnp-\Cr}
    + \frac{\Time[Final]^{\{1\}}}{\precision\mu}\left(
      \precision  (\D + \R) 
      + \recall\Cp
      + (1-\recall)\precision\frac{\Tnp}{2} 
    + \precision\recall \EIf \right ) 
  \end{multline*}
  \begin{multline*}
    \Time[Final]^{\{1\}}\left( 1
      - \frac{1}{\precision\mu}\left(
        \precision  (\D + \R) 
        + \recall\Cp
        + (1-\recall)\precision\frac{\Tnp}{2} 
        + \precision\recall \EIf \right )
    \right)\\
    = \Time[base]\frac{\Tnp}{\Tnp-\Cr}
  \end{multline*}
  \begin{multline*}
    \Time[Final]^{\{1\}}\frac{\Tnp-\Cr}{\Tnp}\left(1
      - \frac{1}{\precision\mu}\left(
        \precision  (\D + \R) 
        + \recall\Cp
        + (1-\recall)\precision\frac{\Tnp}{2} 
        + \precision\recall \EIf \right )\right)\\
    = \Time[base]
  \end{multline*}
  \begin{multline*}
     \frac{\Time[base]}{\Time[Final]^{\{1\}}} = \\
     \frac{\Tnp-\Cr}{\Tnp}\left(1
      - \frac{1}{\precision\mu}\left(
        \precision  (\D + \R) 
        + \recall\Cp
        + (1-\recall)\precision\frac{\Tnp}{2} 
        + \precision\recall \EIf \right )\right)
  \end{multline*}}
\noindent After a little rewriting we obtain:
  \begin{multline*}
     \frac{\Time[base]}{\Time[Final]^{\{1\}}} = 
     \left(1-\frac{\Cr}{\Tnp}\right)\left(1
      \!-\! \frac{1}{\precision\mu}\left(
        \precision  (\D + \R) 
        \!+\! \recall\Cp
        \!+\! (1\!-\!\recall)\precision\frac{\Tnp}{2} 
        \!+\! \precision\recall \EIf \right )\!\right )
  \end{multline*}
  Finally, the waste is equal by definition to
  $\frac{\Time[Final]-\Time[base]}{\Time[Final]}$. Therefore, we have:
  \begin{multline}\label{eq.waste.q1.instant}
     \Waste = 1
    -\left(1\!-\!\frac{\Cr}{\Tnp}\right)\!\!\left(1
      \!-\! \frac{1}{\precision\mu}\left(
        \precision  (\D + \R) 
        \!+\! \recall\Cp
        \!+\! (1\!-\!\recall)\precision\frac{\Tnp}{2} 
        \!+\! \precision\recall \EIf \right )\!\!\right )
  \end{multline}

\subsubsection*{Waste minimization}~\\

When $\trust=0$, the optimal value for $\Tnp$ is obviously the same
than the one we computed for \Withckpt and for \Nockpt in the case
$\trust=0$.  We now minimize the waste of the strategy where
$\trust=1$.  In order to compute the optimal value for $\Tnp$, we
identify the fraction of the waste in
Equation~\eqref{eq.waste.q1.instant} that depends on $\Tnp$. We can
rewrite Equation~\eqref{eq.waste.q1.instant} as:
\begin{multline*}
\Waste^{\{1\}} = \beta +
\frac{\Cr}{\Tnp}\left(1\!-\!\frac{1}{\precision\mu}\left(\!\precision(\D\!+\!\R)
  +
  \recall\Cp+\precision\recall\EIf\!\right)\!\right) + \frac{1\!-\!\recall}{\mu} \frac{\Tnp}{2}  
\end{multline*}
where $\beta$ does not depend on $\Tnp$. Therefore, the value of
$\Tnp$ that minimizes the waste is 
$\Tnp^{\extr}$, where
\begin{equation*}
\Tnp^{\extr} = 
\sqrt{\frac{2\Cr\left(\precision\mu-\left(\precision(\D+\R)   +
        \recall\Cp+\precision\recall\EIf\right)\right)}{\precision(1-\recall)}}
\end{equation*}
Again, recall that we must always enforce that $\Tnp^{\extr}$ is always greater than \Cr.
Finally, if we assume that, on average, fault strikes at the middle of
the prediction window, i.e., $\EIf=\frac{I}{2}$, we have:
\begin{equation*}
\Tnp^{\extr} = 
\sqrt{\frac{2\Cr\left(\precision\mu-\left(\precision(\D+\R)   +
        \recall\Cp+\precision\recall\frac{I}{2}\right)\right)}{\precision(1-\recall)}}
\end{equation*}


\section{Simulation results}
\label{sec.simulations}

We start by presenting the simulation framework
(Section~\ref{sec.simulations.framework}). Then we report results
using the characteristics of two fault predictors from the literature
(Section~\ref{sec.simulations.pred}).

\subsection{Simulation framework}
\label{sec.simulations.framework}

In order to validate the model, we have instantiated it with several
scenarios.  The experiments use parameters that are representative of
current and forthcoming large-scale
platforms~\cite{j116,Ferreira2011}.  We take $\Cr=\R=600$ seconds, and
$\D=60$ seconds. We consider three scenarios
where proactive checkpoints are (i) exactly as expensive as periodic
checkpoints ($\Cp=\Cr$); (ii) ten times cheaper ($\Cp=0.1\Cr$); and (iii) two
times more expensive ($\Cp=2\Cr$).  The individual (processor) MTBF is
$\mu_{\text{ind}} = 125$ years, and the total number of processors $N$
varies from $N=2^{16}=16,384$ to $N=2^{19}=524,288$, so that the platform MTBF $\mu$
varies from $\mu=4,010$ min (about $2.8$ days) down to $\mu=125$ min
(about $2$ hours).  For instance the Jaguar platform, with $N=45,208$
processors, is reported to have experienced about one fault per
day~\cite{6264677}, which leads to $\mu_{\text{ind}} =
\frac{45,208}{365}\approx 125$ years.  The application size is set to
$\Time[base]=10,000$ years/N.

We use Maple to analytically compute and plot the optimal value of the
waste for the three prediction-aware policies, \Instant, \Nockpt, and
\Withckpt, for the prediction-ignoring policy \newdaly (corresponding
to the case $\trust=0$), and for the reference heuristic \daly (Daly's~\cite{daly04} periodic policy). 
In order to check the
accuracy of our model, we have compared the analytical results with results
obtained with a discrete-event simulator.  The simulation engine
generates a random trace of faults, parameterized either by an
Exponential fault distribution or by Weibull distribution laws with
shape parameter $0.5$ or $0.7$.  Note that Exponential faults are
widely used for theoretical studies, while Weibull faults are
representative of the behavior of real-world
platforms~\cite{Weibull1,Weibull2,Heien:2011:MTH:2063384.2063444}.  In
both cases, the distribution is scaled so that its expectation
corresponds to the platform MTBF $\mu$.  With probability \recall, we
decide if a fault is predicted or not.  The simulation engine also
generates a random trace of false predictions, whose distribution is
identical to that of the first trace (in
Figures~\ref{fig.082.085.CpCr.unif} through
\ref{fig.04.07.Cp2Cr.unif}, we also consider the case where false
predictions are generated according to a uniform distribution; results
are quite similar). This second distribution is scaled so that its
expectation is equal to $\frac{\muP}{1-\precision} = \frac{\precision
  \mu}{\recall (1-\precision)}$, the inter-arrival time of false
predictions.  Finally, both traces are merged to produce the final
trace including all events (true predictions, false predictions, and
non predicted faults). Each reported value is the average over $100$
randomly generated instances.

In the simulations, we compare the five checkpointing strategies
listed above. To assess the quality of each strategy, we compare it
with its \bestper counterpart, defined as the same strategy but using
the best possible period $\Tnp$. This latter period is computed via a
brute-force numerical search for the optimal period. Altogether, there
are four  \bestper heuristics, one for each of the three variants with prediction, 
and one for the case where we ignore predictions,
which corresponds to both \daly and \newdaly. Altogether we have a rich set of nine heuristics,
which enables us to comprehensively
assess the actual quality of the proposed strategies. Note that for  computer algebra plots,
obviously we do not need \bestper heuristics, since each period is
already chosen optimally from the equations.

We experiment with two predictors from the literature: one
accurate predictor with high recall and precision~\cite{5958823},
namely with $\precision=0.82$ and $\recall=0.85$, and another
predictor with more limited recall and precision~\cite{5542627},
namely with $\precision=0.4$ and $\recall=0.7$. In both cases, we use
five different prediction windows, of size $\I=300$, 600, 900, 1200, and 3000
seconds. 
Figures~\ref{fig.082.085.CpCr.same} through~\ref{fig.04.07.Cp2Cr.same}
show the average waste degradation of the nine heuristics for both
predictors, as a function of the number of processors $N$. We draw the
plots as a function of the number of processors $N$ rather than of the
platform MTBF $\mu = \mu_{ind}/N$, because it is more natural to see
the waste increase with larger platforms; however, this work is
agnostic of the granularity of the processors and intrinsically
focuses on the impact of the MTBF on the waste.

\subsection{Analysis of the results}
\label{sec.simulations.pred}

We start with a preliminary remark: when the graphs for \Instant and \Withckpt cannot
be seen in the figures, this is because their performance is identical
to that of \Nockpt, and their respective graphs are superposed.

We first compare the analytical results, plotted by the Maple curves,
to the simulations results. There is a good correspondence between the
analytical curves and the simulations, especially those using an
Exponential distribution of failures. However, the larger the
platform (or the smaller the MTBF), the
less realistic our assumption that no two events happen during an
interval of length $\Tnp+\I+\Cp$, and the analytical models
become less accurate for prediction-aware heuristics. Therefore, the analytical results are
overly pessimistic  in the most failure-prone
platforms. Also, recall that an exponential law is a Weibull law of
shape parameter $1$. Therefore, the further the distribution of failures
is from an exponential law, the larger the difference between 
analytical results and simulated ones. However, in all cases, the
analytical results are able to predict the general \emph{trends}. 

A second assessment of the quality of our analysis comes from the
\bestper variants of our heuristics. When predictions are not taken into account,
\daly, and to a lesser extent \newdaly, are not close to the optimal period given by
\bestper (a similar observation was made in~\cite{sc2011}). This gap increases when the distribution is
further apart from an Exponential distribution. However, 
prediction-aware heuristics are very close
to  \bestper in almost all
configurations. The only exception is with heuristics \Instant when
$\Cp=2\Cr$, the total number of processors $N$ is equal to either
$2^{18}$ and $2^{19}$, and $\I$ is large. However, when $\I=3000$
and $N=2^{19}$, the platform MTBF is approximately equal to $6\Cp$
which renders our hypothesis and analysis invalid. The difference in
this case between \Instant and its \bestper should therefore not come
as a surprise.

To better understand why close-to-optimal periods are obtained by prediction-aware heuristics 
(while this is not the case without predictions), we plot the waste as a function of
the period $\Tnp$ for \newdaly and the prediction-aware heuristics
(Figures~\ref{fig.082.085.Tr.16} through \ref{fig.04.07.Tr.19}). On
these figures one can see that, whatever the configuration, periodic
checkpointing policies (ignoring predictions) have well-defined global
optimum. (One should nevertheless remark that the performance is
almost constant in the neighborhood of the optimal period which
explains why policies using different periods can obtain in practice
similar performance, as in~\cite{doi:10.1007/978-3-642-14390-8_22}.)
For prediction-aware heuristics, however, the behavior is quite
different and two scenarios are possible. In the first one, once the
optimum is reached, the waste very slowly increases to reach an
asymptotic value which is close to the optimum waste (e.g., when the
platform MTBF is large and failures follow an exponential
distribution). Therefore, any period chosen close to the optimal one,
or greater than it, will deliver good quality performance. In the
second scenario, the waste decreases until the period becomes larger
than the application size, and the waste stays constant. In other
words, in these configurations, periodic checkpointing is unnecessary,
only proactive actions matter! This striking result can
be explained as follows: a significant fraction of the failures are
predicted, and thus taken care of, by proactive checkpoints. The impact
of unpredicted failures is mitigated by the proactive measures
taken for false predictions. To further mitigate the impact of
unpredicted faults, the period \Tnp should be significantly shorter
than the mean-time between proactive checkpoints, which would induce a
lot of waste due to unnecessary checkpoints if the mean-time between
unpredicted faults is large with respect to the mean-time between
predictions. This greatly restrict the scenarios for which the
periodic checkpointing can lead to a significant decrease of the
waste.

When the prediction window $\I$ is shorter than the duration $\Cp$ of a
proactive checkpoint, there is no difference between \Nockpt
and \Withckpt. When $\I$ is small but greater than $\Cp$ (say, when
$\I$ is around $2\Cp$), \Withckpt spends most of the prediction window
taking a proactive checkpoint and \Nockpt is more efficient. When $\I$
becomes ``large'' with respect to $\Cp$, \Withckpt can become more
efficient than \Nockpt, but becomes significantly more efficient
only if the proactive checkpoints are significantly shorter than
regular ones. \Instant can hardly be seen in the graphs as its
performance is most of the time equivalent to that of \Nockpt.

Figures~\ref{fig.082.085.I.16} through \ref{fig.04.07.I.19} show the
influence of the size of the prediction window \I on the performance
of the heuristics. As expected, the smaller the
prediction window, the more efficient the prediction-aware
heuristics. Also, the smaller the number of processors (or the larger the
platform MTBF), the larger the impact of the size of the prediction
window. A surprising result is that taking prediction
into account is not always beneficial! The analytical results predict
that prediction-aware heuristics would achieve worse performance than
periodic policies in our settings, as soon as the platform includes $2^{18}$ processors.
In simulations, results are not so extreme. For the largest
platforms considered, using predictions has almost no impact on
performance. But when the prediction window is very large, taking
predictions into account can indeed be detrimental. These observations can be
explained as follows. When the platform includes $2^{19}$ processors, the
platform MTBF is equal to $7500$ s. Therefore, any interval of duration
$3000$ has a $40\%$ chance to include a failure: a
prediction window of $3000$  is not very informative, unless
 the precision and recall of the predictor are almost equal to $1$ (which is never the case in practice).
Since the predictor brings almost no knowledge,
trusting it may be detrimental. When comparing the performance of,
say, \Nockpt for the two predictors, one can see that when failures
follow a Weibull distribution with shape parameter $k=0.7$, $I=600$,
and $N=2^{18}$, \Nockpt achieves better performance than \newdaly when
$\recall=0.85$ and $\precision=0.82$, but worse when $\precision=0.4$
and $\recall=0.7$. The latter predictor generates more false
predictions ---each one inducing an unnecessary proactive
checkpoint--- and misses more actual failures ---each one destroying
some work. The drawbacks of trusting the predictor outweigh the advantages. If
failures are few and apart, almost any predictor will be
beneficial. When the platform MTBF is small with respect to the cost
of proactive checkpoints, only almost perfect predictors will be worth
using. For each set of predictor characteristics, there is a threshold for
the platform MTBF under which predictions will be useless or detrimental, but above which 
predictions will be beneficial.

In order to compare the impact of the heuristics ignoring predictions
to those using them, we report job execution times in
Table~\ref{makespan.tab}. For the strategies with prediction, we
compute the gain (expressed in percentage) over \daly, the reference
strategy without prediction. We first remark that \newdaly achieves
lower makespans than \daly with gains ranging from 1\% with $2^{16}$
processors to 18\% with $2^{19}$ processors. Overall, the gain due to
the predictions decreases when the size of the prediction window
increases, and increases with the platform size. This gain is
obviously closely related to the characteristics of the predictor.

When $\I=300$, the three strategies are identical. When $\I$
increases, \Nockpt achieves slightly better results than \Instant. For
low values of \I, \Withckpt is the worst prediction-aware
heuristics. But when \I becomes large and if the predictor is
efficient, then \Withckpt becomes the heuristics of choice ($\I=3000$, $\precision=0.82$, and $\recall=0.85$).

The reductions in the application executions times due to the
predictor can be very significant. With $\precision=0.85$ and
$\recall=0.82$ and $\I=3000$, we save $25\% $ of the total time with
$N=2^{19}$, and $13\%$ with $N=2^{16}$ using strategy \Withckpt.  With
$\I=300$, we save up to $45\%$ with $N=2^{19}$, and $18\%$ with
$N=2^{16}$ using any strategy (though \Nockpt is slightly better than
\Instant).  Then, with $\precision=0.4$ and $\recall = 0.7$, we still
save $33\%$ of the execution time when $\I=300$ and $N=2^{19}$, and
$14\%$ with $N=2^{16}$. The gain gets smaller with $\I=3000$ and
$N=2^{16}$ but remains non negligible since we can save $8\%$. When
$\I=3000$ and $N=2^{19}$, however, the best solution is to ignore
predictions and simply use \newdaly (we fall-back to the case
$\trust=0$). If we now consider a Weibull law with shape parameter 0.5
instead of 0.7, keeping all other parameters identical ($\I=3000$,
$N=2^{19}$, $\precision=0.4$ and $\recall = 0.7$), then the heuristics
of choice is \Withckpt and the gain with respect to \daly is 57.9\%.

\begin{table}
  \centering
  \resizebox{\textwidth}{!}{%
  \begin{tabular}{c||c|c||c|c||c|c||}
    & \multicolumn{2}{|c||}{$\I=300$ s}     &    \multicolumn{2}{|c||}{$\I=1200$ s}     &    \multicolumn{2}{|c||}{$\I=3000$ s}\\
    & $2^{16}$ procs & $2^{19}$ procs & $2^{16}$ procs & $2^{19}$ procs & $2^{16}$ procs & $2^{19}$ procs \\
    \hline
    \daly & 81.3 & 31.0 & 81.3 & 31.0 & 81.3 & 31.0 \\
    \newdaly & 80.2 (1\%) & 25.5 (18\%) & 80.2 (1\%) & 25.5 (18\%) & 80.2 (1\%) & 25.5 (18\%) \\
    \hline
    \multicolumn{1}{c}{}     & \multicolumn{6}{c}{$\precision=0.82$, $\recall=0.85$} \\\hline
    \Nockpt & \textbf{66.4} (18\%)& \textbf{17.0} (45\%) & \textbf{67.9} (16\%) & \textbf{20.2} (35\%) & 71.0 (13\%) & 24.7 (20\%)\\
    \Withckpt & \textbf{66.4} (18\%) & \textbf{17.0} (45\%) & 68.3 (16\%)& 20.6 (33\%) & \textbf{70.6} (13\%)& \textbf{23.1} (25\%)\\
    \Instant & 66.5 (18\%)& \textbf{17.0} (45\%) & 68.0 (16\%)& 20.3 (34\%) & 70.9 (13\%)& 24.1 (22\%)\\\hline
    \multicolumn{1}{c}{}     & \multicolumn{6}{c}{$\precision=0.4$, $\recall=0.7$} \\
    \hline
    \Nockpt & \textbf{70.2} (14\%)& \textbf{20.6} (33\%) & \textbf{71.8} (12\%)& \textbf{24.2} (22\%) & \textbf{75.0} (8\%)& 28.7 (7\%)\\
    \Withckpt & \textbf{70.2} (14\%) & \textbf{20.6} (33\%) & 73.6 (9\%)& 25.5 (18\%) & 75.1 (8\%)& 26.6 (14\%)\\
    \Instant & 70.3 (13\%)& 20.9 (33\%) & 72.0 (11\%)& 24.6 (21\%) & \textbf{75.0} (8\%)& 27.7 (11\%)\\\hline
  \end{tabular}}
  \caption{Job execution times (in days) under the different checkpointing policies, when failures follow a Weibull distribution of shape parameter $0.7$. Gains are reported with respect to \daly.}
  \label{makespan.tab}
\end{table}

\begin{table}
  \centering
  \resizebox{\textwidth}{!}{%
  \begin{tabular}{c||c|c||c|c||c|c||}
    & \multicolumn{2}{|c||}{$\I=300$ s}     &    \multicolumn{2}{|c||}{$\I=1200$ s}     &    \multicolumn{2}{|c||}{$\I=3000$ s}\\
    & $2^{16}$ procs & $2^{19}$ procs & $2^{16}$ procs & $2^{19}$ procs & $2^{16}$ procs & $2^{19}$ procs \\
    \hline
    \daly & 125.7 & 185.0 & 125.7 & 185.0 & 125.7 & 185.0 \\
    \newdaly & 120.1 (4\%) & 114.8 (38\%) & 120.1 (4\%) & 114.8 (38\%) & 120.1 (4\%) & 114.8 (38\%) \\
    \hline
    \multicolumn{1}{c}{}     & \multicolumn{6}{c}{$\precision=0.82$, $\recall=0.85$} \\\hline
    \Nockpt & 77.4 (38\%)& \textbf{44.9} (76\%) & 81.8 (35\%) & 60.7 (67\%) & 90.0 (28\%) & 71.5 (61\%)\\
    \Withckpt & 77.4 (38\%) & \textbf{44.9} (76\%) & 83.6 (33\%)    &    64.4 (65\%)       & 89.8 (29\%)& 66.2 (64\%)\\
    \Instant      &      77.4 (38\%)& \textbf{45.2} (76\%) & 82.0 (35\%)     & 60.8 (67\%)         & 89.7 (29\%)& 70.6 (62\%)\\\hline
    \multicolumn{1}{c}{}     & \multicolumn{6}{c}{$\precision=0.4$, $\recall=0.7$} \\
    \hline
    \Nockpt & 84.4 (33\%)& 58.3 (68\%) & 89.1 (29\%)& 76.8 (58\%) & 97.9 (22\%)& 83.7 (55\%)\\
    \Withckpt & 84.4 (33\%) & 58.3 (68\%) & 93.8 (25\%)    &   75.4 (59\%)    & 97.8 (22\%)& 77.7 (58\%)\\
    \Instant    &     84.5 (33\%)      & 59.6 (68\%)      & 89.4 (29\%)      & 76.64 (58\%)     & 97.7 (22\%)& 81.9 (56\%)\\\hline
  \end{tabular}}
  \caption{Job execution times (in days) under the different checkpointing policies, when failures follow a Weibull distribution of shape parameter $0.5$. Gains are reported with respect to \daly.}
  \label{makespan05.tab}
\end{table}

\section{Related work}
\label{sec.related}

Considerable research has been conducted on fault prediction using different models (system log
analysis~\cite{5958823}, event-driven approach~\cite{GainaruIPDPS12,5958823,5542627}, 
support vector machines~\cite{LiangZXS07,Fulp:2008:PCS:1855886.1855891}), nearest neighbors~\cite{LiangZXS07}, \dots).
In this section we give a brief overview of the results obtained by predictors. We focus on their 
results rather than on their methods of prediction.

The authors of~\cite{5542627} introduce the \emph{lead time}, that is the time between the prediction and the 
actual fault. This time should be sufficient to take proactive actions. They are also able to give 
the location of the fault. While this has a negative impact on the precision (see the low value of 
\precision in Table~\ref{rel.work.tab}), they state that it has a positive impact on the checkpointing
time (from 1500 seconds to 120 seconds).
The authors of~\cite{5958823} also consider a lead time, and  introduce a \emph{prediction window} when 
the predicted fault should happen. The authors of~\cite{LiangZXS07} study the impact of different prediction techniques with different 
prediction window sizes. They also consider a lead time, but do not state its value.
These two latter studies motivate this work,
even though~\cite{5958823} does not provide the size of their prediction window.

\begin{table}
\centering
\begin{tabular}{|c|c|c|c|c|}
Paper & Lead Time & Precision & Recall & Prediction Window \\
\hline
\cite{5542627} & 300 s & 40 \% & 70\% & - \\
\cite{5542627} & 600 s & 35 \% & 60\% & - \\
\cite{5958823} & 2h & 64.8 \% & 65.2\% & yes (size unknown) \\
\cite{5958823} & 0 min & 82.3 \% & 85.4 \% & yes (size unknown) \\
\cite{GainaruIPDPS12} & 32 s & 93 \% & 43 \% & - \\
\cite{Fulp:2008:PCS:1855886.1855891} & NA & 70 \% & 75 \% & - \\
\cite{LiangZXS07} & NA & 20 \% & 30 \%& 1h \\
\cite{LiangZXS07} & NA & 30 \%& 75 \% & 4h \\
\cite{LiangZXS07} & NA & 40 \%& 90 \% & 6h \\
\cite{LiangZXS07} & NA & 50 \% & 30 \% & 6h \\
\cite{LiangZXS07} & NA & 60 \% & 85\% & 12h \\
\end{tabular}
\caption{Comparative study of different parameters returned by some
  predictors.}
\label{rel.work.tab}
\end{table}

Unfortunately, much of the work done on prediction does not provide information that could be 
really useful for the design of efficient algorithms. These informations are those stated above, namely the lead time 
and the size of the prediction window, but other information that could be useful would be: (i) the 
distribution of the faults in the prediction window; (ii) the precision as a function of the recall (see our 
analysis); and (iii) the precision and recall as functions of the prediction window (what happens with 
a larger prediction window). 

While many studies on fault prediction focus on the conception of the predictor, most of them 
consider that the proactive action should simply be a checkpoint or a migration right in time before the fault. 
However, in their paper~\cite{li2009fault}, Li et al. consider the 
mathematical problem to determine when and how to migrate. In order to be able to use migration, 
they stated that at every time, 2\% of the resources are available. This allowed them to conceive a 
Knapsack-based heuristic. Thanks to their algorithm, they were able to save 30\% of the execution 
time compared to an heuristic that does not take the reliability into account, with a precision and 
recall of 70\%, and with a maximum load of 0.7.

In the simpler case where predictions are exact-date predictions, Gainaru et al~\cite{GainaruSC12} 
have shown that the optimal checkpointing period becomes 
$\ttopt = \sqrt{\dfrac{2 \mu\Cr }{1-\recall}}$, but their analysis is valid only if $\mu$ is very large in front of the other parameters.
Our previous work~\cite{rr-journal-prediction} has refined the results of~\cite{GainaruSC12}, focusing
on a more accurate analysis of fault 
prediction with exact dates, and providing a detailed study on the impact of recall and precision
on the waste. As shown in Section~\ref{sec.intervals}, the analysis of the waste is 
dramatically more complicated when using prediction windows than when using exact-date predictions.
To the best of our knowledge, this work is the first to focus on the mathematical 
aspect of  fault prediction with prediction windows, and to provide a model and a detailed analysis of the 
waste due to all three types of events (true and false predictions and unpredicted failures).


\section{Conclusion}
\label{sec.conclusion}

In this work, we have studied the impact of prediction windows on
checkpointing strategies.  We have designed several heuristics that
decide whether to trust these predictions, and when it is worth taking
preventive checkpoints. We have been able to derive a comprehensive
set of results and conclusions:\\
$\bullet$ We have introduced an analytical model to capture the waste
incurred by each strategy, and provided for each optimization problem
a closed-form formula giving its optimal solution. Contrarily to the cases without prediction,
or with exact-date predictions, the computation of the waste requires a sophisticated analysis
of the various events, including the time spent irregular or proactive modes.\\
$\bullet$ The simulations fully validate the model, and the
brute-force computation of the optimal period
guarantees that our prediction-aware strategies are always very close to the optimal. This holds true both for Exponential and Weibull failure distributions.\\
$\bullet$ The model is quite accurate and its validity goes beyond the
conservative assumption that requires a single event per time
interval; even more surprising, the accuracy of the model for
prediction-aware strategies is much better than for the case without
predictions, where \daly can be far from the optimal period in the
case of Weibull failure distributions.\\
$\bullet$ Both the analytical computations and the simulations enable
to characterize when prediction is useful, and which strategy performs
better, given the key parameters of the system: recall \recall,
precision \precision, size of the prediction window \I, size of
proactive checkpoints \Cp versus regular checkpoints \Cr, and
platform MTBF $\mu$.

Altogether, the analytical model and the comprehensive results provided in this work enable to
fully assess the impact of fault prediction with time-windows on optimal checkpointing strategies. 
Future
work will be devoted to refine the assessment of the usefulness of prediction with trace-based failure and prediction logs
from current large-scale supercomputers.

\bigskip
\noindent{\em Acknowledgments.} 
The authors are with Universit\'e de Lyon, France.
Y.~Robert is with the Institut Universitaire de France.
This work was supported in part by the ANR {\em RESCUE} project.
\bigskip

\bibliographystyle{abbrv}
\bibliography{../biblio}

\newlength\SimFigWidth
\setlength{\SimFigWidth}{0.22\linewidth}
\newlength\MapleFigWidth
\setlength{\MapleFigWidth}{0.22\linewidth}
\newlength\SimFigTrWidth
\setlength{\SimFigTrWidth}{0.30\linewidth}

\begin{figure*}
\centering
\includegraphics[scale=0.5]{../fig/legende.fig}\\
\hspace{-1cm}
\subfloat{\rotatebox{90}{\setcounter{subfigure}{0}\qquad\I= 300 s}}
\subfloat[Maple]
{
\includegraphics[width=\MapleFigWidth]{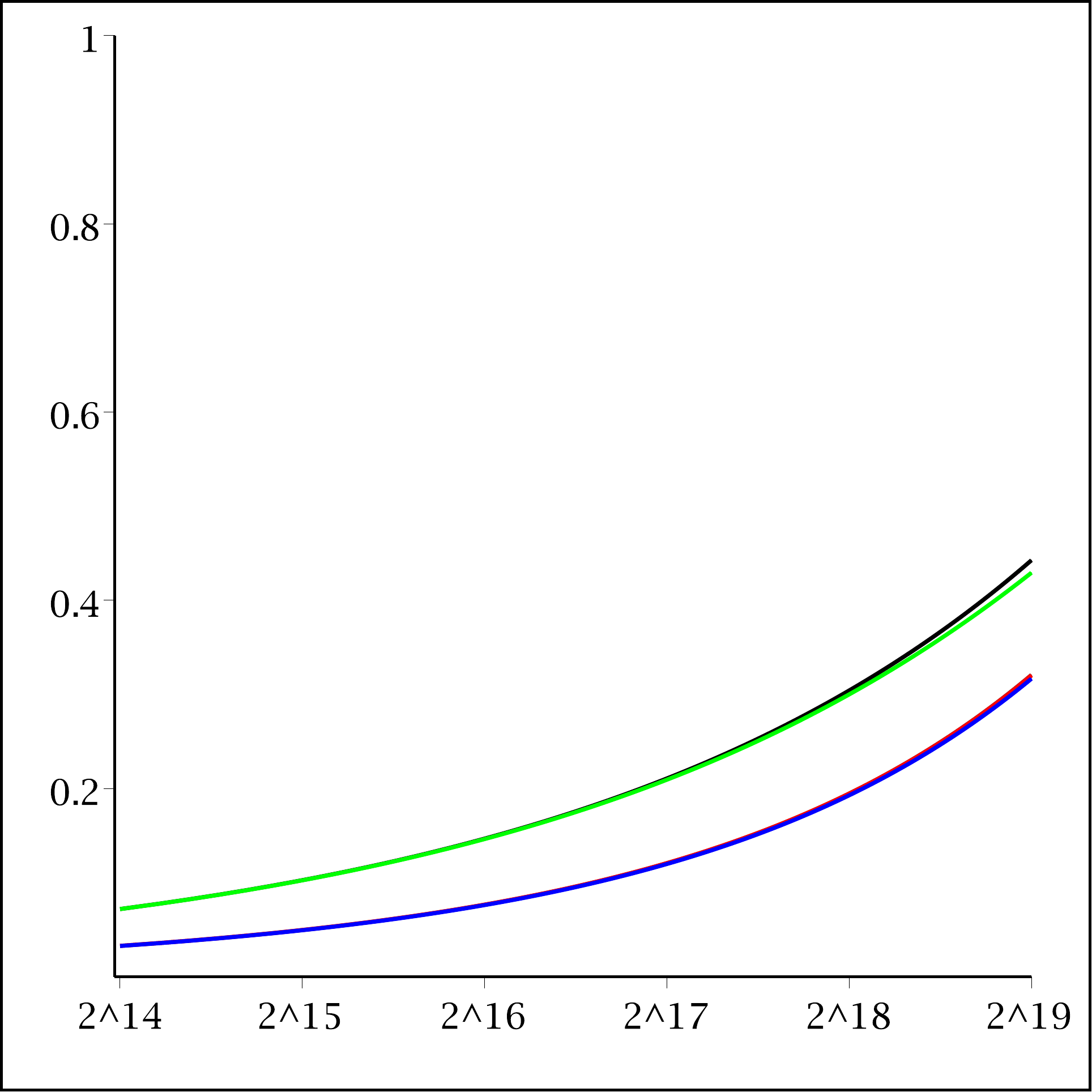}
}
\subfloat[Exponential]
{
\includegraphics[width=\SimFigWidth]{fig/EXP-R85-I300-alpha1-fixedC-appli0-platform-variation.fig}
}	
\subfloat[Weibull $k=0.7$]
{
\includegraphics[width=\SimFigWidth]{fig/WEIBULL-07-R85-I300-alpha1-fixedC-appli0-platform-variation.fig}
}
\subfloat[Weibull $k=0.5$]
{
\includegraphics[width=\SimFigWidth]{fig/WEIBULL-05-R85-I300-alpha1-fixedC-appli0-platform-variation.fig}
}
\\
\hspace{-1cm}
\subfloat{\rotatebox{90}{\setcounter{subfigure}{4}\qquad \I= 600 s}}
\subfloat[Maple]
{
\includegraphics[width=\MapleFigWidth]{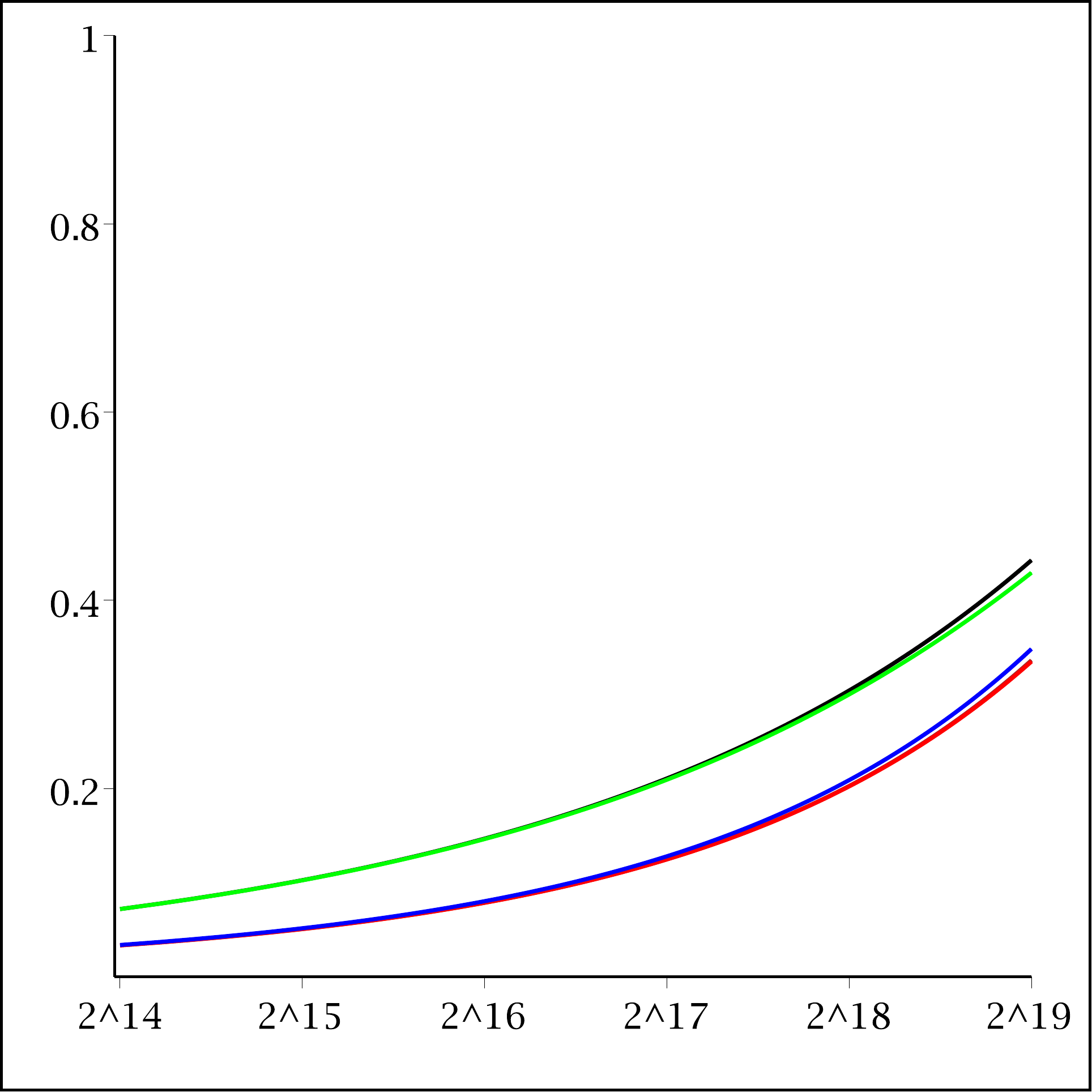}
}
\subfloat[Exponential]
{
\includegraphics[width=\SimFigWidth]{fig/EXP-R85-I600-alpha1-fixedC-appli0-platform-variation.fig}
}	
\subfloat[Weibull $k=0.7$]
{
\includegraphics[width=\SimFigWidth]{fig/WEIBULL-07-R85-I600-alpha1-fixedC-appli0-platform-variation.fig}
}
\subfloat[Weibull $k=0.5$]
{
\includegraphics[width=\SimFigWidth]{fig/WEIBULL-05-R85-I600-alpha1-fixedC-appli0-platform-variation.fig}
}
\\
\hspace{-1cm}
\subfloat{\rotatebox{90}{\setcounter{subfigure}{8}\qquad \I= 900 s}}
\subfloat[Maple]
{
\includegraphics[width=\MapleFigWidth]{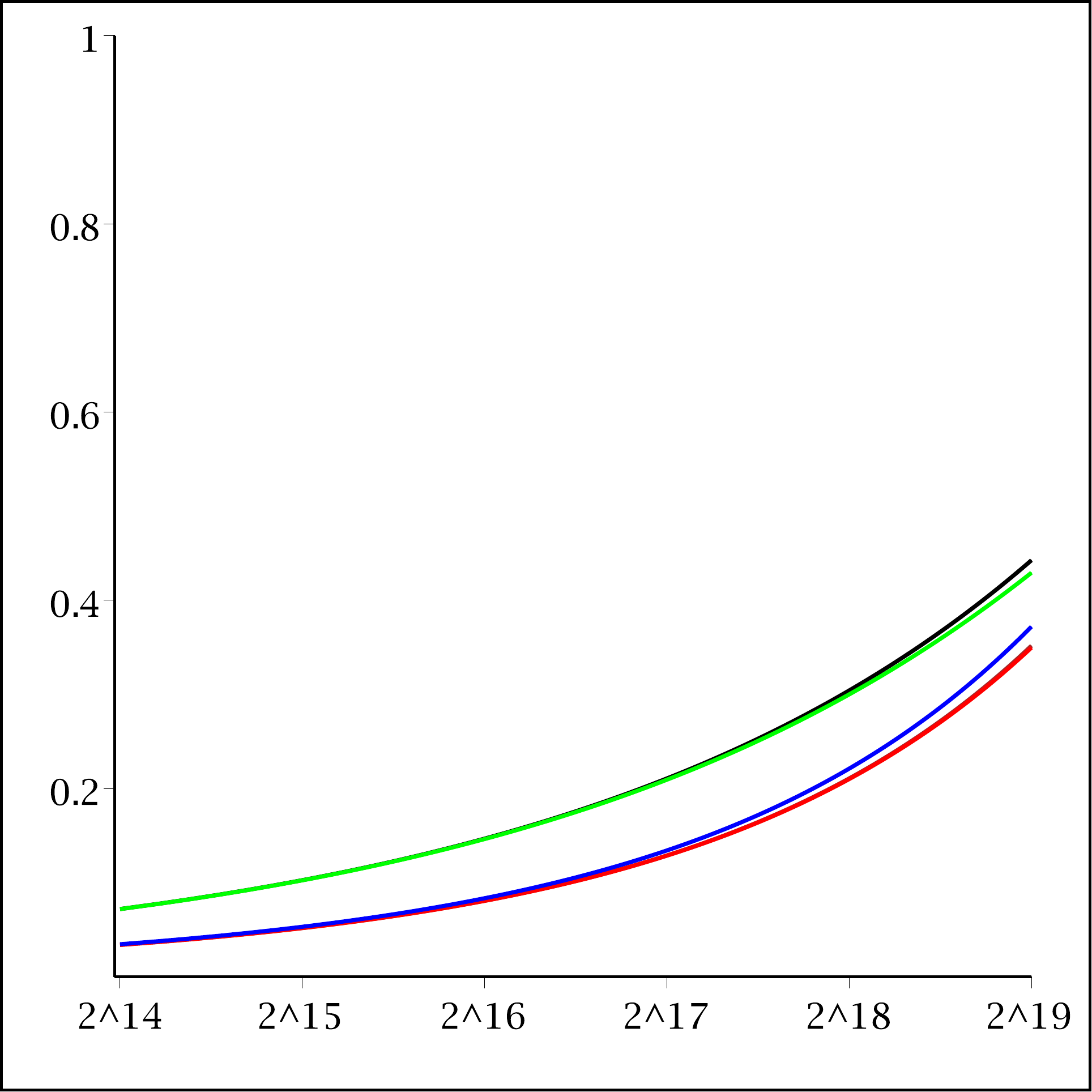}
}
\subfloat[Exponential]
{
\includegraphics[width=\SimFigWidth]{fig/EXP-R85-I900-alpha1-fixedC-appli0-platform-variation.fig}
}	
\subfloat[Weibull $k=0.7$]
{
\includegraphics[width=\SimFigWidth]{fig/WEIBULL-07-R85-I900-alpha1-fixedC-appli0-platform-variation.fig}
}
\subfloat[Weibull $k=0.5$]
{
\includegraphics[width=\SimFigWidth]{fig/WEIBULL-05-R85-I900-alpha1-fixedC-appli0-platform-variation.fig}
}
\\
\hspace{-1cm}
\subfloat{\rotatebox{90}{\setcounter{subfigure}{12}\qquad \I= 1200 s}}
\subfloat[Maple]
{
\includegraphics[width=\MapleFigWidth]{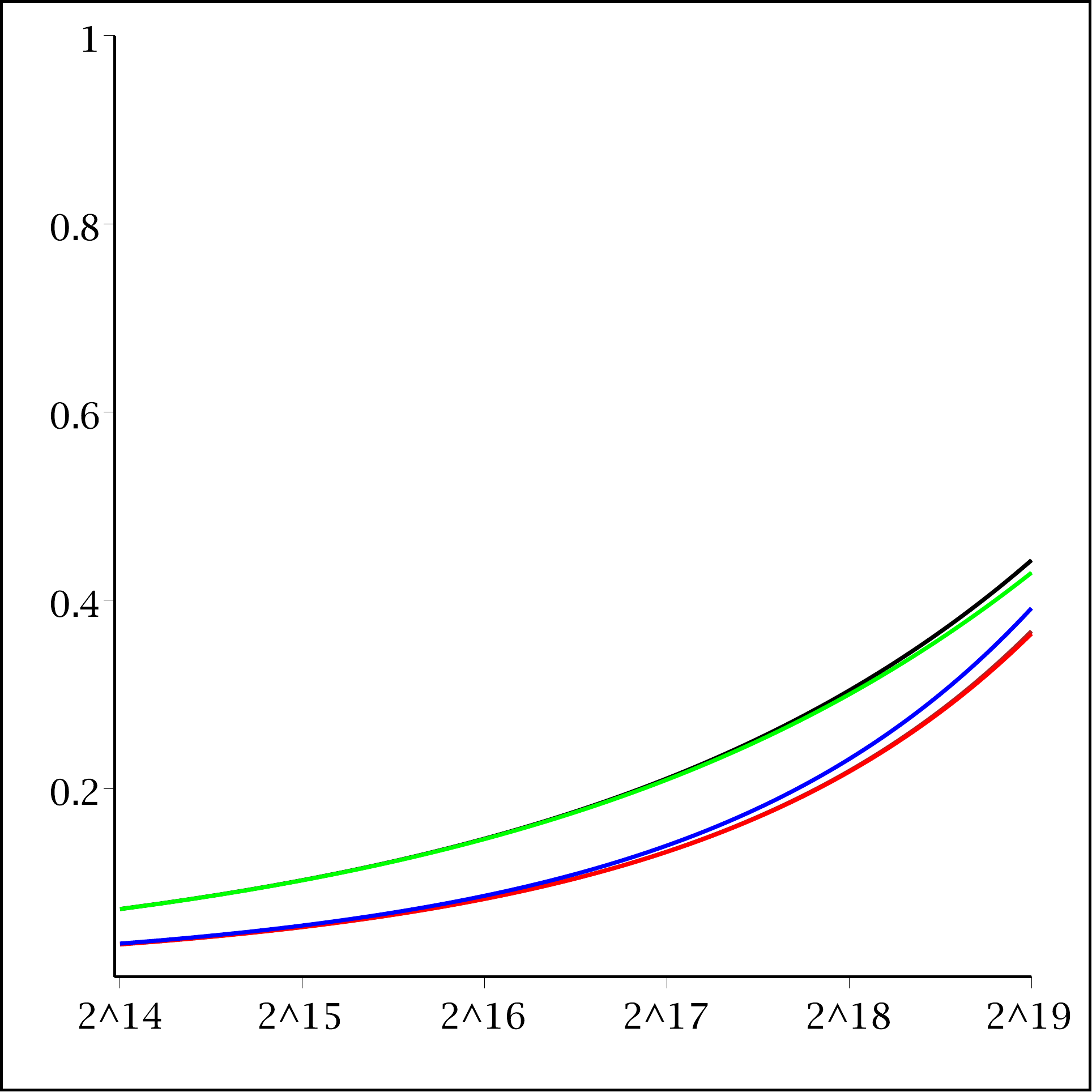}
}
\subfloat[Exponential]
{
\includegraphics[width=\SimFigWidth]{fig/EXP-R85-I1200-alpha1-fixedC-appli0-platform-variation.fig}
}	
\subfloat[Weibull $k=0.7$]
{
\includegraphics[width=\SimFigWidth]{fig/WEIBULL-07-R85-I1200-alpha1-fixedC-appli0-platform-variation.fig}
}
\subfloat[Weibull $k=0.5$]
{
\includegraphics[width=\SimFigWidth]{fig/WEIBULL-05-R85-I1200-alpha1-fixedC-appli0-platform-variation.fig}
}
\\
\hspace{-1cm}
\subfloat{\rotatebox{90}{\setcounter{subfigure}{16}\qquad \I= 3000 s}}
\subfloat[Maple]
{
\includegraphics[width=\MapleFigWidth]{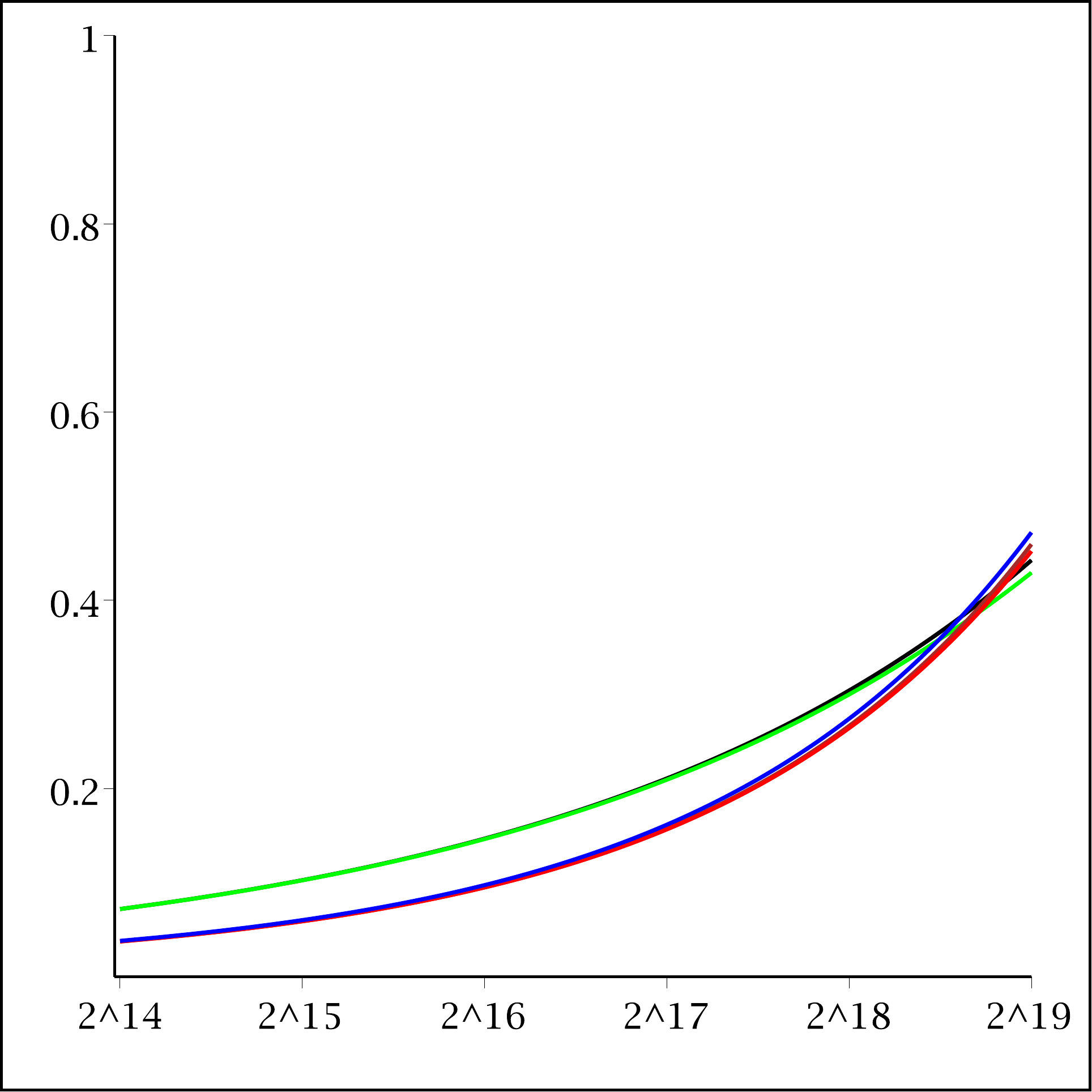}
}
\subfloat[Exponential]
{
\includegraphics[width=\SimFigWidth]{fig/EXP-R85-I3000-alpha1-fixedC-appli0-platform-variation.fig}
}	
\subfloat[Weibull $k=0.7$]
{
\includegraphics[width=\SimFigWidth]{fig/WEIBULL-07-R85-I3000-alpha1-fixedC-appli0-platform-variation.fig}
}
\subfloat[Weibull $k=0.5$]
{
\includegraphics[width=\SimFigWidth]{fig/WEIBULL-05-R85-I3000-alpha1-fixedC-appli0-platform-variation.fig}
}
\caption{Waste for the different heuristics, with $\precision=0.82$,
  $\recall=0.85$, $\Cp=\Cr$, and with a trace of false predictions
  parametrized by a distribution identical to the distribution of the
  trace of failures.}
	\label{fig.082.085.CpCr.same}
\end{figure*}

\begin{figure*}
\centering
\hspace{-1cm}
\subfloat{\rotatebox{90}{\setcounter{subfigure}{0}\qquad \I= 300 s}}
\subfloat[Maple]
{
\includegraphics[width=\MapleFigWidth]{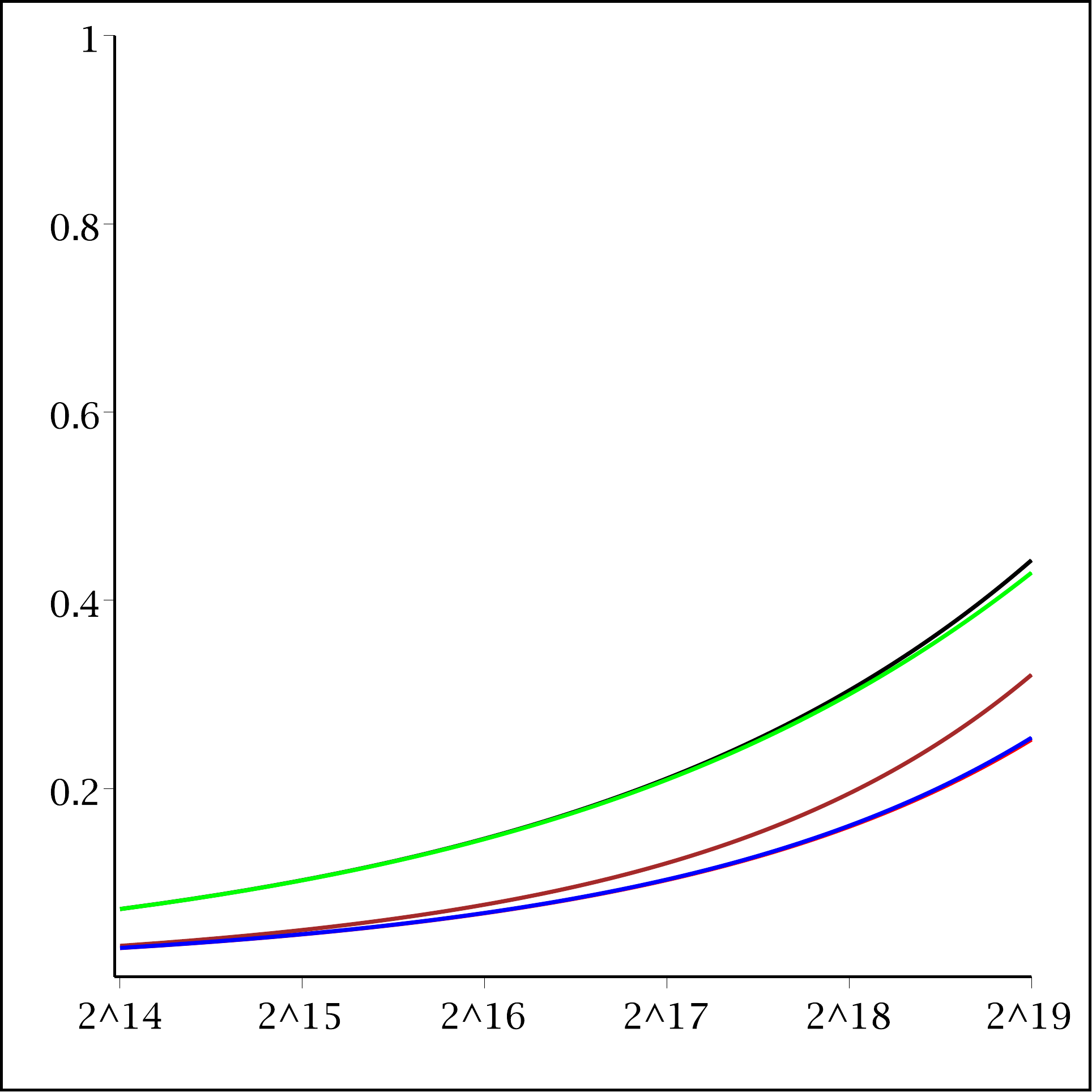}
}
\subfloat[Exponential]
{
\includegraphics[width=\SimFigWidth]{fig/EXP-R85-I300-alpha01-fixedC-appli0-platform-variation.fig}
}	
\subfloat[Weibull $k=0.7$]
{
\includegraphics[width=\SimFigWidth]{fig/WEIBULL-07-R85-I300-alpha01-fixedC-appli0-platform-variation.fig}
}
\subfloat[Weibull $k=0.5$]
{
\includegraphics[width=\SimFigWidth]{fig/WEIBULL-05-R85-I300-alpha01-fixedC-appli0-platform-variation.fig}
}
\\
\hspace{-1cm}
\subfloat{\rotatebox{90}{\setcounter{subfigure}{4}\qquad \I= 600 s}}
\subfloat[Maple]
{
\includegraphics[width=\MapleFigWidth]{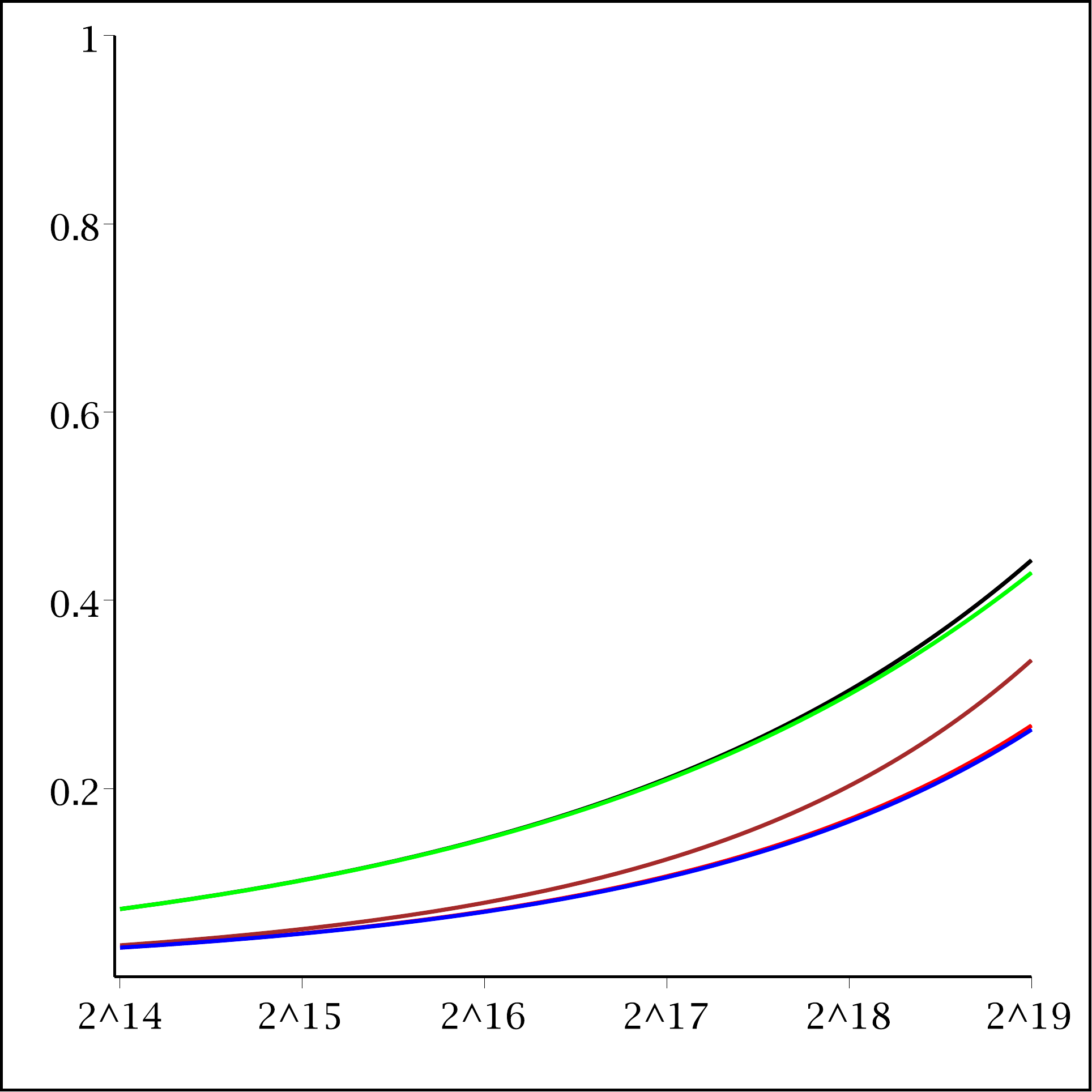}
}\subfloat[Exponential]
{
\includegraphics[width=\SimFigWidth]{fig/EXP-R85-I600-alpha01-fixedC-appli0-platform-variation.fig}
}	
\subfloat[Weibull $k=0.7$]
{
\includegraphics[width=\SimFigWidth]{fig/WEIBULL-07-R85-I600-alpha01-fixedC-appli0-platform-variation.fig}
}
\subfloat[Weibull $k=0.5$]
{
\includegraphics[width=\SimFigWidth]{fig/WEIBULL-05-R85-I600-alpha01-fixedC-appli0-platform-variation.fig}
}
\\
\hspace{-1cm}
\subfloat{\rotatebox{90}{\setcounter{subfigure}{8}\qquad \I= 900 s}}
\subfloat[Maple]
{
\includegraphics[width=\MapleFigWidth]{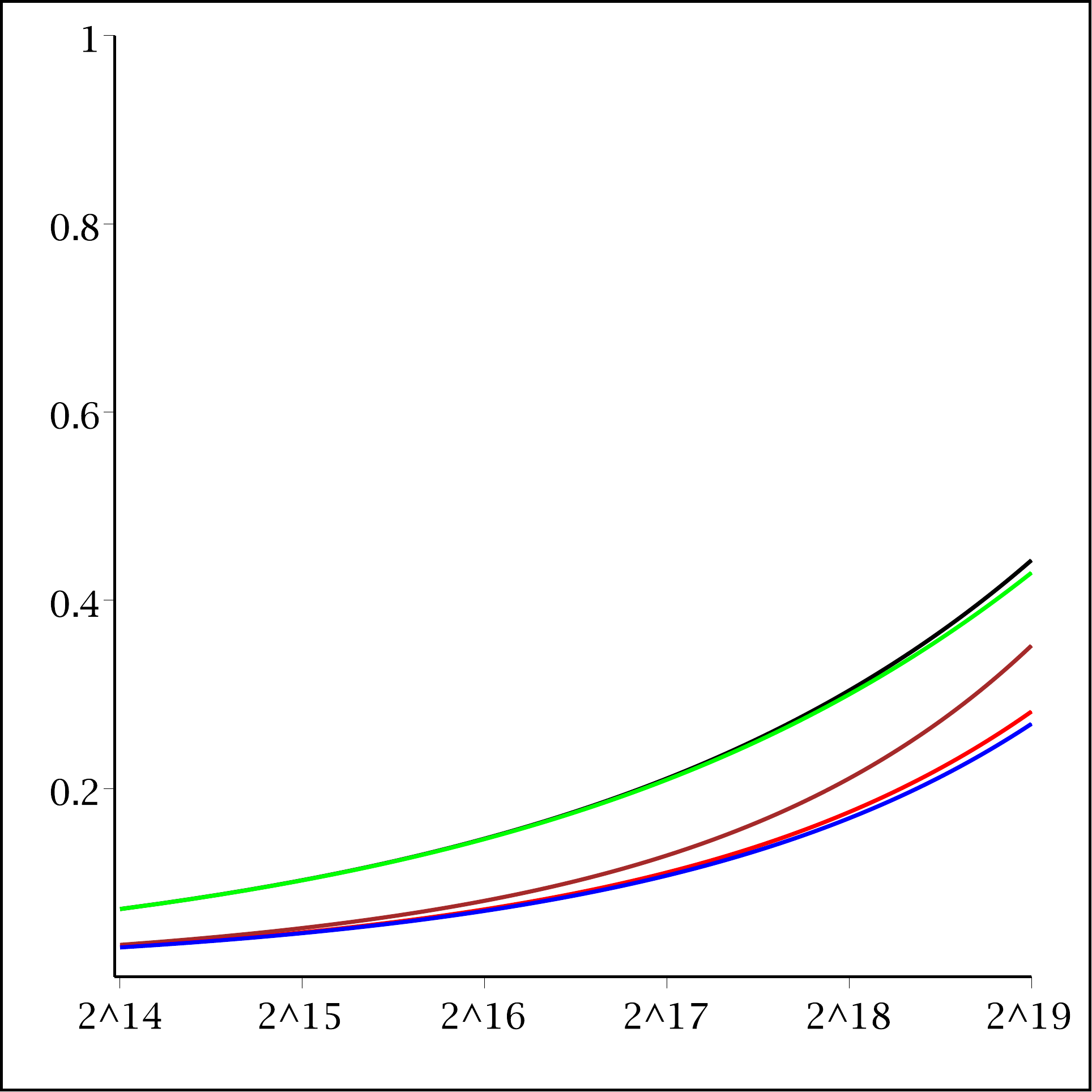}
}
\subfloat[Exponential]
{
\includegraphics[width=\SimFigWidth]{fig/EXP-R85-I900-alpha01-fixedC-appli0-platform-variation.fig}
}	
\subfloat[Weibull $k=0.7$]
{
\includegraphics[width=\SimFigWidth]{fig/WEIBULL-07-R85-I900-alpha01-fixedC-appli0-platform-variation.fig}
}
\subfloat[Weibull $k=0.5$]
{
\includegraphics[width=\SimFigWidth]{fig/WEIBULL-05-R85-I900-alpha01-fixedC-appli0-platform-variation.fig}
}
\\
\hspace{-1cm}
\subfloat{\rotatebox{90}{\setcounter{subfigure}{12}\qquad \I= 1200 s}}
\subfloat[Maple]
{
\includegraphics[width=\MapleFigWidth]{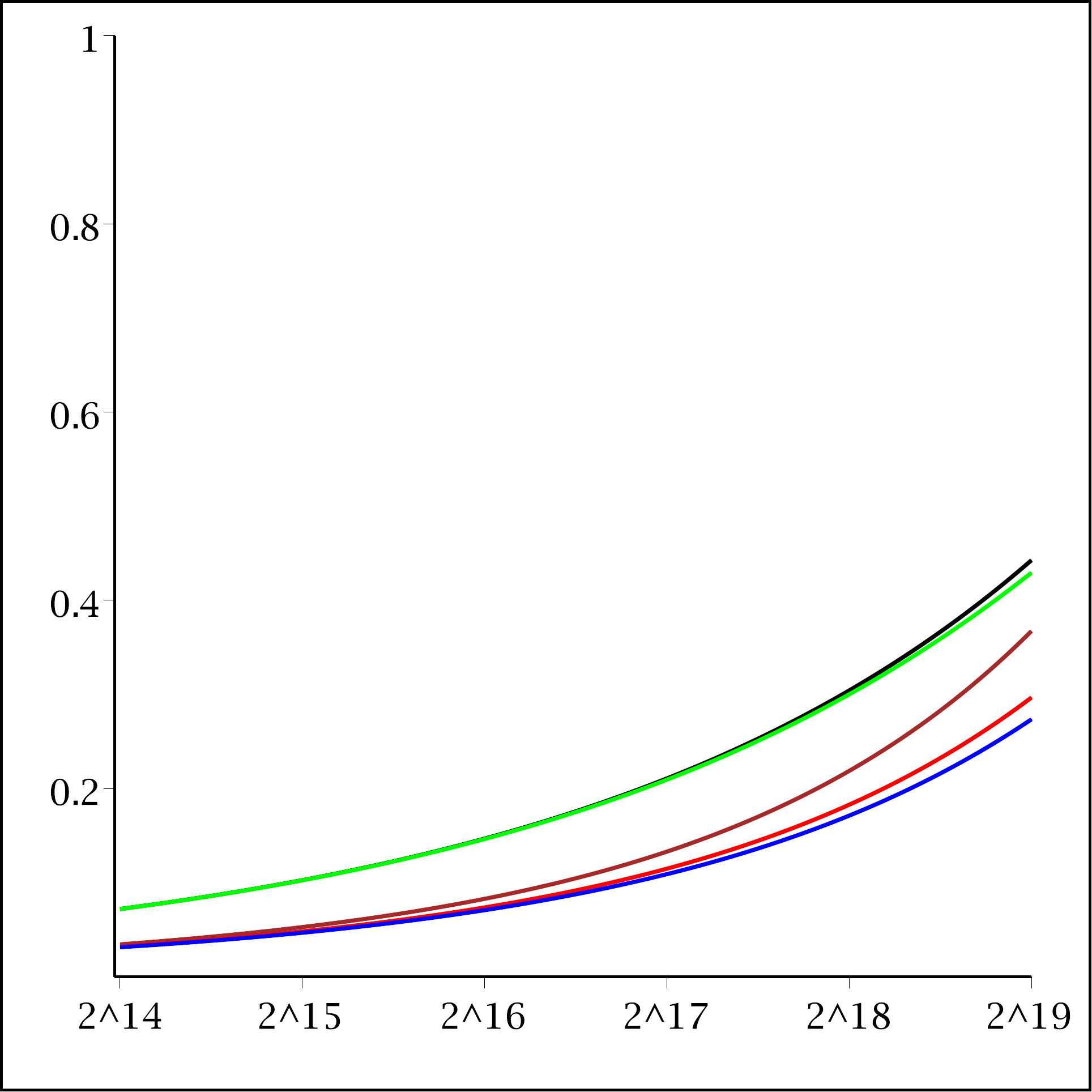}
}
\subfloat[Exponential]
{
\includegraphics[width=\SimFigWidth]{fig/EXP-R85-I1200-alpha01-fixedC-appli0-platform-variation.fig}
}	
\subfloat[Weibull $k=0.7$]
{
\includegraphics[width=\SimFigWidth]{fig/WEIBULL-07-R85-I1200-alpha01-fixedC-appli0-platform-variation.fig}
}
\subfloat[Weibull $k=0.5$]
{
\includegraphics[width=\SimFigWidth]{fig/WEIBULL-05-R85-I1200-alpha01-fixedC-appli0-platform-variation.fig}
}
\\
\hspace{-1cm}
\subfloat{\rotatebox{90}{\setcounter{subfigure}{16}\qquad \I= 3000 s}}
\subfloat[Maple]
{
\includegraphics[width=\MapleFigWidth]{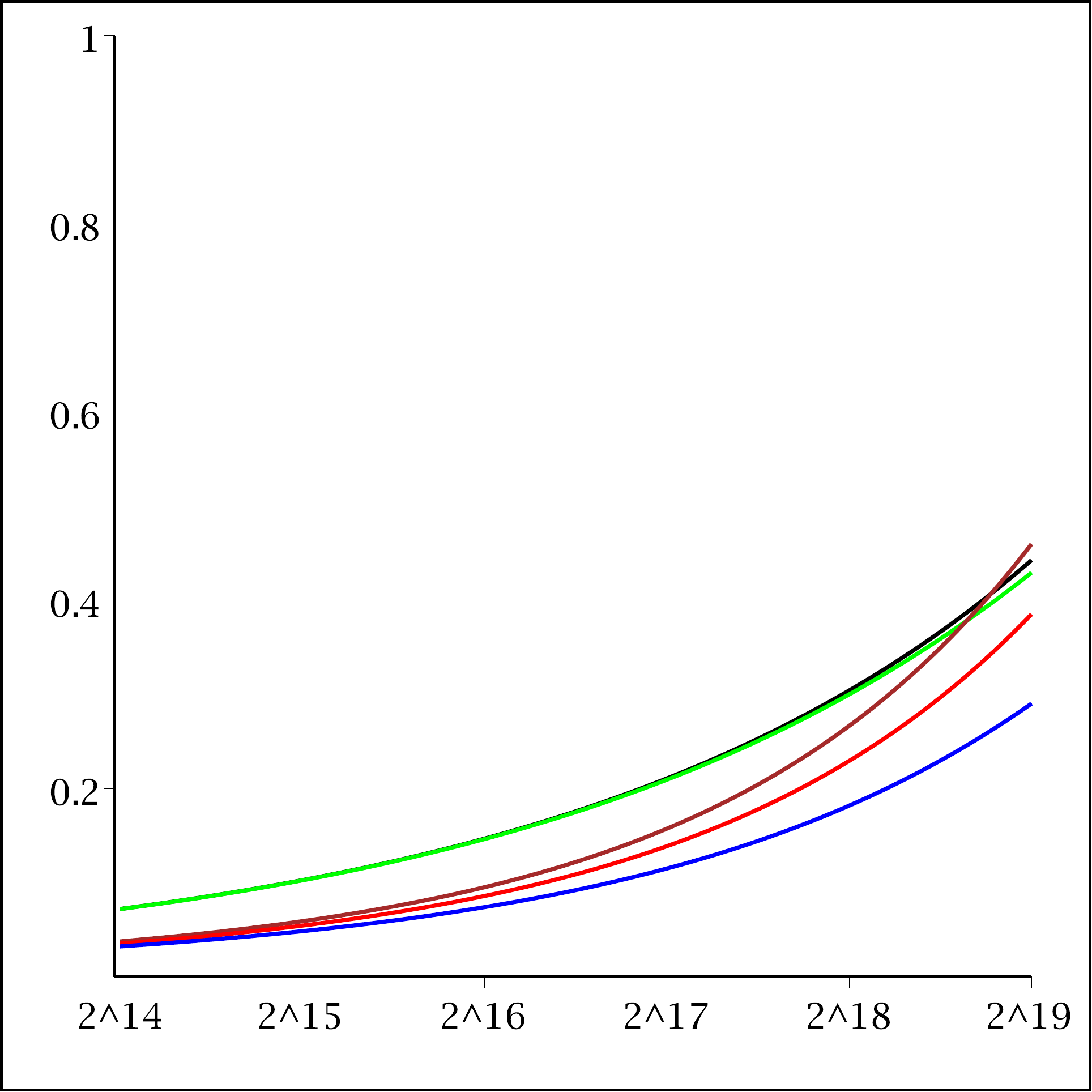}
}
\subfloat[Exponential]
{
\includegraphics[width=\SimFigWidth]{fig/EXP-R85-I3000-alpha01-fixedC-appli0-platform-variation.fig}
}	
\subfloat[Weibull $k=0.7$]
{
\includegraphics[width=\SimFigWidth]{fig/WEIBULL-07-R85-I3000-alpha01-fixedC-appli0-platform-variation.fig}
}
\subfloat[Weibull $k=0.5$]
{
\includegraphics[width=\SimFigWidth]{fig/WEIBULL-05-R85-I3000-alpha01-fixedC-appli0-platform-variation.fig}
}
\caption{Waste for the different heuristics, with $\precision=0.82$,
  $\recall=0.85$, $\Cp=0.1\Cr$, and with a trace of false predictions
  parametrized by a distribution identical to the distribution of the
  trace of failures.}
	\label{fig.082.085.Cp01Cr.same}
\end{figure*}

\begin{figure*}
\centering
\hspace{-1cm}
\subfloat{\rotatebox{90}{\setcounter{subfigure}{0}\qquad \I= 300 s}}
\subfloat[Maple]
{
\includegraphics[width=\MapleFigWidth]{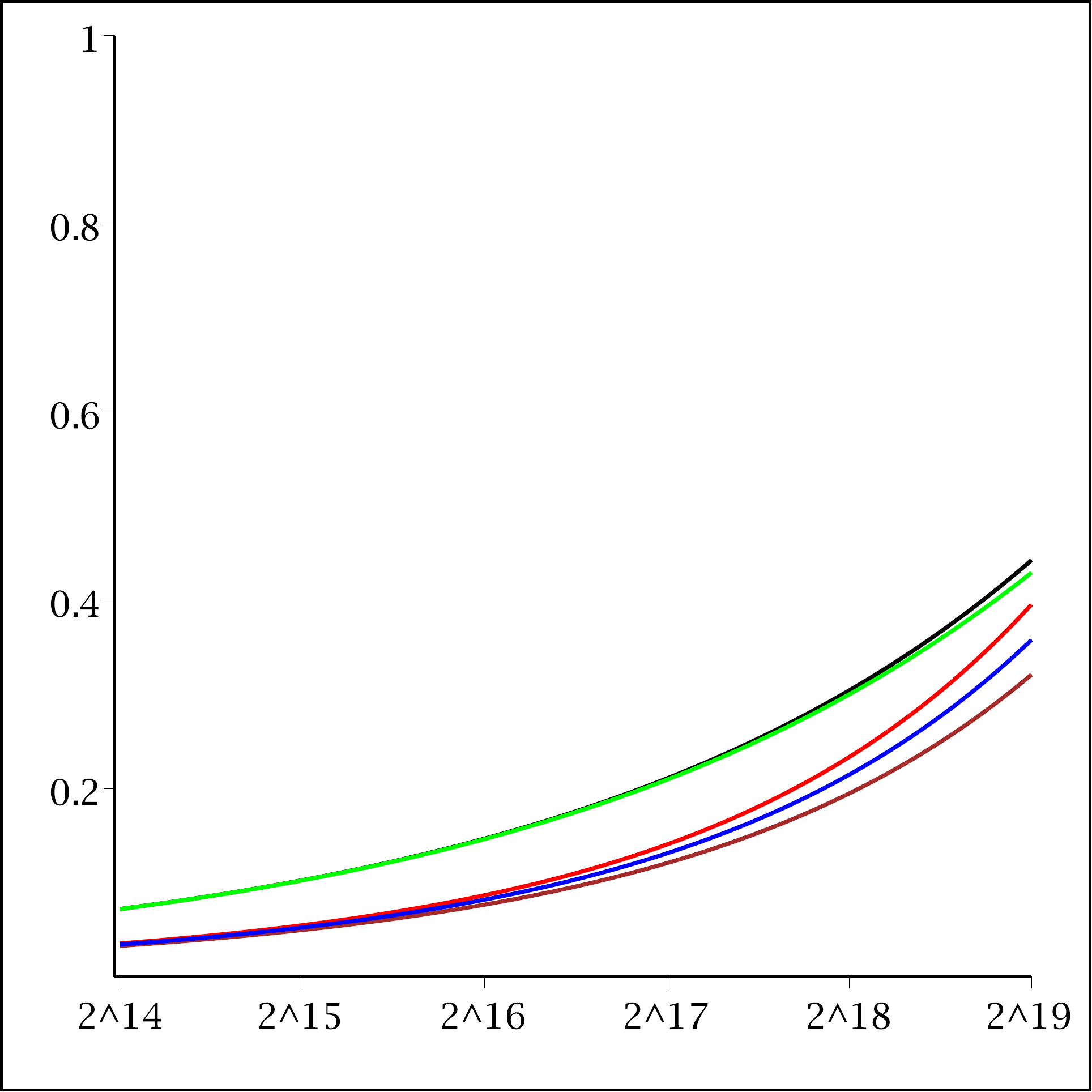}
}
\subfloat[Exponential]
{
\includegraphics[width=\SimFigWidth]{fig/EXP-R85-I300-alpha2-fixedC-appli0-platform-variation.fig}
}	
\subfloat[Weibull $k=0.7$]
{
\includegraphics[width=\SimFigWidth]{fig/WEIBULL-07-R85-I300-alpha2-fixedC-appli0-platform-variation.fig}
}
\subfloat[Weibull $k=0.5$]
{
\includegraphics[width=\SimFigWidth]{fig/WEIBULL-05-R85-I300-alpha2-fixedC-appli0-platform-variation.fig}
}
\\
\hspace{-1cm}
\subfloat{\rotatebox{90}{\setcounter{subfigure}{4}\qquad \I= 600 s}}
\subfloat[Maple]
{
\includegraphics[width=\MapleFigWidth]{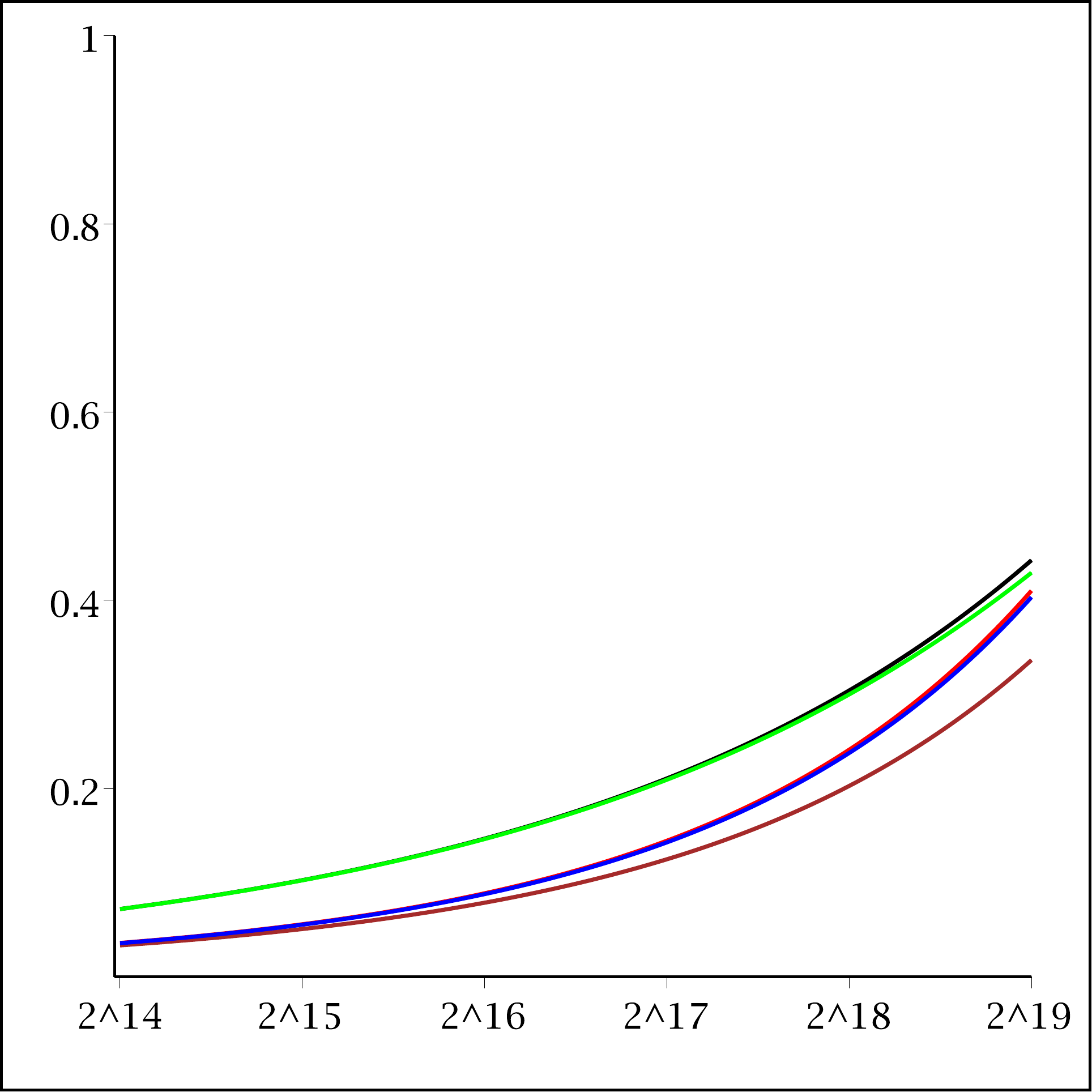}
}
\subfloat[Exponential]
{
\includegraphics[width=\SimFigWidth]{fig/EXP-R85-I600-alpha2-fixedC-appli0-platform-variation.fig}
}	
\subfloat[Weibull $k=0.7$]
{
\includegraphics[width=\SimFigWidth]{fig/WEIBULL-07-R85-I600-alpha2-fixedC-appli0-platform-variation.fig}
}
\subfloat[Weibull $k=0.5$]
{
\includegraphics[width=\SimFigWidth]{fig/WEIBULL-05-R85-I600-alpha2-fixedC-appli0-platform-variation.fig}
}
\\
\hspace{-1cm}
\subfloat{\rotatebox{90}{\setcounter{subfigure}{8}\qquad \I= 900 s}}
\subfloat[Maple]
{
\includegraphics[width=\MapleFigWidth]{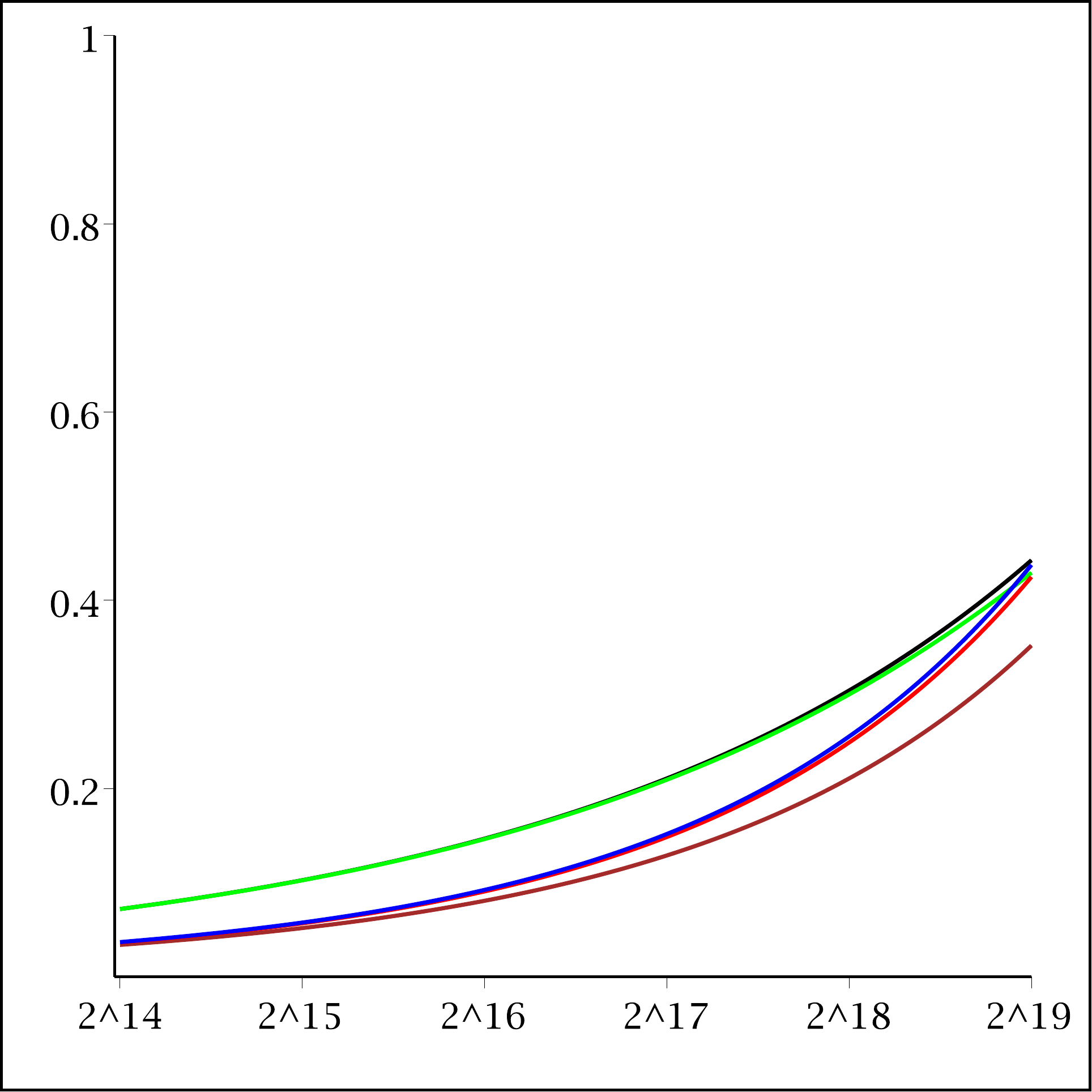}
}
\subfloat[Exponential]
{
\includegraphics[width=\SimFigWidth]{fig/EXP-R85-I900-alpha2-fixedC-appli0-platform-variation.fig}
}	
\subfloat[Weibull $k=0.7$]
{
\includegraphics[width=\SimFigWidth]{fig/WEIBULL-07-R85-I900-alpha2-fixedC-appli0-platform-variation.fig}
}
\subfloat[Weibull $k=0.5$]
{
\includegraphics[width=\SimFigWidth]{fig/WEIBULL-05-R85-I900-alpha2-fixedC-appli0-platform-variation.fig}
}
\\
\hspace{-1cm}
\subfloat{\rotatebox{90}{\setcounter{subfigure}{12}\qquad \I= 1200 s}}
\subfloat[Maple]
{
\includegraphics[width=\MapleFigWidth]{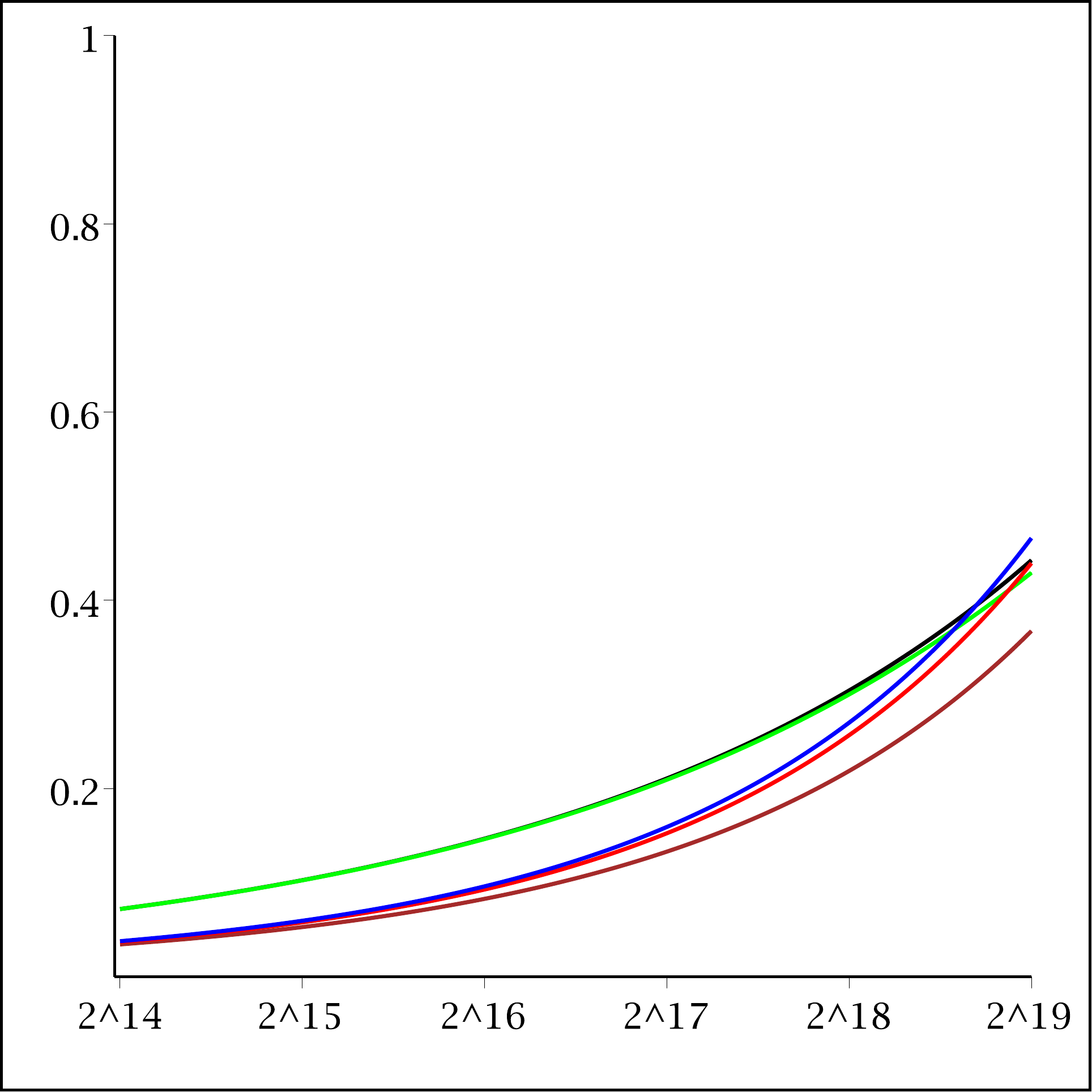}
}
\subfloat[Exponential]
{
\includegraphics[width=\SimFigWidth]{fig/EXP-R85-I1200-alpha2-fixedC-appli0-platform-variation.fig}
}	
\subfloat[Weibull $k=0.7$]
{
\includegraphics[width=\SimFigWidth]{fig/WEIBULL-07-R85-I1200-alpha2-fixedC-appli0-platform-variation.fig}
}
\subfloat[Weibull $k=0.5$]
{
\includegraphics[width=\SimFigWidth]{fig/WEIBULL-05-R85-I1200-alpha2-fixedC-appli0-platform-variation.fig}
}
\\
\hspace{-1cm}
\subfloat{\rotatebox{90}{\setcounter{subfigure}{16}\qquad \I= 3000 s}}
\subfloat[Maple]
{
\includegraphics[width=\MapleFigWidth]{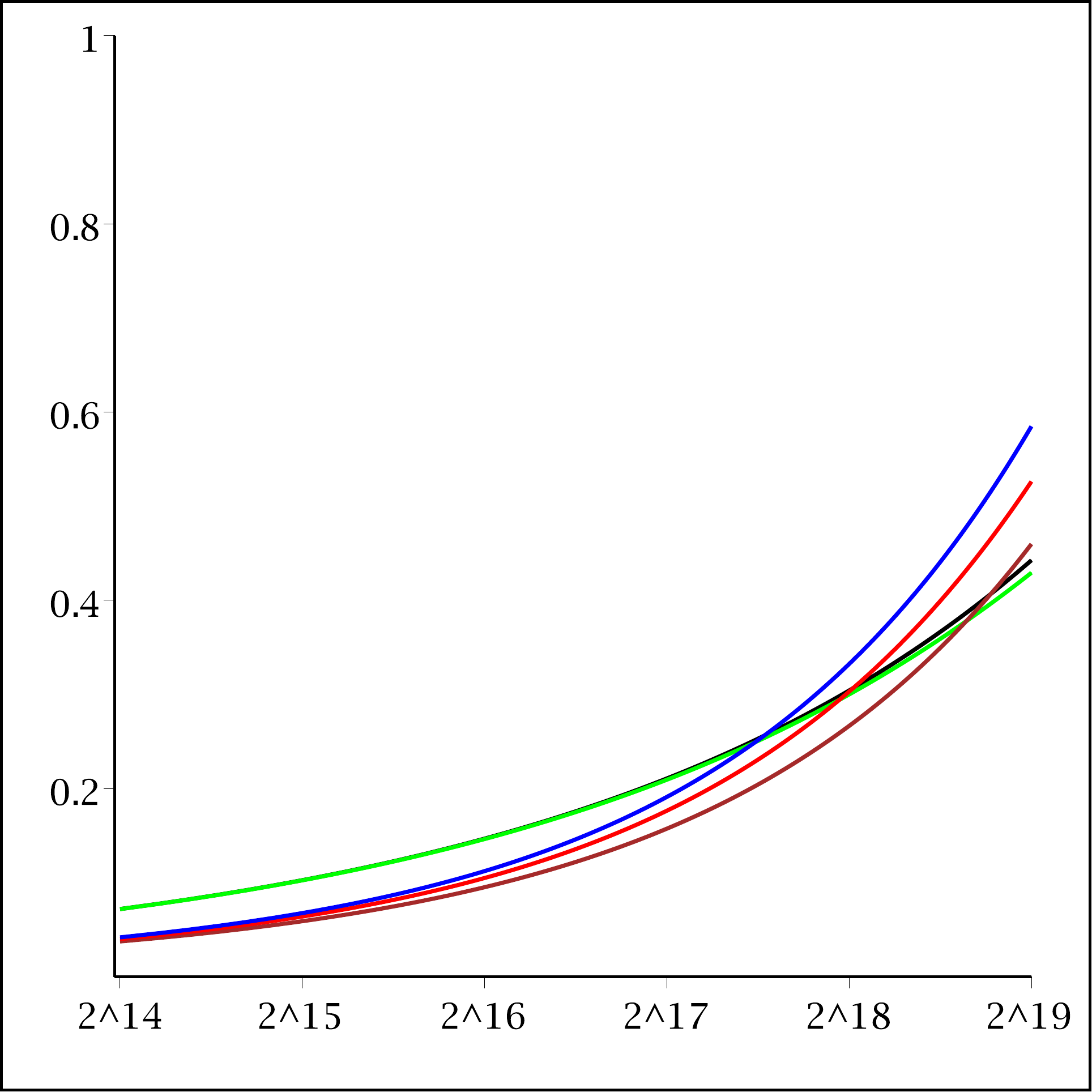}
}
\subfloat[Exponential]
{
\includegraphics[width=\SimFigWidth]{fig/EXP-R85-I3000-alpha2-fixedC-appli0-platform-variation.fig}
}	
\subfloat[Weibull $k=0.7$]
{
\includegraphics[width=\SimFigWidth]{fig/WEIBULL-07-R85-I3000-alpha2-fixedC-appli0-platform-variation.fig}
}
\subfloat[Weibull $k=0.5$]
{
\includegraphics[width=\SimFigWidth]{fig/WEIBULL-05-R85-I3000-alpha2-fixedC-appli0-platform-variation.fig}
}
\caption{Waste for the different heuristics, with $\precision=0.82$,
  $\recall=0.85$, $\Cp=2\Cr$, and with a trace of false predictions
  parametrized by a distribution identical to the distribution of the
  trace of failures.}
	\label{fig.082.085.Cp2Cr.same}
\end{figure*}

\begin{figure*}
\centering
\hspace{-1cm}
\subfloat{\rotatebox{90}{\setcounter{subfigure}{0}\qquad \I= 300 s}}
\subfloat[Maple]
{
\includegraphics[width=\MapleFigWidth]{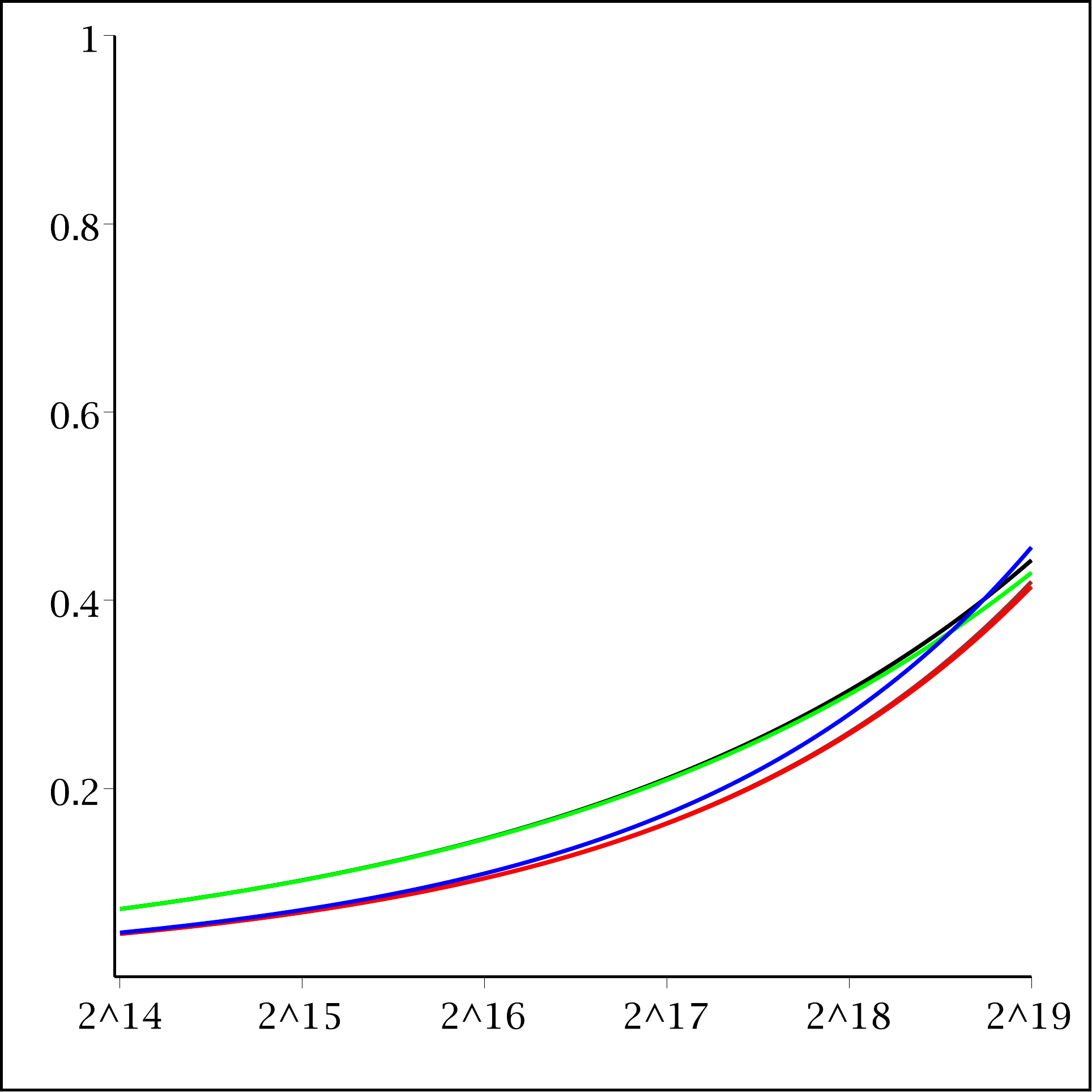}
}
\subfloat[Exponential]
{
\includegraphics[width=\SimFigWidth]{fig/EXP-R07-I300-alpha1-fixedC-appli0-platform-variation.fig}
}	
\subfloat[Weibull $k=0.7$]
{
\includegraphics[width=\SimFigWidth]{fig/WEIBULL-07-R07-I300-alpha1-fixedC-appli0-platform-variation.fig}
}
\subfloat[Weibull $k=0.5$]
{
\includegraphics[width=\SimFigWidth]{fig/WEIBULL-05-R07-I300-alpha1-fixedC-appli0-platform-variation.fig}
}
\\
\hspace{-1cm}
\subfloat{\rotatebox{90}{\setcounter{subfigure}{4}\qquad \I= 600 s}}
\subfloat[Maple]
{
\includegraphics[width=\MapleFigWidth]{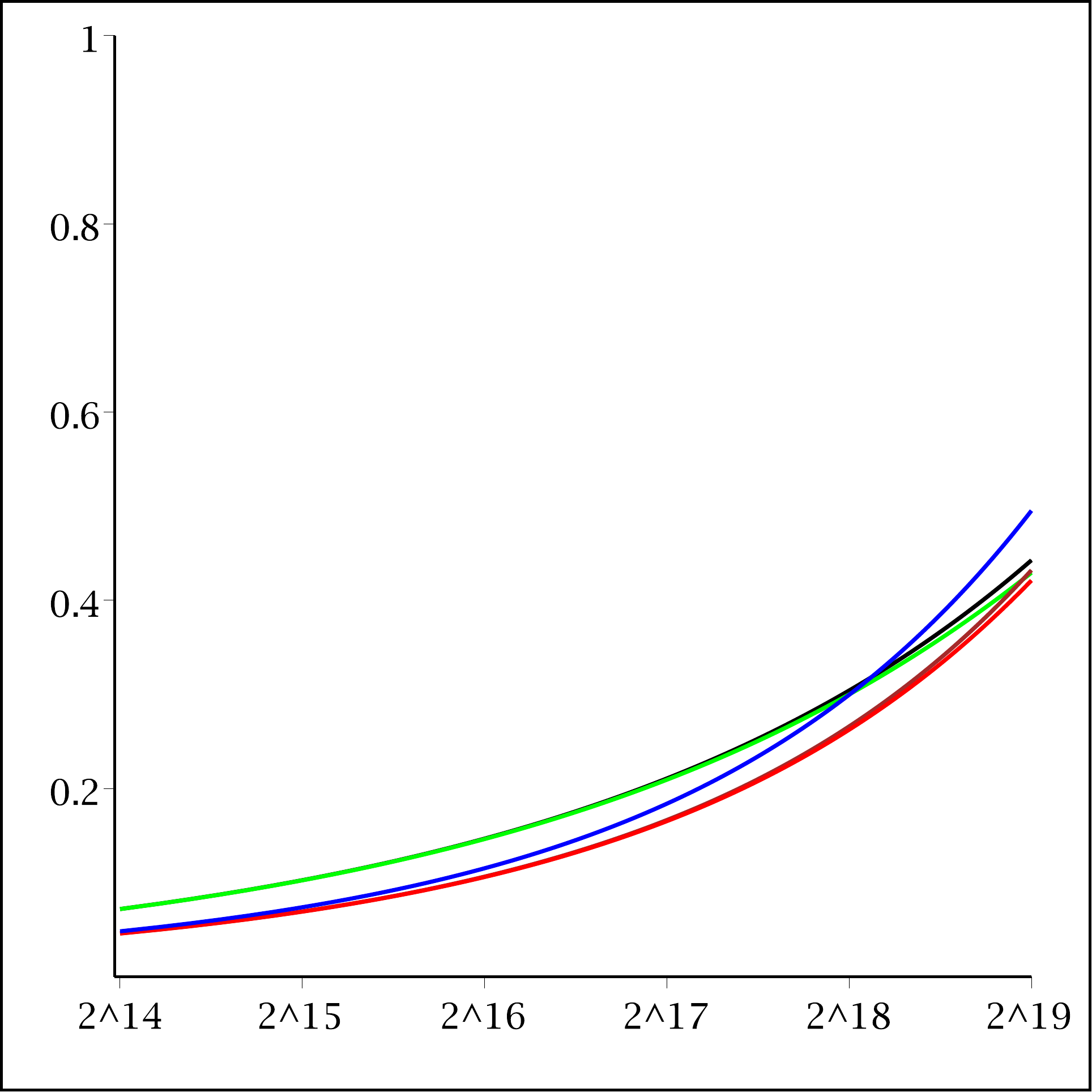}
}
\subfloat[Exponential]
{
\includegraphics[width=\SimFigWidth]{fig/EXP-R07-I600-alpha1-fixedC-appli0-platform-variation.fig}
}	
\subfloat[Weibull $k=0.7$]
{
\includegraphics[width=\SimFigWidth]{fig/WEIBULL-07-R07-I600-alpha1-fixedC-appli0-platform-variation.fig}
}
\subfloat[Weibull $k=0.5$]
{
\includegraphics[width=\SimFigWidth]{fig/WEIBULL-05-R07-I600-alpha1-fixedC-appli0-platform-variation.fig}
}
\\
\hspace{-1cm}
\subfloat{\rotatebox{90}{\setcounter{subfigure}{8}\qquad \I= 900 s}}
\subfloat[Maple]
{
\includegraphics[width=\MapleFigWidth]{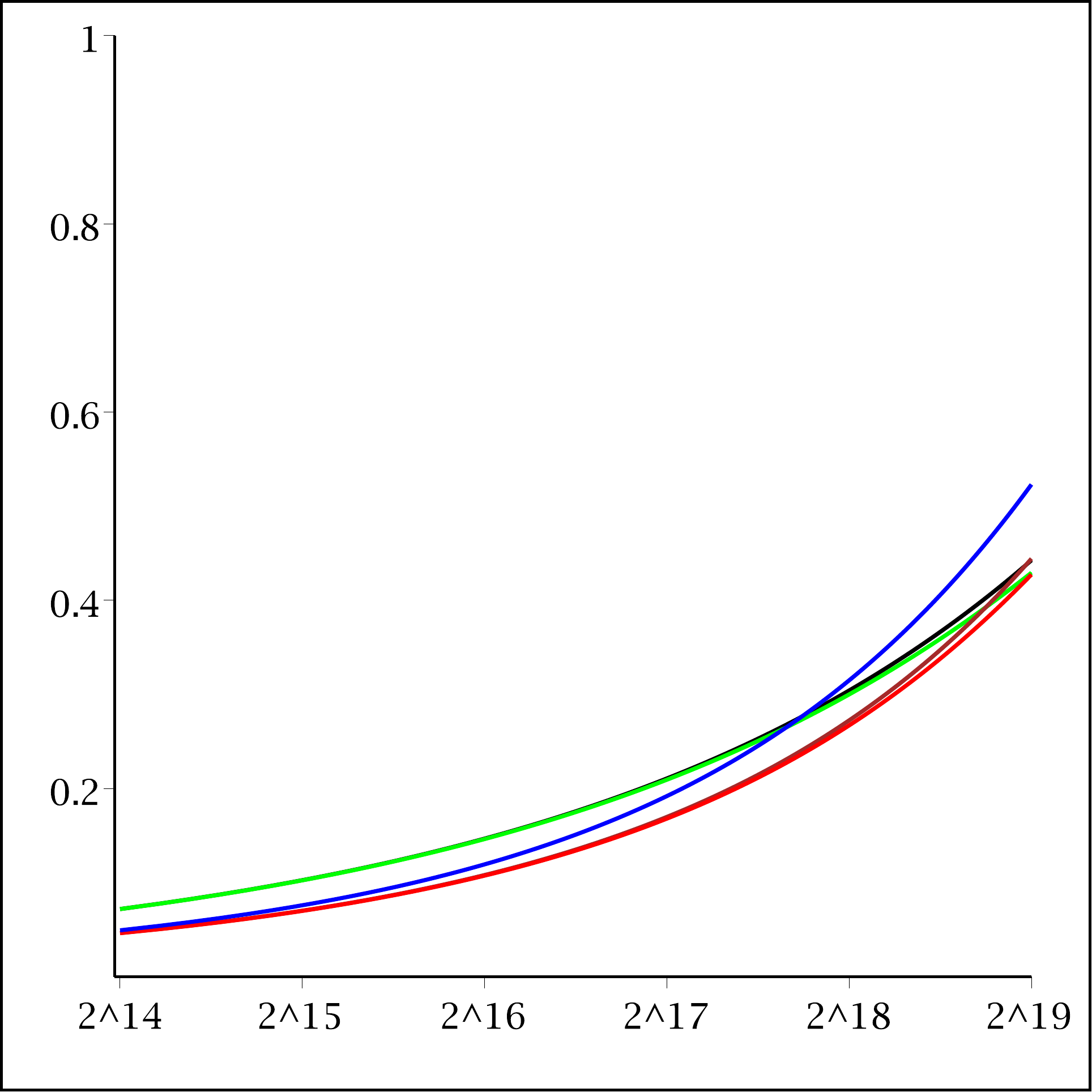}
}
\subfloat[Exponential]
{
\includegraphics[width=\SimFigWidth]{fig/EXP-R07-I900-alpha1-fixedC-appli0-platform-variation.fig}
}	
\subfloat[Weibull $k=0.7$]
{
\includegraphics[width=\SimFigWidth]{fig/WEIBULL-07-R07-I900-alpha1-fixedC-appli0-platform-variation.fig}
}
\subfloat[Weibull $k=0.5$]
{
\includegraphics[width=\SimFigWidth]{fig/WEIBULL-05-R07-I900-alpha1-fixedC-appli0-platform-variation.fig}
}
\\
\hspace{-1cm}
\subfloat{\rotatebox{90}{\setcounter{subfigure}{12}\qquad \I= 1200 s}}
\subfloat[Maple]
{
\includegraphics[width=\MapleFigWidth]{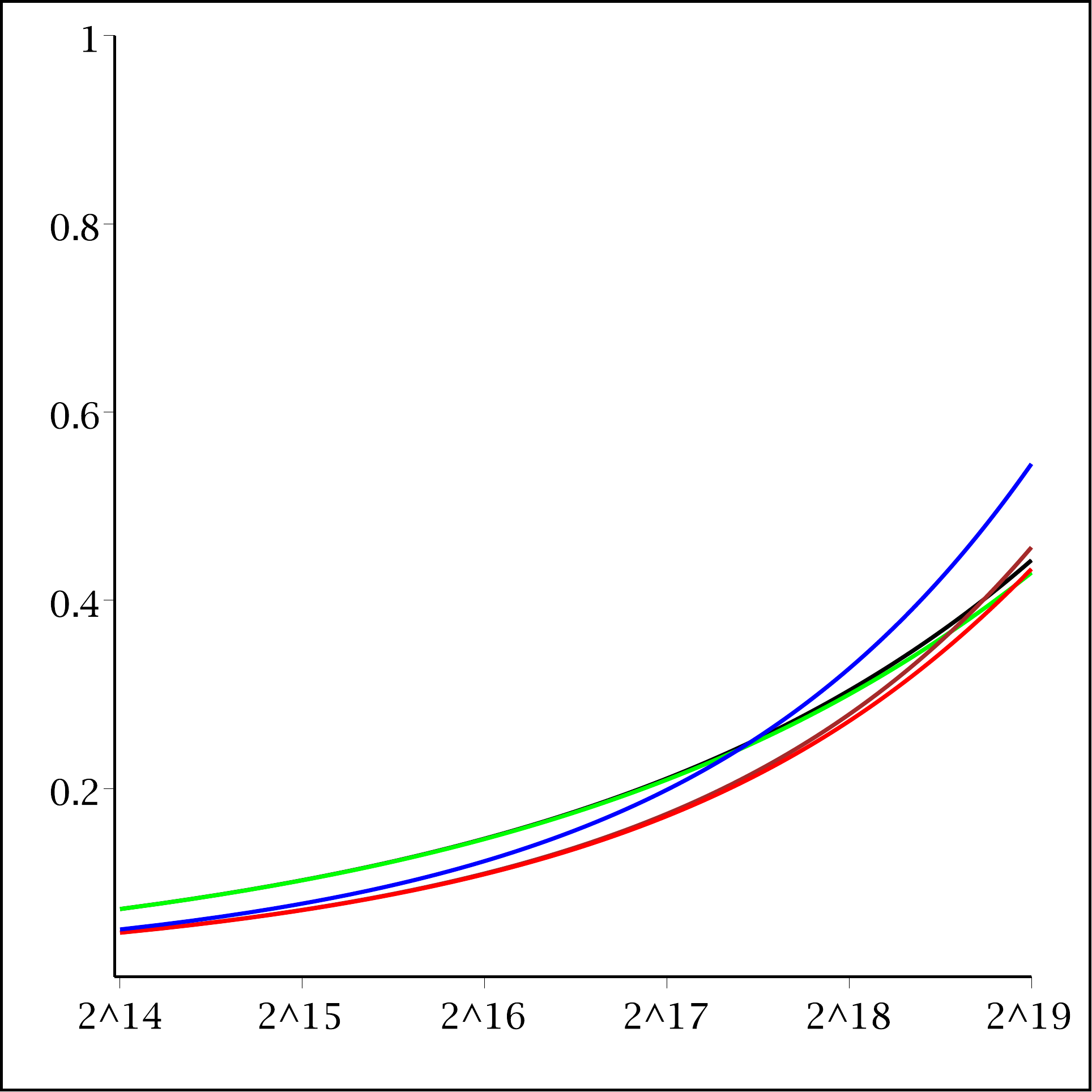}
}
\subfloat[Exponential]
{
\includegraphics[width=\SimFigWidth]{fig/EXP-R07-I1200-alpha1-fixedC-appli0-platform-variation.fig}
}	
\subfloat[Weibull $k=0.7$]
{
\includegraphics[width=\SimFigWidth]{fig/WEIBULL-07-R07-I1200-alpha1-fixedC-appli0-platform-variation.fig}
}
\subfloat[Weibull $k=0.5$]
{
\includegraphics[width=\SimFigWidth]{fig/WEIBULL-05-R07-I1200-alpha1-fixedC-appli0-platform-variation.fig}
}
\\
\hspace{-1cm}
\subfloat{\rotatebox{90}{\setcounter{subfigure}{16}\qquad \I= 3000 s}}
\subfloat[Maple]
{
\includegraphics[width=\MapleFigWidth]{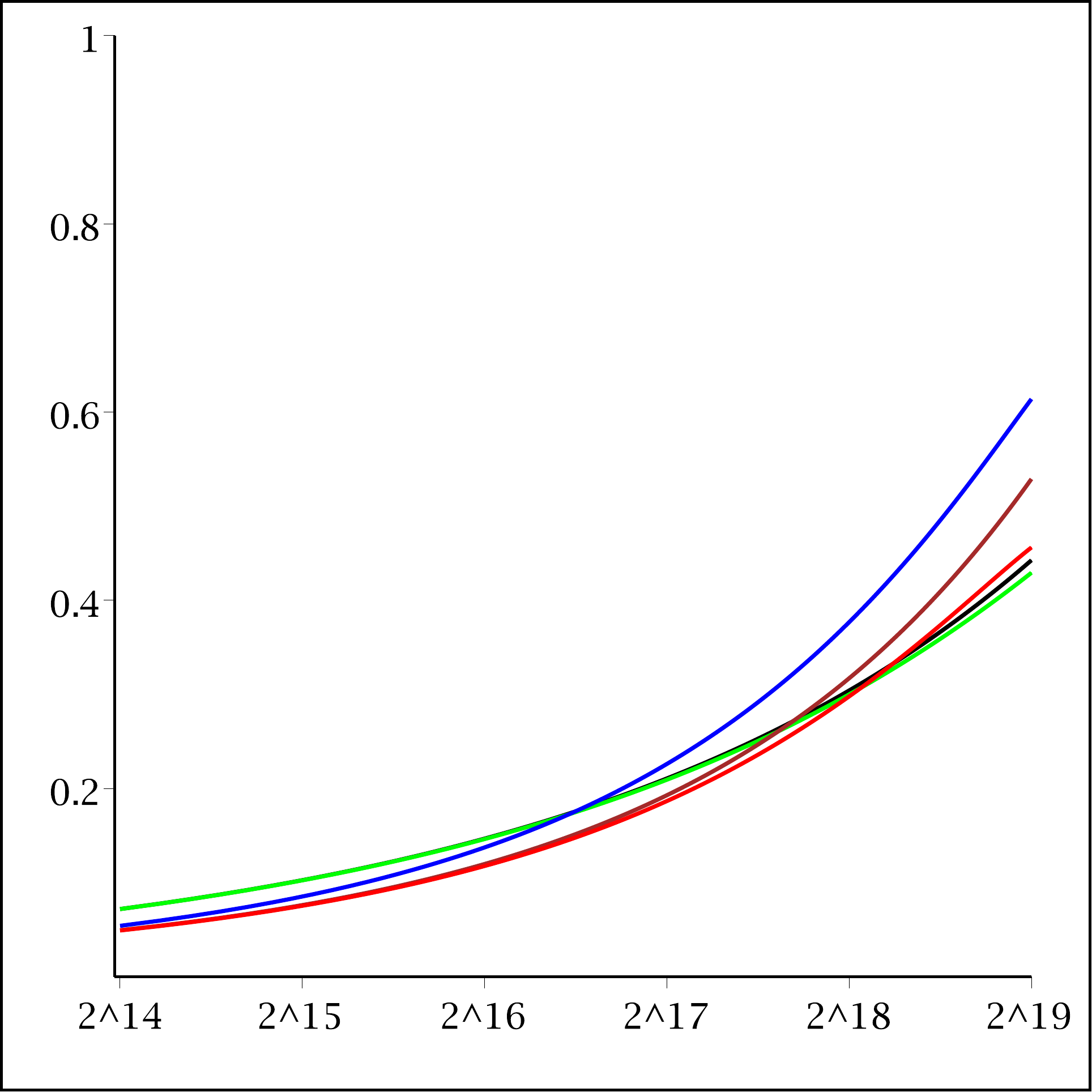}
}
\subfloat[Exponential]
{
\includegraphics[width=\SimFigWidth]{fig/EXP-R07-I3000-alpha1-fixedC-appli0-platform-variation.fig}
}	
\subfloat[Weibull $k=0.7$]
{
\includegraphics[width=\SimFigWidth]{fig/WEIBULL-07-R07-I3000-alpha1-fixedC-appli0-platform-variation.fig}
}
\subfloat[Weibull $k=0.5$]
{
\includegraphics[width=\SimFigWidth]{fig/WEIBULL-05-R07-I3000-alpha1-fixedC-appli0-platform-variation.fig}
}
\caption{Waste for the different heuristics, with $\precision=0.4$,
  $\recall=0.7$, $\Cp=\Cr$, and with a trace of false predictions
  parametrized by a distribution identical to the distribution of the
  trace of failures.}
	\label{fig.04.07.CpCr.same}
\end{figure*}

\begin{figure*}
\centering
\hspace{-1cm}
\subfloat{\rotatebox{90}{\setcounter{subfigure}{0}\qquad \I= 300 s}}
\subfloat[Maple]
{
\includegraphics[width=\MapleFigWidth]{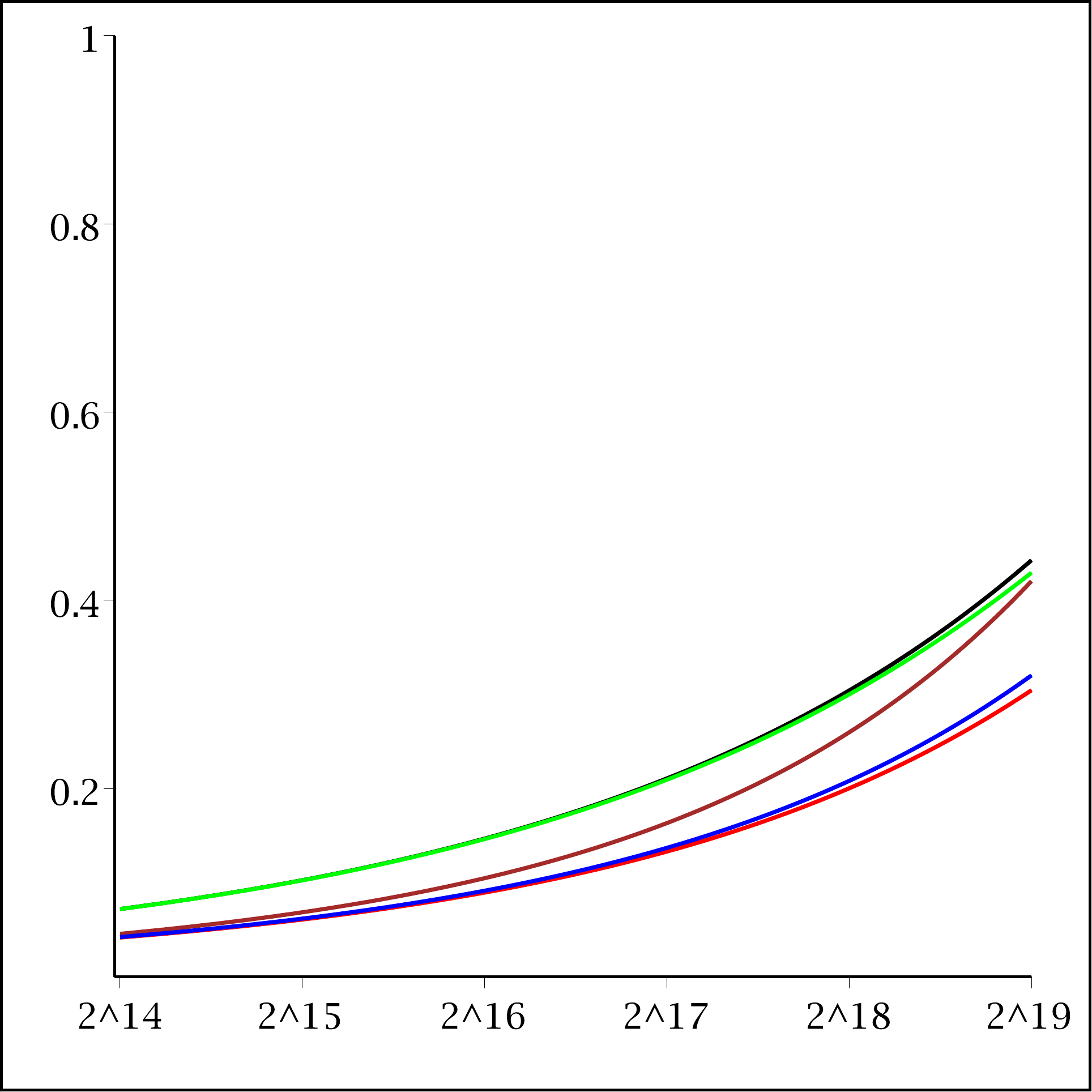}
}
\subfloat[Exponential]
{
\includegraphics[width=\SimFigWidth]{fig/EXP-R07-I300-alpha01-fixedC-appli0-platform-variation.fig}
}	
\subfloat[Weibull $k=0.7$]
{
\includegraphics[width=\SimFigWidth]{fig/WEIBULL-07-R07-I300-alpha01-fixedC-appli0-platform-variation.fig}
}
\subfloat[Weibull $k=0.5$]
{
\includegraphics[width=\SimFigWidth]{fig/WEIBULL-05-R07-I300-alpha01-fixedC-appli0-platform-variation.fig}
}
\\
\hspace{-1cm}
\subfloat{\rotatebox{90}{\setcounter{subfigure}{4}\qquad \I= 600 s}}
\subfloat[Maple]
{
\includegraphics[width=\MapleFigWidth]{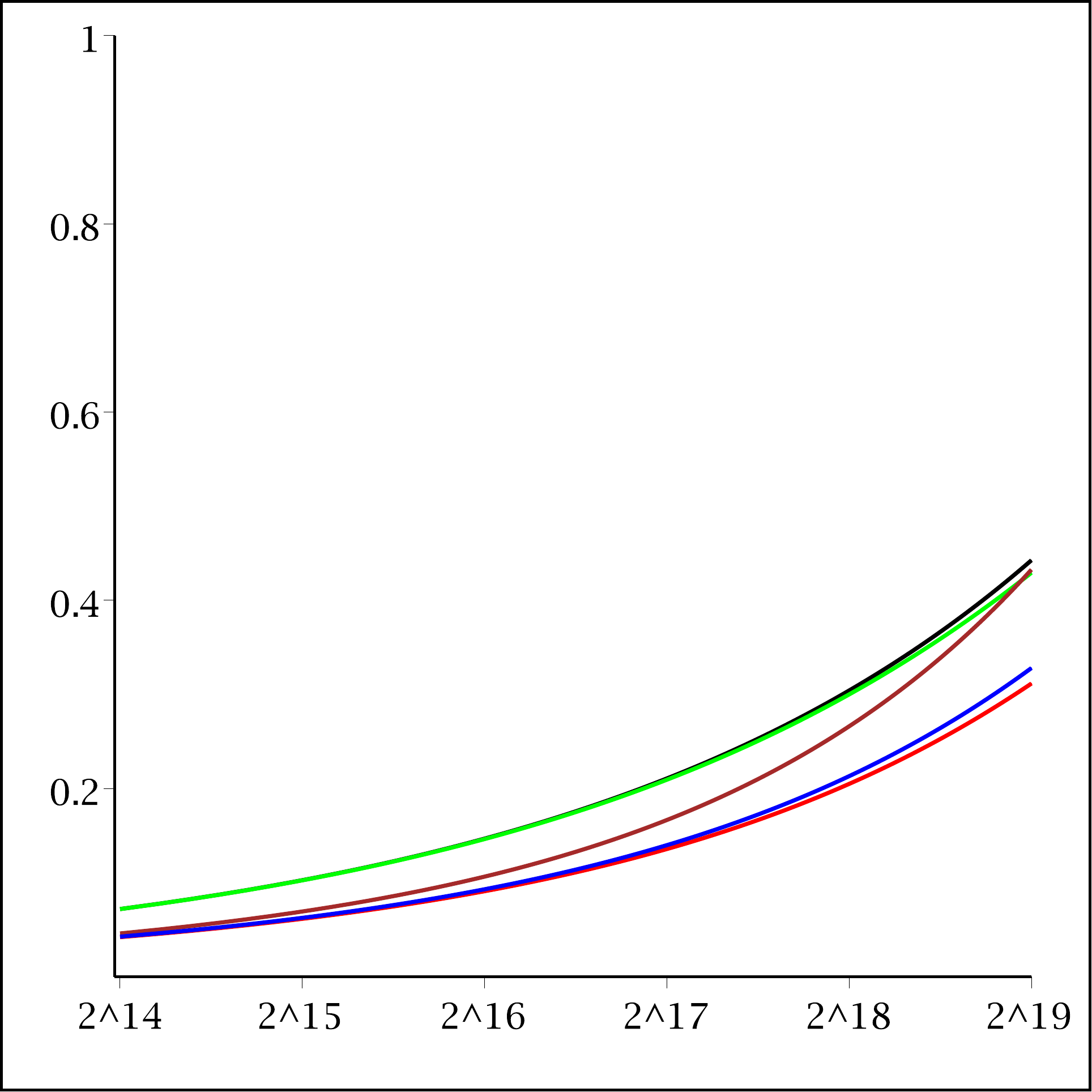}
}
\subfloat[Exponential]
{
\includegraphics[width=\SimFigWidth]{fig/EXP-R07-I600-alpha01-fixedC-appli0-platform-variation.fig}
}	
\subfloat[Weibull $k=0.7$]
{
\includegraphics[width=\SimFigWidth]{fig/WEIBULL-07-R07-I600-alpha01-fixedC-appli0-platform-variation.fig}
}
\subfloat[Weibull $k=0.5$]
{
\includegraphics[width=\SimFigWidth]{fig/WEIBULL-05-R07-I600-alpha01-fixedC-appli0-platform-variation.fig}
}
\\
\hspace{-1cm}
\subfloat{\rotatebox{90}{\setcounter{subfigure}{8}\qquad \I= 900 s}}
\subfloat[Maple]
{
\includegraphics[width=\MapleFigWidth]{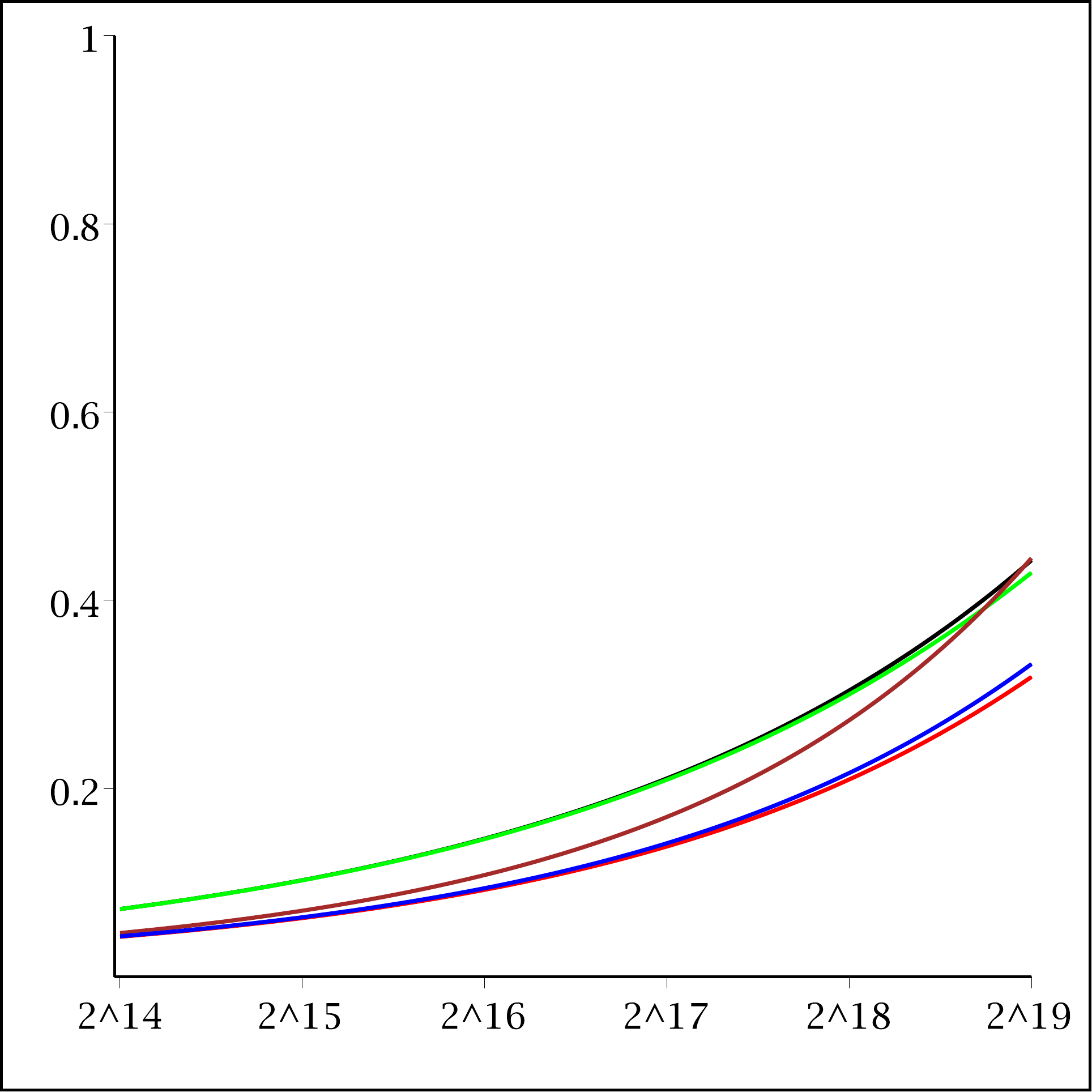}
}
\subfloat[Exponential]
{
\includegraphics[width=\SimFigWidth]{fig/EXP-R07-I900-alpha01-fixedC-appli0-platform-variation.fig}
}	
\subfloat[Weibull $k=0.7$]
{
\includegraphics[width=\SimFigWidth]{fig/WEIBULL-07-R07-I900-alpha01-fixedC-appli0-platform-variation.fig}
}
\subfloat[Weibull $k=0.5$]
{
\includegraphics[width=\SimFigWidth]{fig/WEIBULL-05-R07-I900-alpha01-fixedC-appli0-platform-variation.fig}
}
\\
\hspace{-1cm}
\subfloat{\rotatebox{90}{\setcounter{subfigure}{12}\qquad \I= 1200 s}}
\subfloat[Maple]
{
\includegraphics[width=\MapleFigWidth]{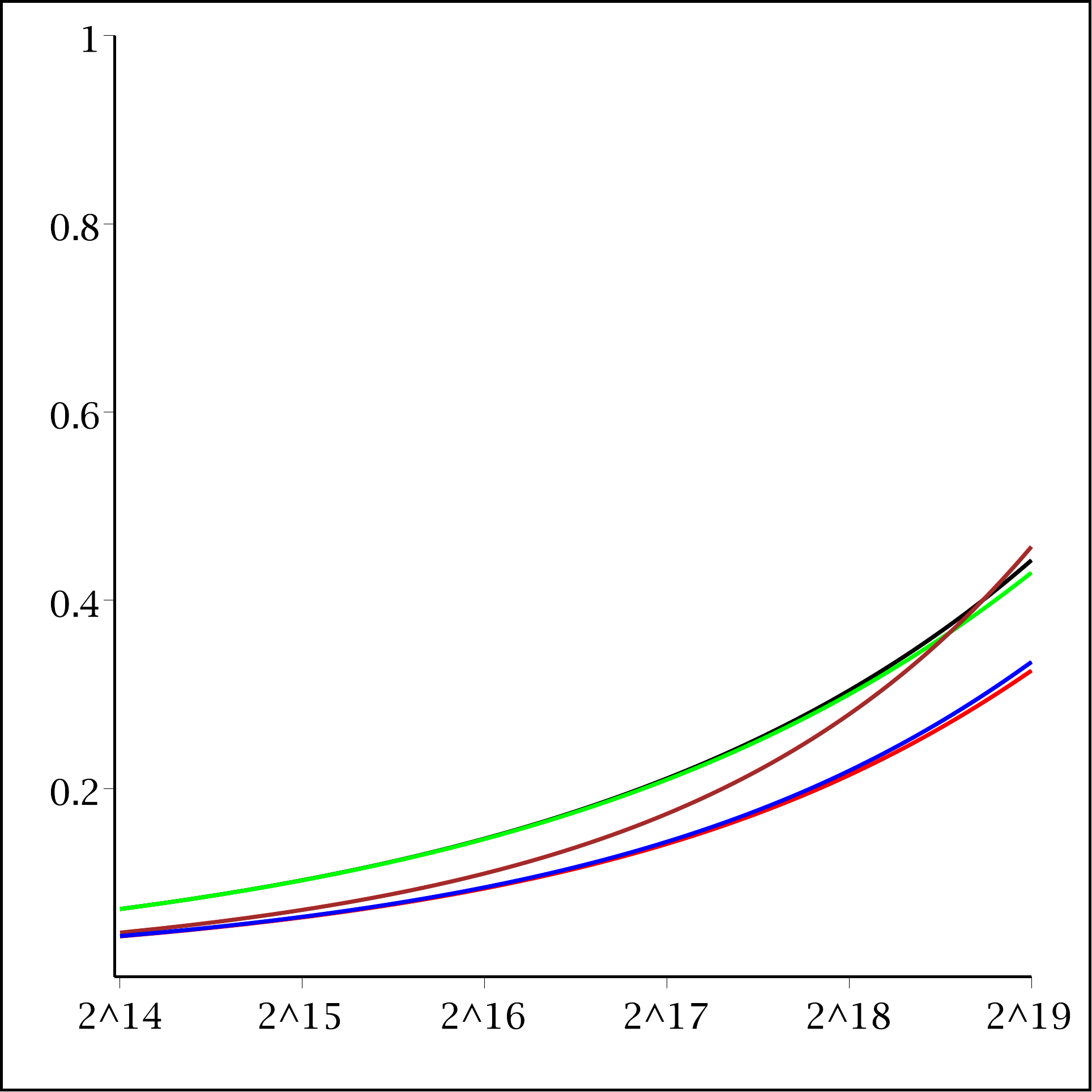}
}
\subfloat[Exponential]
{
\includegraphics[width=\SimFigWidth]{fig/EXP-R07-I1200-alpha01-fixedC-appli0-platform-variation.fig}
}	
\subfloat[Weibull $k=0.7$]
{
\includegraphics[width=\SimFigWidth]{fig/WEIBULL-07-R07-I1200-alpha01-fixedC-appli0-platform-variation.fig}
}
\subfloat[Weibull $k=0.5$]
{
\includegraphics[width=\SimFigWidth]{fig/WEIBULL-05-R07-I1200-alpha01-fixedC-appli0-platform-variation.fig}
}
\\
\hspace{-1cm}
\subfloat{\rotatebox{90}{\setcounter{subfigure}{16}\qquad \I= 3000 s}}
\subfloat[Maple]
{
\includegraphics[width=\MapleFigWidth]{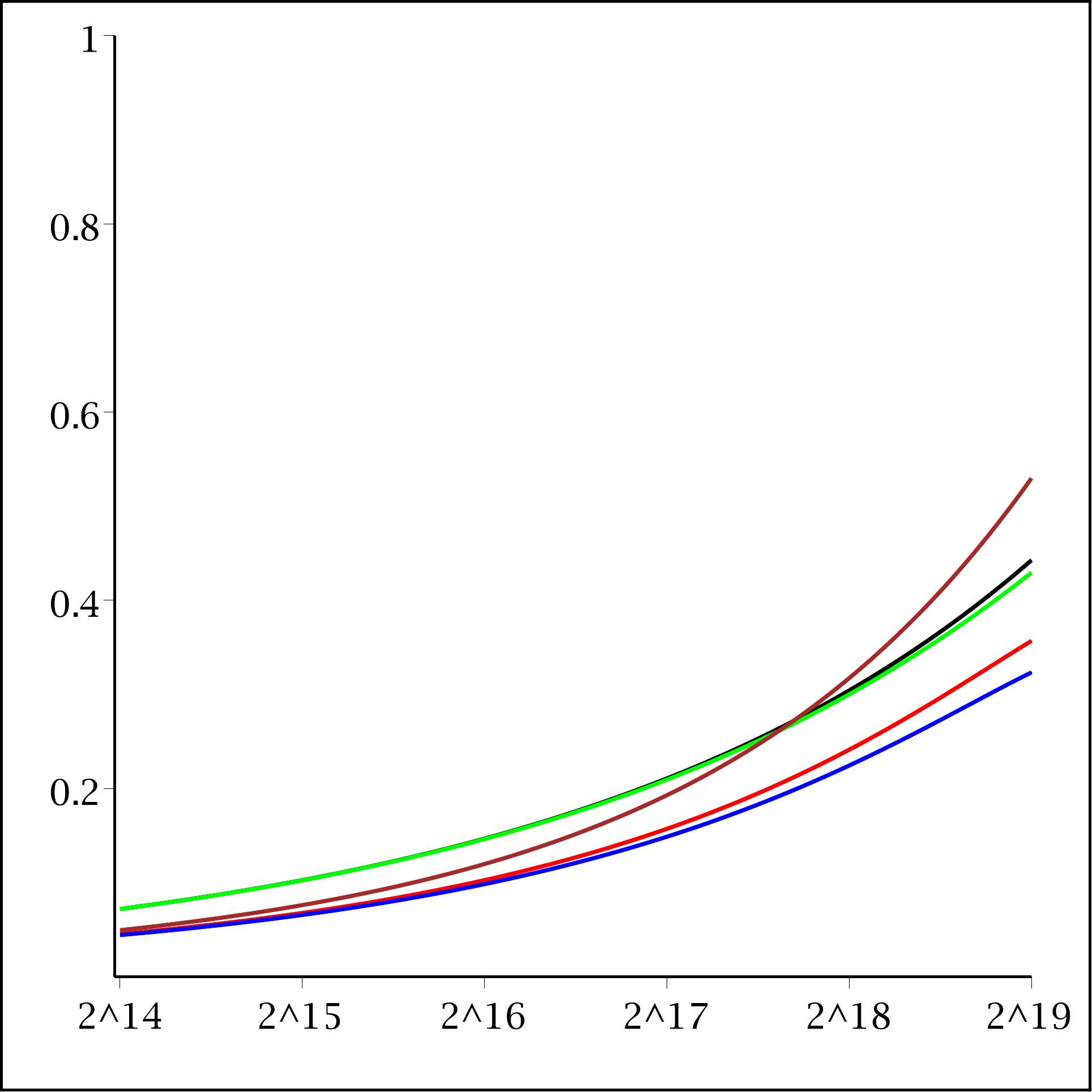}
}
\subfloat[Exponential]
{
\includegraphics[width=\SimFigWidth]{fig/EXP-R07-I3000-alpha01-fixedC-appli0-platform-variation.fig}
}	
\subfloat[Weibull $k=0.7$]
{
\includegraphics[width=\SimFigWidth]{fig/WEIBULL-07-R07-I3000-alpha01-fixedC-appli0-platform-variation.fig}
}
\subfloat[Weibull $k=0.5$]
{
\includegraphics[width=\SimFigWidth]{fig/WEIBULL-05-R07-I3000-alpha01-fixedC-appli0-platform-variation.fig}
}
\caption{Waste for the different heuristics, with\,$\precision=0.4$,
  $\recall=0.7$, $\Cp=0.1\Cr$, and with a trace of false predictions
  parametrized by a distribution identical to the distribution of the
  trace of failures.}
\label{fig.04.07.Cp01Cr.same}
\end{figure*}

\begin{figure*}
\centering
\hspace{-1cm}
\subfloat{\rotatebox{90}{\setcounter{subfigure}{0}\qquad \I= 300 s}}
\subfloat[Maple]
{
\includegraphics[width=\MapleFigWidth]{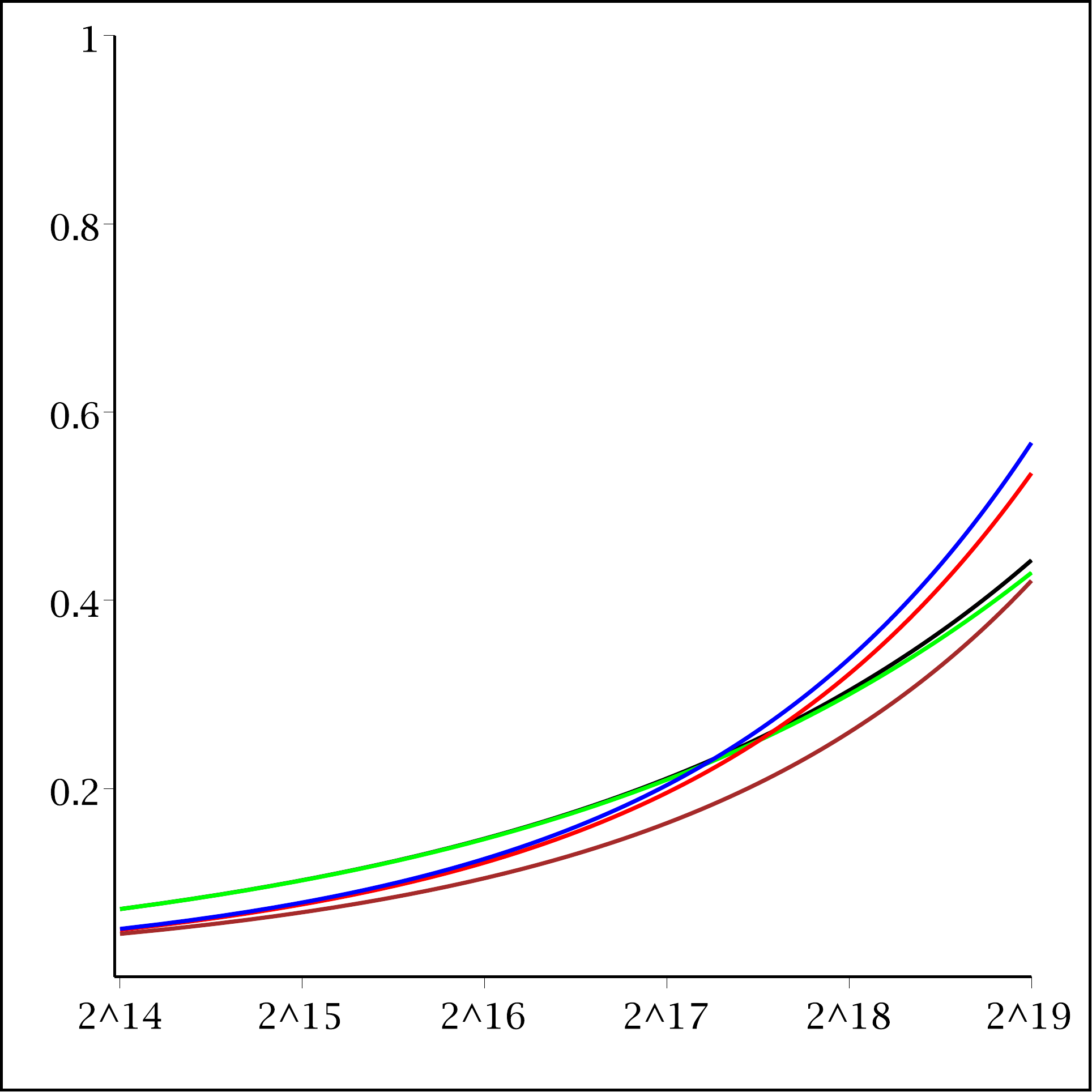}
}
\subfloat[Exponential]
{
\includegraphics[width=\SimFigWidth]{fig/EXP-R07-I300-alpha2-fixedC-appli0-platform-variation.fig}
}	
\subfloat[Weibull $k=0.7$]
{
\includegraphics[width=\SimFigWidth]{fig/WEIBULL-07-R07-I300-alpha2-fixedC-appli0-platform-variation.fig}
}
\subfloat[Weibull $k=0.5$]
{
\includegraphics[width=\SimFigWidth]{fig/WEIBULL-05-R07-I300-alpha2-fixedC-appli0-platform-variation.fig}
}
\\
\hspace{-1cm}
\subfloat{\rotatebox{90}{\setcounter{subfigure}{4}\qquad \I= 600 s}}
\subfloat[Maple]
{
\includegraphics[width=\MapleFigWidth]{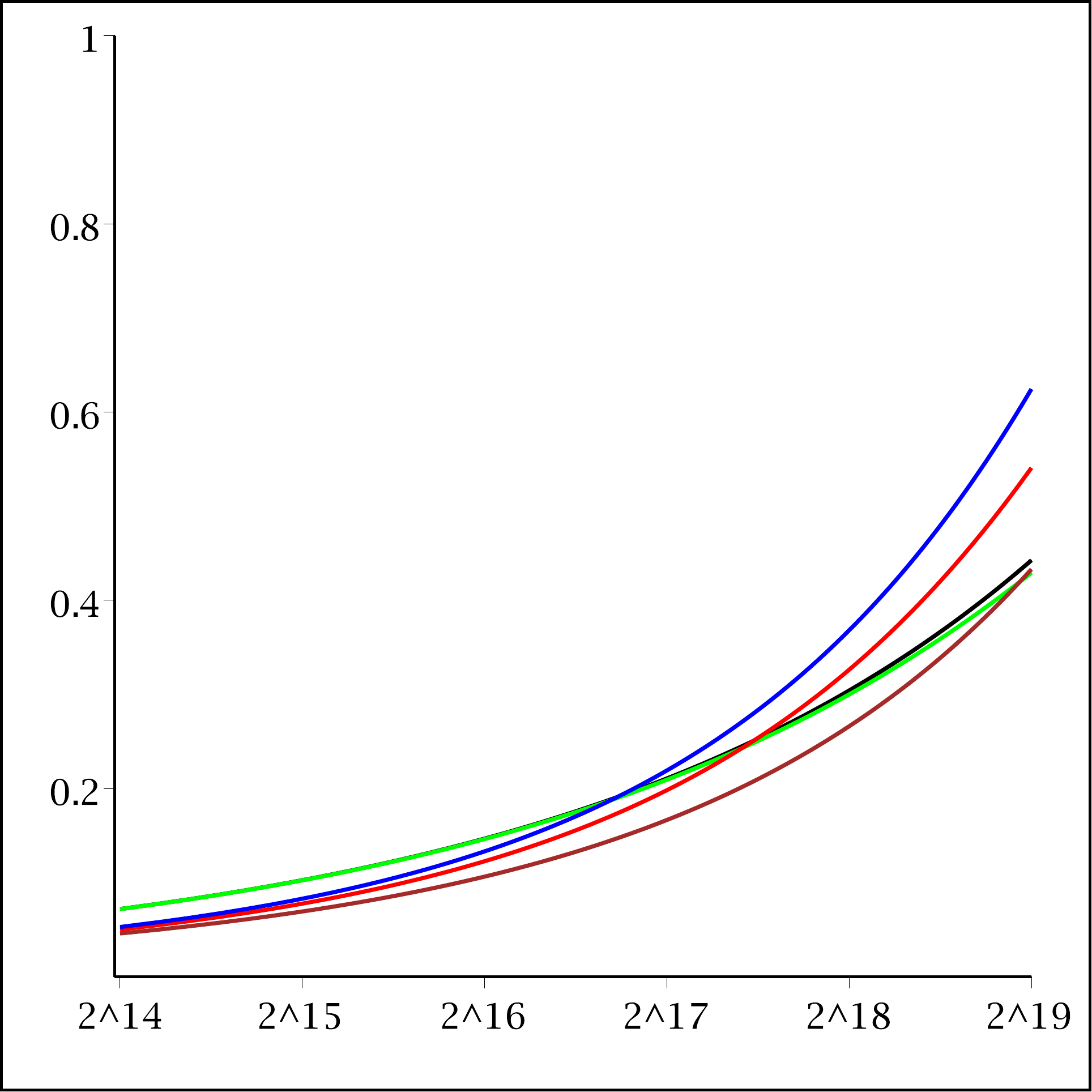}
}
\subfloat[Exponential]
{
\includegraphics[width=\SimFigWidth]{fig/EXP-R07-I600-alpha2-fixedC-appli0-platform-variation.fig}
}	
\subfloat[Weibull $k=0.7$]
{
\includegraphics[width=\SimFigWidth]{fig/WEIBULL-07-R07-I600-alpha2-fixedC-appli0-platform-variation.fig}
}
\subfloat[Weibull $k=0.5$]
{
\includegraphics[width=\SimFigWidth]{fig/WEIBULL-05-R07-I600-alpha2-fixedC-appli0-platform-variation.fig}
}
\\
\hspace{-1cm}
\subfloat{\rotatebox{90}{\setcounter{subfigure}{8}\qquad \I= 900 s}}
\subfloat[Maple]
{
\includegraphics[width=\MapleFigWidth]{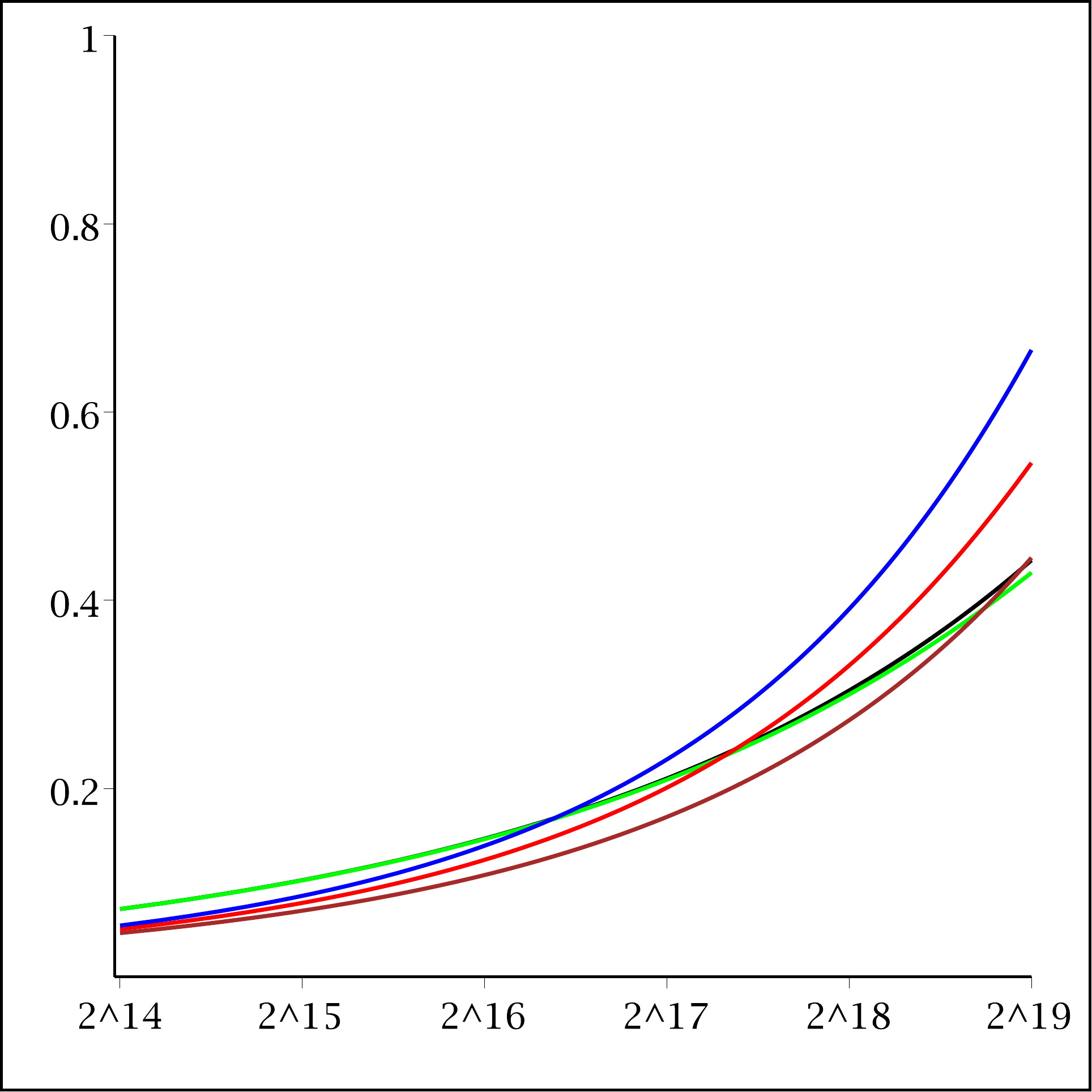}
}
\subfloat[Exponential]
{
\includegraphics[width=\SimFigWidth]{fig/EXP-R07-I900-alpha2-fixedC-appli0-platform-variation.fig}
}	
\subfloat[Weibull $k=0.7$]
{
\includegraphics[width=\SimFigWidth]{fig/WEIBULL-07-R07-I900-alpha2-fixedC-appli0-platform-variation.fig}
}
\subfloat[Weibull $k=0.5$]
{
\includegraphics[width=\SimFigWidth]{fig/WEIBULL-05-R07-I900-alpha2-fixedC-appli0-platform-variation.fig}
}
\\
\hspace{-1cm}
\subfloat{\rotatebox{90}{\setcounter{subfigure}{12}\qquad \I= 1200 s}}
\subfloat[Maple]
{
\includegraphics[width=\MapleFigWidth]{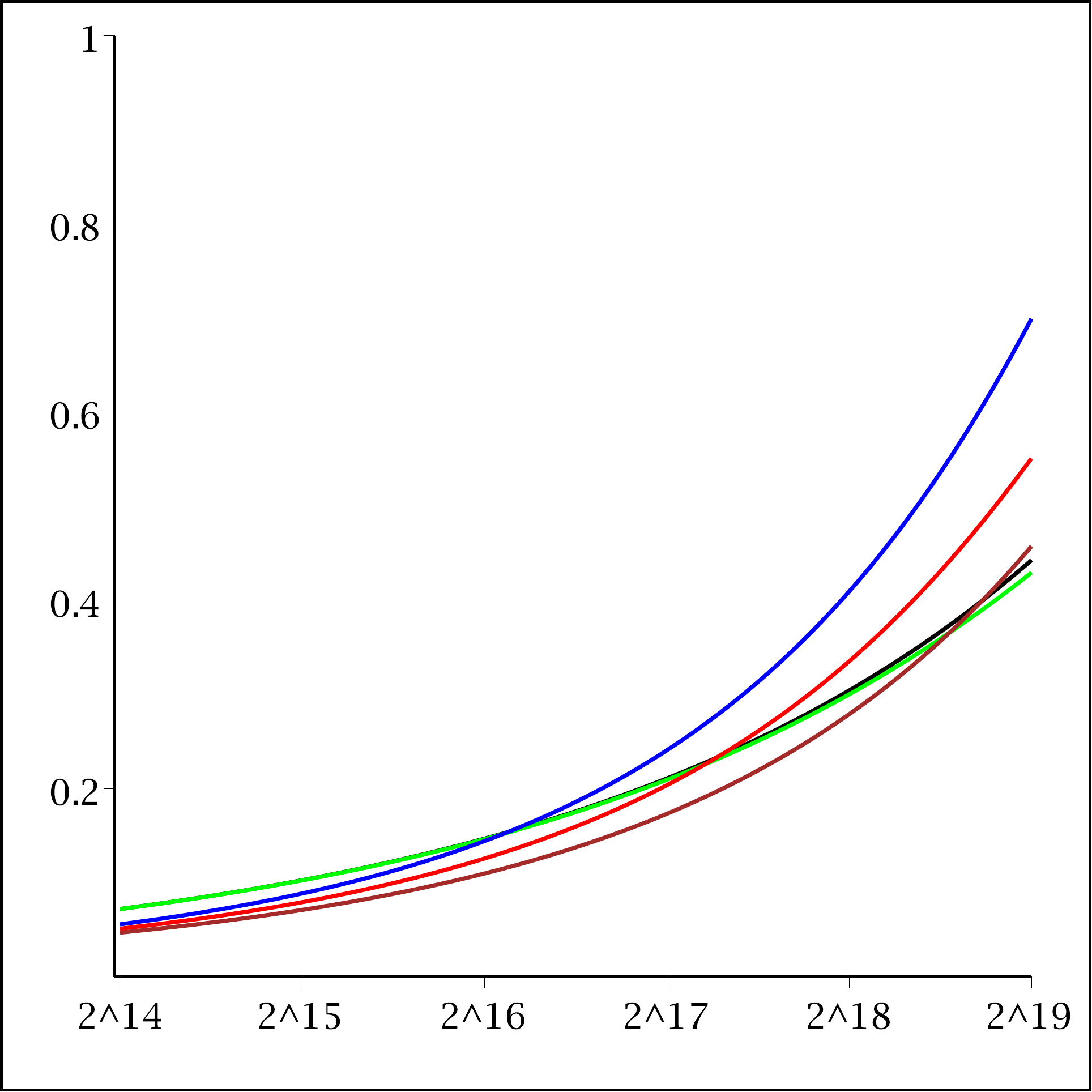}
}
\subfloat[Exponential]
{
\includegraphics[width=\SimFigWidth]{fig/EXP-R07-I1200-alpha2-fixedC-appli0-platform-variation.fig}
}	
\subfloat[Weibull $k=0.7$]
{
\includegraphics[width=\SimFigWidth]{fig/WEIBULL-07-R07-I1200-alpha2-fixedC-appli0-platform-variation.fig}
}
\subfloat[Weibull $k=0.5$]
{
\includegraphics[width=\SimFigWidth]{fig/WEIBULL-05-R07-I1200-alpha2-fixedC-appli0-platform-variation.fig}
}
\\
\hspace{-1cm}
\subfloat{\rotatebox{90}{\setcounter{subfigure}{16}\qquad \I= 3000 s}}
\subfloat[Maple]
{
\includegraphics[width=\MapleFigWidth]{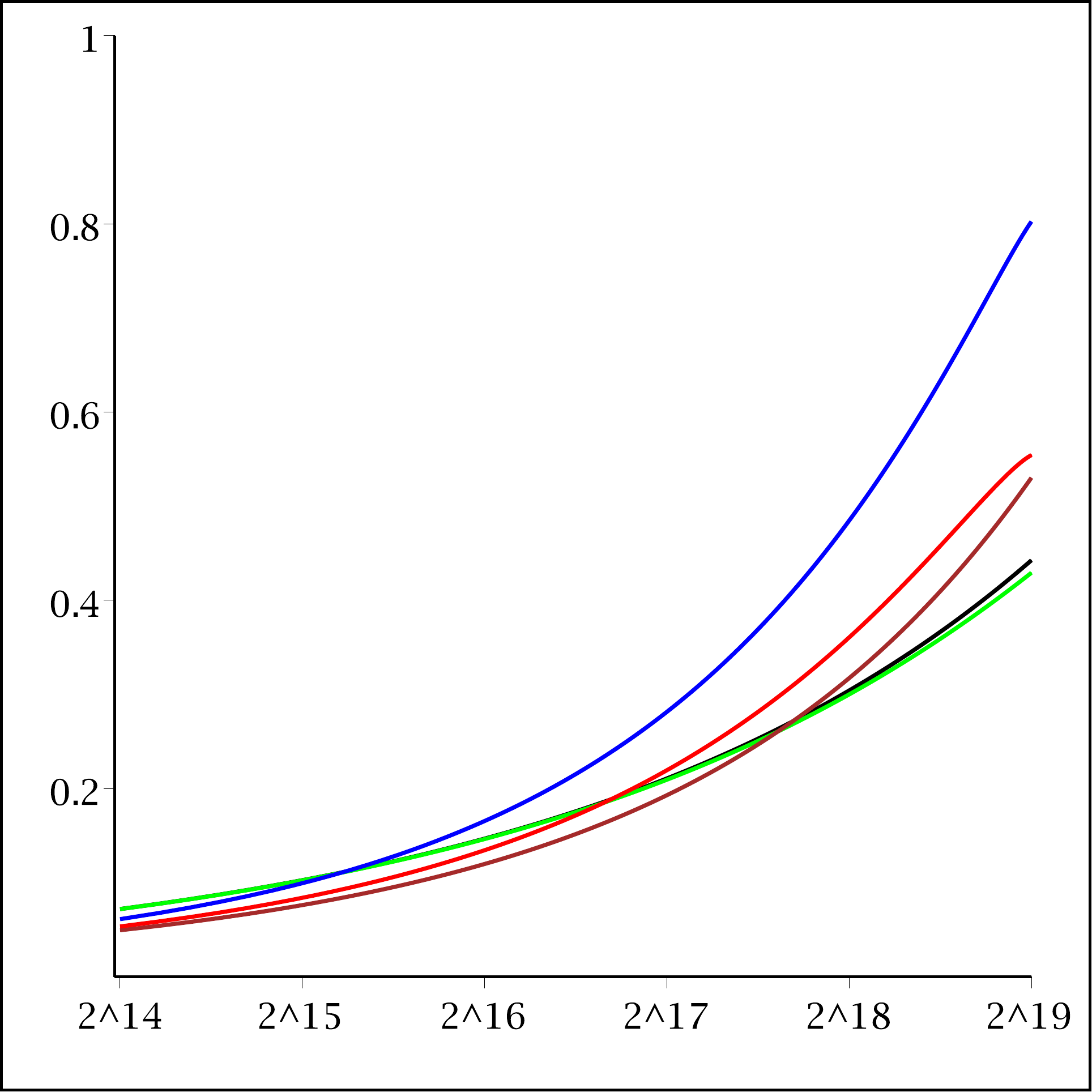}
}
\subfloat[Exponential]
{
\includegraphics[width=\SimFigWidth]{fig/EXP-R07-I3000-alpha2-fixedC-appli0-platform-variation.fig}
}	
\subfloat[Weibull $k=0.7$]
{
\includegraphics[width=\SimFigWidth]{fig/WEIBULL-07-R07-I3000-alpha2-fixedC-appli0-platform-variation.fig}
}
\subfloat[Weibull $k=0.5$]
{
\includegraphics[width=\SimFigWidth]{fig/WEIBULL-05-R07-I3000-alpha2-fixedC-appli0-platform-variation.fig}
}
\caption{Waste for the different heuristics, with\,$\precision=0.4$,
  $\recall=0.7$, $\Cp=2\Cr$, and with a trace of false predictions
  parametrized by a distribution identical to the distribution of the
  trace of failures.}
\label{fig.04.07.Cp2Cr.same}
\end{figure*}

\begin{figure*}
\centering
\hspace{-1cm}
\subfloat{\rotatebox{90}{\setcounter{subfigure}{0}\qquad \I= 300 s}}
\subfloat[Exponential]
{
\includegraphics[width=\SimFigWidth]{fig/Unif-EXP-R85-I300-alpha1-fixedC-appli0-platform-variation.fig}
}	
\subfloat[Weibull $k=0.7$]
{
\includegraphics[width=\SimFigWidth]{fig/Unif-WEIBULL-07-R85-I300-alpha1-fixedC-appli0-platform-variation.fig}
}
\subfloat[Weibull $k=0.5$]
{
\includegraphics[width=\SimFigWidth]{fig/Unif-WEIBULL-05-R85-I300-alpha1-fixedC-appli0-platform-variation.fig}
}
\\
\hspace{-1cm}
\subfloat{\rotatebox{90}{\setcounter{subfigure}{4}\qquad \I= 600 s}}
\subfloat[Exponential]
{
\includegraphics[width=\SimFigWidth]{fig/Unif-EXP-R85-I600-alpha1-fixedC-appli0-platform-variation.fig}
}	
\subfloat[Weibull $k=0.7$]
{
\includegraphics[width=\SimFigWidth]{fig/Unif-WEIBULL-07-R85-I600-alpha1-fixedC-appli0-platform-variation.fig}
}
\subfloat[Weibull $k=0.5$]
{
\includegraphics[width=\SimFigWidth]{fig/Unif-WEIBULL-05-R85-I600-alpha1-fixedC-appli0-platform-variation.fig}
}
\\
\hspace{-1cm}
\subfloat{\rotatebox{90}{\setcounter{subfigure}{8}\qquad \I= 900 s}}
\subfloat[Exponential]
{
\includegraphics[width=\SimFigWidth]{fig/Unif-EXP-R85-I900-alpha1-fixedC-appli0-platform-variation.fig}
}	
\subfloat[Weibull $k=0.7$]
{
\includegraphics[width=\SimFigWidth]{fig/Unif-WEIBULL-07-R85-I900-alpha1-fixedC-appli0-platform-variation.fig}
}
\subfloat[Weibull $k=0.5$]
{
\includegraphics[width=\SimFigWidth]{fig/Unif-WEIBULL-05-R85-I900-alpha1-fixedC-appli0-platform-variation.fig}
}
\\
\hspace{-1cm}
\subfloat{\rotatebox{90}{\setcounter{subfigure}{12}\qquad \I= 1200 s}}
\subfloat[Exponential]
{
\includegraphics[width=\SimFigWidth]{fig/Unif-EXP-R85-I1200-alpha1-fixedC-appli0-platform-variation.fig}
}	
\subfloat[Weibull $k=0.7$]
{
\includegraphics[width=\SimFigWidth]{fig/Unif-WEIBULL-07-R85-I1200-alpha1-fixedC-appli0-platform-variation.fig}
}
\subfloat[Weibull $k=0.5$]
{
\includegraphics[width=\SimFigWidth]{fig/Unif-WEIBULL-05-R85-I1200-alpha1-fixedC-appli0-platform-variation.fig}
}
\\
\hspace{-1cm}
\subfloat{\rotatebox{90}{\setcounter{subfigure}{16}\qquad \I= 3000 s}}
\subfloat[Exponential]
{
\includegraphics[width=\SimFigWidth]{fig/Unif-EXP-R85-I3000-alpha1-fixedC-appli0-platform-variation.fig}
}	
\subfloat[Weibull $k=0.7$]
{
\includegraphics[width=\SimFigWidth]{fig/Unif-WEIBULL-07-R85-I3000-alpha1-fixedC-appli0-platform-variation.fig}
}
\subfloat[Weibull $k=0.5$]
{
\includegraphics[width=\SimFigWidth]{fig/Unif-WEIBULL-05-R85-I3000-alpha1-fixedC-appli0-platform-variation.fig}
}
\caption{Waste for the different heuristics, with $\precision=0.82$,
  $\recall=0.85$, $\Cp=\Cr$, and with a trace of false predictions
  parametrized by a uniform distribution.}
	\label{fig.082.085.CpCr.unif}
\end{figure*}

\begin{figure*}
\centering
\hspace{-1cm}
\subfloat{\rotatebox{90}{\setcounter{subfigure}{0}\qquad \I= 300 s}}
\subfloat[Exponential]
{
\includegraphics[width=\SimFigWidth]{fig/Unif-EXP-R85-I300-alpha01-fixedC-appli0-platform-variation.fig}
}	
\subfloat[Weibull $k=0.7$]
{
\includegraphics[width=\SimFigWidth]{fig/Unif-WEIBULL-07-R85-I300-alpha01-fixedC-appli0-platform-variation.fig}
}
\subfloat[Weibull $k=0.5$]
{
\includegraphics[width=\SimFigWidth]{fig/Unif-WEIBULL-05-R85-I300-alpha01-fixedC-appli0-platform-variation.fig}
}
\\
\hspace{-1cm}
\subfloat{\rotatebox{90}{\setcounter{subfigure}{4}\qquad \I= 600 s}}
\subfloat[Exponential]
{
\includegraphics[width=\SimFigWidth]{fig/Unif-EXP-R85-I600-alpha01-fixedC-appli0-platform-variation.fig}
}	
\subfloat[Weibull $k=0.7$]
{
\includegraphics[width=\SimFigWidth]{fig/Unif-WEIBULL-07-R85-I600-alpha01-fixedC-appli0-platform-variation.fig}
}
\subfloat[Weibull $k=0.5$]
{
\includegraphics[width=\SimFigWidth]{fig/Unif-WEIBULL-05-R85-I600-alpha01-fixedC-appli0-platform-variation.fig}
}
\\
\hspace{-1cm}
\subfloat{\rotatebox{90}{\setcounter{subfigure}{8}\qquad \I= 900 s}}
\subfloat[Exponential]
{
\includegraphics[width=\SimFigWidth]{fig/Unif-EXP-R85-I900-alpha01-fixedC-appli0-platform-variation.fig}
}	
\subfloat[Weibull $k=0.7$]
{
\includegraphics[width=\SimFigWidth]{fig/Unif-WEIBULL-07-R85-I900-alpha01-fixedC-appli0-platform-variation.fig}
}
\subfloat[Weibull $k=0.5$]
{
\includegraphics[width=\SimFigWidth]{fig/Unif-WEIBULL-05-R85-I900-alpha01-fixedC-appli0-platform-variation.fig}
}
\\
\hspace{-1cm}
\subfloat{\rotatebox{90}{\setcounter{subfigure}{12}\qquad \I= 1200 s}}
\subfloat[Exponential]
{
\includegraphics[width=\SimFigWidth]{fig/Unif-EXP-R85-I1200-alpha01-fixedC-appli0-platform-variation.fig}
}	
\subfloat[Weibull $k=0.7$]
{
\includegraphics[width=\SimFigWidth]{fig/Unif-WEIBULL-07-R85-I1200-alpha01-fixedC-appli0-platform-variation.fig}
}
\subfloat[Weibull $k=0.5$]
{
\includegraphics[width=\SimFigWidth]{fig/Unif-WEIBULL-05-R85-I1200-alpha01-fixedC-appli0-platform-variation.fig}
}
\\
\hspace{-1cm}
\subfloat{\rotatebox{90}{\setcounter{subfigure}{16}\qquad \I= 3000 s}}
\subfloat[Exponential]
{
\includegraphics[width=\SimFigWidth]{fig/Unif-EXP-R85-I3000-alpha01-fixedC-appli0-platform-variation.fig}
}	
\subfloat[Weibull $k=0.7$]
{
\includegraphics[width=\SimFigWidth]{fig/Unif-WEIBULL-07-R85-I3000-alpha01-fixedC-appli0-platform-variation.fig}
}
\subfloat[Weibull $k=0.5$]
{
\includegraphics[width=\SimFigWidth]{fig/Unif-WEIBULL-05-R85-I3000-alpha01-fixedC-appli0-platform-variation.fig}
}
\caption{Waste for the different heuristics, with $\precision=0.82$,
  $\recall=0.85$, $\Cp=0.1\Cr$, and with a trace of false predictions
  parametrized by a uniform distribution.}
	\label{fig.082.085.Cp01Cr.unif}
\end{figure*}

\begin{figure*}
\centering
\hspace{-1cm}
\subfloat{\rotatebox{90}{\setcounter{subfigure}{0}\qquad \I= 300 s}}
\subfloat[Exponential]
{
\includegraphics[width=\SimFigWidth]{fig/Unif-EXP-R85-I300-alpha2-fixedC-appli0-platform-variation.fig}
}	
\subfloat[Weibull $k=0.7$]
{
\includegraphics[width=\SimFigWidth]{fig/Unif-WEIBULL-07-R85-I300-alpha2-fixedC-appli0-platform-variation.fig}
}
\subfloat[Weibull $k=0.5$]
{
\includegraphics[width=\SimFigWidth]{fig/Unif-WEIBULL-05-R85-I300-alpha2-fixedC-appli0-platform-variation.fig}
}
\\
\hspace{-1cm}
\subfloat{\rotatebox{90}{\setcounter{subfigure}{4}\qquad \I= 600 s}}
\subfloat[Exponential]
{
\includegraphics[width=\SimFigWidth]{fig/Unif-EXP-R85-I600-alpha2-fixedC-appli0-platform-variation.fig}
}	
\subfloat[Weibull $k=0.7$]
{
\includegraphics[width=\SimFigWidth]{fig/Unif-WEIBULL-07-R85-I600-alpha2-fixedC-appli0-platform-variation.fig}
}
\subfloat[Weibull $k=0.5$]
{
\includegraphics[width=\SimFigWidth]{fig/Unif-WEIBULL-05-R85-I600-alpha2-fixedC-appli0-platform-variation.fig}
}
\\
\hspace{-1cm}
\subfloat{\rotatebox{90}{\setcounter{subfigure}{8}\qquad \I= 900 s}}
\subfloat[Exponential]
{
\includegraphics[width=\SimFigWidth]{fig/Unif-EXP-R85-I900-alpha2-fixedC-appli0-platform-variation.fig}
}	
\subfloat[Weibull $k=0.7$]
{
\includegraphics[width=\SimFigWidth]{fig/Unif-WEIBULL-07-R85-I900-alpha2-fixedC-appli0-platform-variation.fig}
}
\subfloat[Weibull $k=0.5$]
{
\includegraphics[width=\SimFigWidth]{fig/Unif-WEIBULL-05-R85-I900-alpha2-fixedC-appli0-platform-variation.fig}
}
\\
\hspace{-1cm}
\subfloat{\rotatebox{90}{\setcounter{subfigure}{12}\qquad \I= 1200 s}}
\subfloat[Exponential]
{
\includegraphics[width=\SimFigWidth]{fig/Unif-EXP-R85-I1200-alpha2-fixedC-appli0-platform-variation.fig}
}	
\subfloat[Weibull $k=0.7$]
{
\includegraphics[width=\SimFigWidth]{fig/Unif-WEIBULL-07-R85-I1200-alpha2-fixedC-appli0-platform-variation.fig}
}
\subfloat[Weibull $k=0.5$]
{
\includegraphics[width=\SimFigWidth]{fig/Unif-WEIBULL-05-R85-I1200-alpha2-fixedC-appli0-platform-variation.fig}
}
\\
\hspace{-1cm}
\subfloat{\rotatebox{90}{\setcounter{subfigure}{16}\qquad \I= 3000 s}}
\subfloat[Exponential]
{
\includegraphics[width=\SimFigWidth]{fig/Unif-EXP-R85-I3000-alpha2-fixedC-appli0-platform-variation.fig}
}	
\subfloat[Weibull $k=0.7$]
{
\includegraphics[width=\SimFigWidth]{fig/Unif-WEIBULL-07-R85-I3000-alpha2-fixedC-appli0-platform-variation.fig}
}
\subfloat[Weibull $k=0.5$]
{
\includegraphics[width=\SimFigWidth]{fig/Unif-WEIBULL-05-R85-I3000-alpha2-fixedC-appli0-platform-variation.fig}
}
\caption{Waste for the different heuristics, with $\precision=0.82$,
  $\recall=0.85$, $\Cp=2\Cr$, and with a trace of false predictions
  parametrized by a uniform distribution.}
	\label{fig.082.085.Cp2Cr.unif}
\end{figure*}


\begin{figure*}
\centering
\hspace{-1cm}
\subfloat{\rotatebox{90}{\setcounter{subfigure}{0}\qquad \I= 300 s}}
\subfloat[Exponential]
{
\includegraphics[width=\SimFigWidth]{fig/Unif-EXP-R07-I300-alpha1-fixedC-appli0-platform-variation.fig}
}	
\subfloat[Weibull $k=0.7$]
{
\includegraphics[width=\SimFigWidth]{fig/Unif-WEIBULL-07-R07-I300-alpha1-fixedC-appli0-platform-variation.fig}
}
\subfloat[Weibull $k=0.5$]
{
\includegraphics[width=\SimFigWidth]{fig/Unif-WEIBULL-05-R07-I300-alpha1-fixedC-appli0-platform-variation.fig}
}
\\
\hspace{-1cm}
\subfloat{\rotatebox{90}{\setcounter{subfigure}{4}\qquad \I= 600 s}}
\subfloat[Exponential]
{
\includegraphics[width=\SimFigWidth]{fig/Unif-EXP-R07-I600-alpha1-fixedC-appli0-platform-variation.fig}
}	
\subfloat[Weibull $k=0.7$]
{
\includegraphics[width=\SimFigWidth]{fig/Unif-WEIBULL-07-R07-I600-alpha1-fixedC-appli0-platform-variation.fig}
}
\subfloat[Weibull $k=0.5$]
{
\includegraphics[width=\SimFigWidth]{fig/Unif-WEIBULL-05-R07-I600-alpha1-fixedC-appli0-platform-variation.fig}
}
\\
\hspace{-1cm}
\subfloat{\rotatebox{90}{\setcounter{subfigure}{8}\qquad \I= 900 s}}
\subfloat[Exponential]
{
\includegraphics[width=\SimFigWidth]{fig/Unif-EXP-R07-I900-alpha1-fixedC-appli0-platform-variation.fig}
}	
\subfloat[Weibull $k=0.7$]
{
\includegraphics[width=\SimFigWidth]{fig/Unif-WEIBULL-07-R07-I900-alpha1-fixedC-appli0-platform-variation.fig}
}
\subfloat[Weibull $k=0.5$]
{
\includegraphics[width=\SimFigWidth]{fig/Unif-WEIBULL-05-R07-I900-alpha1-fixedC-appli0-platform-variation.fig}
}
\\
\hspace{-1cm}
\subfloat{\rotatebox{90}{\setcounter{subfigure}{12}\qquad \I= 1200 s}}
\subfloat[Exponential]
{
\includegraphics[width=\SimFigWidth]{fig/Unif-EXP-R07-I1200-alpha1-fixedC-appli0-platform-variation.fig}
}	
\subfloat[Weibull $k=0.7$]
{
\includegraphics[width=\SimFigWidth]{fig/Unif-WEIBULL-07-R07-I1200-alpha1-fixedC-appli0-platform-variation.fig}
}
\subfloat[Weibull $k=0.5$]
{
\includegraphics[width=\SimFigWidth]{fig/Unif-WEIBULL-05-R07-I1200-alpha1-fixedC-appli0-platform-variation.fig}
}
\\
\hspace{-1cm}
\subfloat{\rotatebox{90}{\setcounter{subfigure}{16}\qquad \I= 3000 s}}
\subfloat[Exponential]
{
\includegraphics[width=\SimFigWidth]{fig/Unif-EXP-R07-I3000-alpha1-fixedC-appli0-platform-variation.fig}
}	
\subfloat[Weibull $k=0.7$]
{
\includegraphics[width=\SimFigWidth]{fig/Unif-WEIBULL-07-R07-I3000-alpha1-fixedC-appli0-platform-variation.fig}
}
\subfloat[Weibull $k=0.5$]
{
\includegraphics[width=\SimFigWidth]{fig/Unif-WEIBULL-05-R07-I3000-alpha1-fixedC-appli0-platform-variation.fig}
}
\caption{Waste for the different heuristics, with $\precision=0.4$,
  $\recall=0.7$, $\Cp=\Cr$, and with a trace of false predictions
  parametrized by a uniform distribution.}
	\label{fig.04.07.CpCr.unif}
\end{figure*}


\begin{figure*}
\centering
\hspace{-1cm}
\subfloat{\rotatebox{90}{\setcounter{subfigure}{0}\qquad \I= 300 s}}
\subfloat[Exponential]
{
\includegraphics[width=\SimFigWidth]{fig/Unif-EXP-R07-I300-alpha01-fixedC-appli0-platform-variation.fig}
}	
\subfloat[Weibull $k=0.7$]
{
\includegraphics[width=\SimFigWidth]{fig/Unif-WEIBULL-07-R07-I300-alpha01-fixedC-appli0-platform-variation.fig}
}
\subfloat[Weibull $k=0.5$]
{
\includegraphics[width=\SimFigWidth]{fig/Unif-WEIBULL-05-R07-I300-alpha01-fixedC-appli0-platform-variation.fig}
}
\\
\hspace{-1cm}
\subfloat{\rotatebox{90}{\setcounter{subfigure}{4}\qquad \I= 600 s}}
\subfloat[Exponential]
{
\includegraphics[width=\SimFigWidth]{fig/Unif-EXP-R07-I600-alpha01-fixedC-appli0-platform-variation.fig}
}	
\subfloat[Weibull $k=0.7$]
{
\includegraphics[width=\SimFigWidth]{fig/Unif-WEIBULL-07-R07-I600-alpha01-fixedC-appli0-platform-variation.fig}
}
\subfloat[Weibull $k=0.5$]
{
\includegraphics[width=\SimFigWidth]{fig/Unif-WEIBULL-05-R07-I600-alpha01-fixedC-appli0-platform-variation.fig}
}
\\
\hspace{-1cm}
\subfloat{\rotatebox{90}{\setcounter{subfigure}{8}\qquad \I= 900 s}}
\subfloat[Exponential]
{
\includegraphics[width=\SimFigWidth]{fig/Unif-EXP-R07-I900-alpha01-fixedC-appli0-platform-variation.fig}
}	
\subfloat[Weibull $k=0.7$]
{
\includegraphics[width=\SimFigWidth]{fig/Unif-WEIBULL-07-R07-I900-alpha01-fixedC-appli0-platform-variation.fig}
}
\subfloat[Weibull $k=0.5$]
{
\includegraphics[width=\SimFigWidth]{fig/Unif-WEIBULL-05-R07-I900-alpha01-fixedC-appli0-platform-variation.fig}
}
\\
\hspace{-1cm}
\subfloat{\rotatebox{90}{\setcounter{subfigure}{12}\qquad \I= 1200 s}}
\subfloat[Exponential]
{
\includegraphics[width=\SimFigWidth]{fig/Unif-EXP-R07-I1200-alpha01-fixedC-appli0-platform-variation.fig}
}	
\subfloat[Weibull $k=0.7$]
{
\includegraphics[width=\SimFigWidth]{fig/Unif-WEIBULL-07-R07-I1200-alpha01-fixedC-appli0-platform-variation.fig}
}
\subfloat[Weibull $k=0.5$]
{
\includegraphics[width=\SimFigWidth]{fig/Unif-WEIBULL-05-R07-I1200-alpha01-fixedC-appli0-platform-variation.fig}
}
\\
\hspace{-1cm}
\subfloat{\rotatebox{90}{\setcounter{subfigure}{16}\qquad \I= 3000 s}}
\subfloat[Exponential]
{
\includegraphics[width=\SimFigWidth]{fig/Unif-EXP-R07-I3000-alpha01-fixedC-appli0-platform-variation.fig}
}	
\subfloat[Weibull $k=0.7$]
{
\includegraphics[width=\SimFigWidth]{fig/Unif-WEIBULL-07-R07-I3000-alpha01-fixedC-appli0-platform-variation.fig}
}
\subfloat[Weibull $k=0.5$]
{
\includegraphics[width=\SimFigWidth]{fig/Unif-WEIBULL-05-R07-I3000-alpha01-fixedC-appli0-platform-variation.fig}
}
\caption{Waste for the different heuristics, with\,$\precision=0.4$,
  $\recall=0.7$, $\Cp=0.1\Cr$, and with a trace of false predictions
  parametrized by a uniform distribution.}
\label{fig.04.07.Cp01Cr.unif}
\end{figure*}

\begin{figure*}
\centering
\hspace{-1cm}
\subfloat{\rotatebox{90}{\setcounter{subfigure}{0}\qquad \I= 300 s}}
\subfloat[Exponential]
{
\includegraphics[width=\SimFigWidth]{fig/Unif-EXP-R07-I300-alpha2-fixedC-appli0-platform-variation.fig}
}	
\subfloat[Weibull $k=0.7$]
{
\includegraphics[width=\SimFigWidth]{fig/Unif-WEIBULL-07-R07-I300-alpha2-fixedC-appli0-platform-variation.fig}
}
\subfloat[Weibull $k=0.5$]
{
\includegraphics[width=\SimFigWidth]{fig/Unif-WEIBULL-05-R07-I300-alpha2-fixedC-appli0-platform-variation.fig}
}
\\
\hspace{-1cm}
\subfloat{\rotatebox{90}{\setcounter{subfigure}{4}\qquad \I= 600 s}}
\subfloat[Exponential]
{
\includegraphics[width=\SimFigWidth]{fig/Unif-EXP-R07-I600-alpha2-fixedC-appli0-platform-variation.fig}
}	
\subfloat[Weibull $k=0.7$]
{
\includegraphics[width=\SimFigWidth]{fig/Unif-WEIBULL-07-R07-I600-alpha2-fixedC-appli0-platform-variation.fig}
}
\subfloat[Weibull $k=0.5$]
{
\includegraphics[width=\SimFigWidth]{fig/Unif-WEIBULL-05-R07-I600-alpha2-fixedC-appli0-platform-variation.fig}
}
\\
\hspace{-1cm}
\subfloat{\rotatebox{90}{\setcounter{subfigure}{8}\qquad \I= 900 s}}
\subfloat[Exponential]
{
\includegraphics[width=\SimFigWidth]{fig/Unif-EXP-R07-I900-alpha2-fixedC-appli0-platform-variation.fig}
}	
\subfloat[Weibull $k=0.7$]
{
\includegraphics[width=\SimFigWidth]{fig/Unif-WEIBULL-07-R07-I900-alpha2-fixedC-appli0-platform-variation.fig}
}
\subfloat[Weibull $k=0.5$]
{
\includegraphics[width=\SimFigWidth]{fig/Unif-WEIBULL-05-R07-I900-alpha2-fixedC-appli0-platform-variation.fig}
}
\\
\hspace{-1cm}
\subfloat{\rotatebox{90}{\setcounter{subfigure}{12}\qquad \I= 1200 s}}
\subfloat[Exponential]
{
\includegraphics[width=\SimFigWidth]{fig/Unif-EXP-R07-I1200-alpha2-fixedC-appli0-platform-variation.fig}
}	
\subfloat[Weibull $k=0.7$]
{
\includegraphics[width=\SimFigWidth]{fig/Unif-WEIBULL-07-R07-I1200-alpha2-fixedC-appli0-platform-variation.fig}
}
\subfloat[Weibull $k=0.5$]
{
\includegraphics[width=\SimFigWidth]{fig/Unif-WEIBULL-05-R07-I1200-alpha2-fixedC-appli0-platform-variation.fig}
}
\\
\hspace{-1cm}
\subfloat{\rotatebox{90}{\setcounter{subfigure}{16}\qquad \I= 3000 s}}
\subfloat[Exponential]
{
\includegraphics[width=\SimFigWidth]{fig/Unif-EXP-R07-I3000-alpha2-fixedC-appli0-platform-variation.fig}
}	
\subfloat[Weibull $k=0.7$]
{
\includegraphics[width=\SimFigWidth]{fig/Unif-WEIBULL-07-R07-I3000-alpha2-fixedC-appli0-platform-variation.fig}
}
\subfloat[Weibull $k=0.5$]
{
\includegraphics[width=\SimFigWidth]{fig/Unif-WEIBULL-05-R07-I3000-alpha2-fixedC-appli0-platform-variation.fig}
}

\caption{Waste for the different heuristics, with\,$\precision=0.4$,
  $\recall=0.7$, $\Cp=2\Cr$, and with a trace of false predictions
  parametrized by a uniform distribution.}
\label{fig.04.07.Cp2Cr.unif}
\end{figure*}

\begin{figure*}
\subfloat{\rotatebox{90}{\setcounter{subfigure}{0}\qquad\I= 300 s}}
\subfloat[Exponential]
{
\includegraphics[width=\SimFigTrWidth]{fig/Tr-EXP-R85-I300-alpha1-fixedC-appli0-platform16.fig}
}	
\subfloat[Weibull $k=0.7$]
{
\includegraphics[width=\SimFigTrWidth]{fig/Tr-WEIBULL-07-R85-I300-alpha1-fixedC-appli0-platform16.fig}
}
\subfloat[Weibull $k=0.5$]
{
\includegraphics[width=\SimFigTrWidth]{fig/Tr-WEIBULL-05-R85-I300-alpha1-fixedC-appli0-platform16.fig}
}
\\
\hspace{-1cm}
\subfloat{\rotatebox{90}{\setcounter{subfigure}{12}\qquad \I= 1200 s}}
\subfloat[Exponential]
{
\includegraphics[width=\SimFigTrWidth]{fig/Tr-EXP-R85-I1200-alpha1-fixedC-appli0-platform16.fig}
}	
\subfloat[Weibull $k=0.7$]
{
\includegraphics[width=\SimFigTrWidth]{fig/Tr-WEIBULL-07-R85-I1200-alpha1-fixedC-appli0-platform16.fig}
}
\subfloat[Weibull $k=0.5$]
{
\includegraphics[width=\SimFigTrWidth]{fig/Tr-WEIBULL-05-R85-I1200-alpha1-fixedC-appli0-platform16.fig}
}
\\
\hspace{-1cm}
\subfloat{\rotatebox{90}{\setcounter{subfigure}{16}\qquad \I= 3000 s}}
\subfloat[Exponential]
{
\includegraphics[width=\SimFigTrWidth]{fig/Tr-EXP-R85-I3000-alpha1-fixedC-appli0-platform16.fig}
}	
\subfloat[Weibull $k=0.7$]
{
\includegraphics[width=\SimFigTrWidth]{fig/Tr-WEIBULL-07-R85-I3000-alpha1-fixedC-appli0-platform16.fig}
}
\subfloat[Weibull $k=0.5$]
{
\includegraphics[width=\SimFigTrWidth]{fig/Tr-WEIBULL-05-R85-I3000-alpha1-fixedC-appli0-platform16.fig}
}
\caption{Waste as function of the period $\Tnp$ for the different heuristics, with $\precision=0.82$,
  $\recall=0.85$, $\Cp=\Cr$, and with a platform of $2^{16}$ processors.}
	\label{fig.082.085.Tr.16}
\end{figure*}

\begin{figure*}
\subfloat{\rotatebox{90}{\setcounter{subfigure}{0}\qquad\I= 300 s}}
\subfloat[Exponential]
{
\includegraphics[width=\SimFigTrWidth]{fig/Tr-EXP-R85-I300-alpha1-fixedC-appli0-platform-variation.fig}
}	
\subfloat[Weibull $k=0.7$]
{
\includegraphics[width=\SimFigTrWidth]{fig/Tr-WEIBULL-07-R85-I300-alpha1-fixedC-appli0-platform-variation.fig}
}
\subfloat[Weibull $k=0.5$]
{
\includegraphics[width=\SimFigTrWidth]{fig/Tr-WEIBULL-05-R85-I300-alpha1-fixedC-appli0-platform-variation.fig}
}
\\
\hspace{-1cm}
\subfloat{\rotatebox{90}{\setcounter{subfigure}{12}\qquad \I= 1200 s}}
\subfloat[Exponential]
{
\includegraphics[width=\SimFigTrWidth]{fig/Tr-EXP-R85-I1200-alpha1-fixedC-appli0-platform-variation.fig}
}	
\subfloat[Weibull $k=0.7$]
{
\includegraphics[width=\SimFigTrWidth]{fig/Tr-WEIBULL-07-R85-I1200-alpha1-fixedC-appli0-platform-variation.fig}
}
\subfloat[Weibull $k=0.5$]
{
\includegraphics[width=\SimFigTrWidth]{fig/Tr-WEIBULL-05-R85-I1200-alpha1-fixedC-appli0-platform-variation.fig}
}
\\
\hspace{-1cm}
\subfloat{\rotatebox{90}{\setcounter{subfigure}{16}\qquad \I= 3000 s}}
\subfloat[Exponential]
{
\includegraphics[width=\SimFigTrWidth]{fig/Tr-EXP-R85-I3000-alpha1-fixedC-appli0-platform-variation.fig}
}	
\subfloat[Weibull $k=0.7$]
{
\includegraphics[width=\SimFigTrWidth]{fig/Tr-WEIBULL-07-R85-I3000-alpha1-fixedC-appli0-platform-variation.fig}
}
\subfloat[Weibull $k=0.5$]
{
\includegraphics[width=\SimFigTrWidth]{fig/Tr-WEIBULL-05-R85-I3000-alpha1-fixedC-appli0-platform-variation.fig}
}
\caption{Waste as function of the period $\Tnp$ for the different heuristics, with $\precision=0.82$,
  $\recall=0.85$, $\Cp=\Cr$, and with a platform of $2^{19}$ processors.}
	\label{fig.082.085.Tr.19}
\end{figure*}

\begin{figure*}
\centering
\hspace{-1cm}
\subfloat{\rotatebox{90}{\setcounter{subfigure}{0}\qquad \I= 300 s}}
\subfloat[Exponential]
{
\includegraphics[width=\SimFigTrWidth]{fig/Tr-EXP-R07-I300-alpha1-fixedC-appli0-platform16.fig}
}	
\subfloat[Weibull $k=0.7$]
{
\includegraphics[width=\SimFigTrWidth]{fig/Tr-WEIBULL-07-R07-I300-alpha1-fixedC-appli0-platform16.fig}
}
\subfloat[Weibull $k=0.5$]
{
\includegraphics[width=\SimFigTrWidth]{fig/Tr-WEIBULL-05-R07-I300-alpha1-fixedC-appli0-platform16.fig}
}
\\
\hspace{-1cm}
\subfloat{\rotatebox{90}{\setcounter{subfigure}{12}\qquad \I= 1200 s}}
\subfloat[Exponential]
{
\includegraphics[width=\SimFigTrWidth]{fig/Tr-EXP-R07-I1200-alpha1-fixedC-appli0-platform16.fig}
}	
\subfloat[Weibull $k=0.7$]
{
\includegraphics[width=\SimFigTrWidth]{fig/Tr-WEIBULL-07-R07-I1200-alpha1-fixedC-appli0-platform16.fig}
}
\subfloat[Weibull $k=0.5$]
{
\includegraphics[width=\SimFigTrWidth]{fig/Tr-WEIBULL-05-R07-I1200-alpha1-fixedC-appli0-platform16.fig}
}
\\
\hspace{-1cm}
\subfloat{\rotatebox{90}{\setcounter{subfigure}{16}\qquad \I= 3000 s}}
\subfloat[Exponential]
{
\includegraphics[width=\SimFigTrWidth]{fig/Tr-EXP-R07-I3000-alpha1-fixedC-appli0-platform16.fig}
}	
\subfloat[Weibull $k=0.7$]
{
\includegraphics[width=\SimFigTrWidth]{fig/Tr-WEIBULL-07-R07-I3000-alpha1-fixedC-appli0-platform16.fig}
}
\subfloat[Weibull $k=0.5$]
{
\includegraphics[width=\SimFigTrWidth]{fig/Tr-WEIBULL-05-R07-I3000-alpha1-fixedC-appli0-platform16.fig}
}
\caption{Waste as function of the period $\Tnp$ for the different heuristics, with $\precision=0.4$,
  $\recall=0.7$, $\Cp=\Cr$, and with a platform of $2^{16}$ processors.}
	\label{fig.04.07.Tr.16}
\end{figure*}

\begin{figure*}
\centering
\hspace{-1cm}
\subfloat{\rotatebox{90}{\setcounter{subfigure}{0}\qquad \I= 300 s}}
\subfloat[Exponential]
{
\includegraphics[width=\SimFigTrWidth]{fig/Tr-EXP-R07-I300-alpha1-fixedC-appli0-platform-variation.fig}
}	
\subfloat[Weibull $k=0.7$]
{
\includegraphics[width=\SimFigTrWidth]{fig/Tr-WEIBULL-07-R07-I300-alpha1-fixedC-appli0-platform-variation.fig}
}
\subfloat[Weibull $k=0.5$]
{
\includegraphics[width=\SimFigTrWidth]{fig/Tr-WEIBULL-05-R07-I300-alpha1-fixedC-appli0-platform-variation.fig}
}
\\
\hspace{-1cm}
\subfloat{\rotatebox{90}{\setcounter{subfigure}{12}\qquad \I= 1200 s}}
\subfloat[Exponential]
{
\includegraphics[width=\SimFigTrWidth]{fig/Tr-EXP-R07-I1200-alpha1-fixedC-appli0-platform-variation.fig}
}	
\subfloat[Weibull $k=0.7$]
{
\includegraphics[width=\SimFigTrWidth]{fig/Tr-WEIBULL-07-R07-I1200-alpha1-fixedC-appli0-platform-variation.fig}
}
\subfloat[Weibull $k=0.5$]
{
\includegraphics[width=\SimFigTrWidth]{fig/Tr-WEIBULL-05-R07-I1200-alpha1-fixedC-appli0-platform-variation.fig}
}
\\
\hspace{-1cm}
\subfloat{\rotatebox{90}{\setcounter{subfigure}{16}\qquad \I= 3000 s}}
\subfloat[Exponential]
{
\includegraphics[width=\SimFigTrWidth]{fig/Tr-EXP-R07-I3000-alpha1-fixedC-appli0-platform-variation.fig}
}	
\subfloat[Weibull $k=0.7$]
{
\includegraphics[width=\SimFigTrWidth]{fig/Tr-WEIBULL-07-R07-I3000-alpha1-fixedC-appli0-platform-variation.fig}
}
\subfloat[Weibull $k=0.5$]
{
\includegraphics[width=\SimFigTrWidth]{fig/Tr-WEIBULL-05-R07-I3000-alpha1-fixedC-appli0-platform-variation.fig}
}
\caption{Waste as function of the period $\Tnp$ for the different heuristics, with $\precision=0.4$,
  $\recall=0.7$, $\Cp=\Cr$, and with a platform of $2^{19}$ processors.}
	\label{fig.04.07.Tr.19}
\end{figure*}

\begin{figure*}
\subfloat[Exponential]
{
\includegraphics[width=\SimFigTrWidth]{fig/I-EXP-R85-alpha1-fixedC-appli0-platform16.fig}
}	
\subfloat[Weibull $k=0.7$]
{
\includegraphics[width=\SimFigTrWidth]{fig/I-WEIBULL-07-R85-alpha1-fixedC-appli0-platform16.fig}
}
\subfloat[Weibull $k=0.5$]
{
\includegraphics[width=\SimFigTrWidth]{fig/I-WEIBULL-05-R85-alpha1-fixedC-appli0-platform16.fig}
}
\caption{Waste as function of the prediction window I for the different heuristics, with $\precision=0.82$,
  $\recall=0.85$, $\Cp=\Cr$, and with a platform of $2^{16}$ processors.}
	\label{fig.082.085.I.16}
\end{figure*}

\begin{figure*}
\subfloat[Exponential]
{
\includegraphics[width=\SimFigTrWidth]{fig/I-EXP-R85-alpha1-fixedC-appli0-platform19.fig}
}	
\subfloat[Weibull $k=0.7$]
{
\includegraphics[width=\SimFigTrWidth]{fig/I-WEIBULL-07-R85-alpha1-fixedC-appli0-platform19.fig}
}
\subfloat[Weibull $k=0.5$]
{
\includegraphics[width=\SimFigTrWidth]{fig/I-WEIBULL-05-R85-alpha1-fixedC-appli0-platform19.fig}
}
\caption{Waste as function of the prediction window I for the different heuristics, with $\precision=0.82$,
  $\recall=0.85$, $\Cp=\Cr$, and with a platform of $2^{19}$ processors.}
	\label{fig.082.085.I.19}
\end{figure*}

\begin{figure*}
\centering
\hspace{-1cm}
\subfloat[Exponential]
{
\includegraphics[width=\SimFigTrWidth]{fig/I-EXP-R07-alpha1-fixedC-appli0-platform16.fig}
}	
\subfloat[Weibull $k=0.7$]
{
\includegraphics[width=\SimFigTrWidth]{fig/I-WEIBULL-07-R07-alpha1-fixedC-appli0-platform16.fig}
}
\subfloat[Weibull $k=0.5$]
{
\includegraphics[width=\SimFigTrWidth]{fig/I-WEIBULL-05-R07-alpha1-fixedC-appli0-platform16.fig}
}
\caption{Waste as function of the prediction window I for the different heuristics, with $\precision=0.4$,
  $\recall=0.7$, $\Cp=\Cr$, and with a platform of $2^{16}$ processors.}
	\label{fig.04.07.I.16}
\end{figure*}

\begin{figure*}
\centering
\hspace{-1cm}
\subfloat[Exponential]
{
\includegraphics[width=\SimFigTrWidth]{fig/I-EXP-R07-alpha1-fixedC-appli0-platform19.fig}
}	
\subfloat[Weibull $k=0.7$]
{
\includegraphics[width=\SimFigTrWidth]{fig/I-WEIBULL-07-R07-alpha1-fixedC-appli0-platform19.fig}
}
\subfloat[Weibull $k=0.5$]
{
\includegraphics[width=\SimFigTrWidth]{fig/I-WEIBULL-05-R07-alpha1-fixedC-appli0-platform19.fig}
}
\caption{Waste as function of the prediction window I for the different heuristics, with $\precision=0.4$,
  $\recall=0.7$, $\Cp=\Cr$, and with a platform of $2^{19}$ processors.}
\label{fig.04.07.I.19}
\end{figure*}

\end{document}